\documentclass{pas}

\usepackage{multirow}
\usepackage{float}
\usepackage{aas_macros}
\usepackage{placeins}

\begin{document}

\lefttitle{Compact Remnant Binary Populations in the Space Telescope Era}
\righttitle{Eastep, Eldridge, \& Bray}

\jnlPage{1}{4}
\jnlDoiYr{2021}
\doival{10.1017/pasa.xxxx.xx}

\articletitt{Research Paper}

\title{Quiescent and Interacting Compact Remnant Binary Populations in the Space Telescope Era}

\author{\sn{Erin} \gn{Eastep}$^{1}$, \sn{Jan J.} \gn{Eldridge}$^{1}$, and \sn{J. C.} \gn{Bray}$^{1}$}

\affil{$^1$Department of Physics, University of Auckland, Private Bag 92019, Auckland, New Zealand}

\corresp{E. Eastep, Email: erin.eastep.astro@gmail.com}

\citeauth{}
% Author1 C and Author2 C, an open-source python tool for simulations of source recovery and completeness in galaxy surveys. {\it Publications of the Astronomical Society of Australia} {\bf 00}, 1--12. https://doi.org/10.1017/pasa.xxxx.xx

\history{(Received xx xx xxxx; revised xx xx xxxx; accepted xx xx xxxx)}

\begin{abstract}
Since the discovery of Gaia BH1, interest in quiescent binary systems has grown. Two additional black-hole binaries have since been discovered alongside several candidate neutron-star binaries, yet the formation pathways of these systems remain poorly understood. Theoretical studies of these systems, together with the populations into which they evolve, such as X-ray binaries (XRBs), can help constrain their formation channels. We present predictions for the Galactic population of quiescent binary systems and their later evolutionary phases, including XRBs and post-interaction quiescent binaries. We combine stellar models from the Binary Population and Spectral Synthesis code (BPASS) with a synthetic Milky Way generated using Feedback In Realistic Environments (FIRE) simulations. We compare our results with existing observations to assess how well the synthetic population reproduces the observations. We find generally good agreement, with observed donor masses, compact-object masses, and orbital periods largely reproduced, except for the Gaia BH systems. Our results also suggest that some accreting compact remnants with low-mass donors and wide orbital periods may not be X-ray bright. Remaining discrepancies indicate that the \textsc{BPASS} initial parameter space should be expanded, the common-envelope evolution prescription revised, and magnetic braking included in low-mass donor models. We also predict the number of systems detectable in future Gaia data releases and by the Nancy Grace Roman Space Telescope (hereafter referred to as Roman). In particular, the Galactic Bulge Time Domain Survey is expected to yield a significant population of quiescent black-hole binaries. We predict that $\sim200-600$ candidate quiescent binary systems will be detected in the Galactic Bulge Time Domain Survey, while $\sim2500-4000$ candidate systems will be identified in Gaia DR5.
\end{abstract}


\maketitle

\section{Introduction}\label{INTRO}

It is observationally demonstrated that \textgreater50\% of stars are in binary or higher order systems and two thirds of high mass stars are affected by some sort of interaction in their lifetime \citep{Sana2012, Ps&Qs}. Given their population dominance, binary systems have a profound impact on the evolution of our Galaxy. The X-ray binaries (XRBs) formed by stars and compact objects are an opportunity to study an exciting, yet transitory, moment in the life of interacting binary systems. These are bright objects that have been studied observationally and theoretically extensively for some time \citep{Corral_Santana_2016,2024MNRAS.527.5023L,2023ApJ...954..212S,2003ApJ...597.1036P,2005MNRAS.356..401R}

The progenitors of XRBs are important as these binaries spend a considerable amount of their earlier evolution in quiescence. Starting with Gaia BH1, binary systems containing a luminous companion orbiting a black hole (BH) with little to no mass transfer and no X-ray flux have been observed (henceforth referred to as quiescent binaries), giving us insight into the birth masses of the compact remnants before they start accreting \citep[e.g][]{El_Badry2023-1, 2024PASP..136a4202N}. Thus, quiescent binary systems, are an important subgroup to study to understand the formation of compact remnants to further our understanding of these objects and their complete lifecycle. 

Furthermore, it is also useful to consider the evolution of systems after their time as an XRB. For example, a Wolf-Rayet (WR) star, around a black hole or neutron star are likely post-XRB objects. It should be noted that one WR XRB is known, however it is extragalactic. A more common end-phase for low mass and intermediate mass X-ray binaries (LMXRBs, IMXRBs) would be formation of a white dwarf-neutron star binary (WDNS). These typically take the form of millisecond-pulsars (MPSs) with a helium or carbon-oxygen white dwarf, or in some extreme cases an ultra-compact X-ray binary (UCXRB) \citep{2002ApJ...565.1107P, 1988A&A...191...57P, 1989A&A...208...52P}.

In this study, our aim is to create synthetic populations of binaries including compact remnants, including their quiescent and post-interaction phases as a resource to understand the evolution of compact remnants and their masses. To do this we use the Binary Population and Spectral Synthesis (\textsc{BPASS}) code suite \citep{2017PASA...34...58E,se2018} and the Feedback in Realistic Environment (\textsc{FIRE}) \citep{2014MNRAS.445..581H} software suite. The synthesis of this theoretical stellar population builds on earlier work concerning gravitational wave populations \citep{tang2024predictinggravitationalwavesignals} but now with a focus on electromagnetic observations from current and future space telescopes.

\textsc{BPASS} is a suite of codes that models the evolution of binary star systems and is used for the study of stellar populations and various properties related to them \citep[see ][for details]{2017PASA...34...58E,se2018}. \textsc{BPASS} uses detailed stellar models evolved with a custom version of the Cambridge STARS code. \citep{Eldridge_2008, 2017PASA...34...58E}. This allows for the effects of binary interactions on the donor star to be modelled in detail and thus an accurate understanding of stability of mass transfer. Over the past two decades, \textsc{BPASS} has been validated against a slew of different observations both within our Galaxy, and out to the furthest reaches of the observable universe \citep[e.g.][and references therein]{2017PASA...34...58E,2022EldridgeStanway,Kobayashi_2023}.

\textsc{BPASS} alone only models the evolution of stars. To predict Galactic populations, we need a model of the Galaxy. We make use of a Milky Way-like galaxy model from the Feedback in Realistic Environment (FIRE) software suite \citep{2014MNRAS.445..581H}. Specifically, we utilise the \textsc{m12i}  simulation from the `Latte' model. \citep{2023ApJS..265...44W} Based on the \textsc{FIRE-2} code from \citet{2018MNRAS.480..800H} with the code \textsc{GIZMO}\footnote{\url{http://www.tapir.caltech.edu/ phopkins/Site/GIZMO.html}} \citep{2015MNRAS.450...53H}, which is used to solve the hydrodynamics equations using the mesh-free Lagrangian Godunov `MFM' method. By combining the \textsc{BPASS} stellar models with the \textsc{FIRE} simulation, we are able to synthesise populations of different compact remnant binaries that host a BH or neutron star (NS) with a luminous stellar companion. We then compare them to known compact object binaries to gain insight into the validity of the stellar models and identify where the underlying physics may need to be refined.

The extant observations have complex and difficult to describe observational selection effects, however, our synthetic populations effectively provide the underlying populations allowing us to predict future yields of observational studies and surveys where the selection effects are better constrained. Future discoveries of quiescent binaries will complement the current known systems and increase our understanding of these systems. Via Gaia DR3 and preliminary DR4 data, three quiescent black-holes binaries \citep{El_Badry2023-1, El_Badry2023, BH3_2024} and 20+ quiescent neutron star binary candidates \citep{elbadry2024populationneutronstarcandidates} have been discovered to date. While Gaia observations ceased on the 15th of January 2025, the full mission data has yet to be analysed and released. Gaia DR4 will contain data collected from 25 July 2014 through 20 January 2020 while Gaia DR5 will contain all mission data. In this project we predict the possible populations that will be observed in Gaia DR5 in anticipation of future yields. 

Another exciting future mission is the Roman, NASA's latest infrared space telescope. The primary science instrument on board is the Wide Field Instrument (WFI), a 300-megapixel near infrared instrument (0.5 to 2.3 microns) capable of imaging and slitless spectroscopy. This detector, along with the planned high-cadence surveys allows Roman to achieve sub-milliarcsecond precision with comparable sensitivity and resolution to the Hubble Space Telescope (HST) \citep{committee2025romanobservationstimeallocation}. Roman's main science goals include the study of exoplanet demographics as well as dark matter and dark energy. To achieve these goals, the mission consists of four Core Community Surveys that account for roughly 75\% of observing time over Roman's primary five-year mission; The Galactic Plane Survey, The High Latitude Wide Area Survey, The High Latitude Time Domain Survey, and The Galactic Bulge Time Domain Survey.
\begin{figure}
\includegraphics[width=\columnwidth]{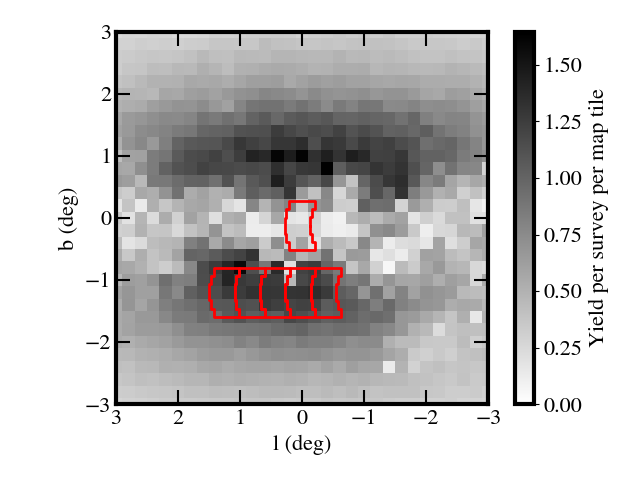}
\caption{The recommended field for the GBTDS from \protect\citet{committee2025romanobservationstimeallocation}. The contour and colour bar represent the expected yields for $\rm{1M_{Earth}}$ planets.}
\label{fig:GBTDSField}
\end{figure}
Of particular interest to the authors is the Galactic Bulge Time-Domain Survey (GBTDS). The survey will observe a $\sim0.2 \, {\rm deg}^2$ field at the centre of the Milky Way to study microlensing events, primarily in search of exoplanets. The GBTDS will consist of six seasons, each of which lasts $72$ days where the entire field will be observed with a cadence of $\sim12.1$ minutes, recording astrometry and photometry simultaneously. High-cadence surveys such as these collect data useful for other time-domain phenomena. One such phenomenon is the `wobble' observed by quiescent binaries like the Gaia BH systems. While not the sole method of detection, an astrometric wobble serves as an important indicator to an invisible companion to a star. \citet{RomanWP} investigated Roman's potential for detecting these quiescent compact object binaries via the GBTDS and found that its capabilities put it on par with, or even exceed the capabilities of Gaia. Given the exciting capabilities of this mission we also aim to predict the number of quiescent compact remnant binaries that it might discover.

The structure of this paper is as follows: in Section \ref{methods} we outline the method of combining \textsc{BPASS} and FIRE, we describe how we identify the different phases of compact remnant binary evolution, outline the extant observational data we use to compare to the synthetic populations and finally we describe our methodology for simulating the space telescope surveys of our synthetic populations. In Section \ref{results} we go through the different evolutionary phases and discuss how the synthetic and observed populations compare. In Section \ref{SpaceTeleResults} we describe the results of our simulated surveys on the synthetic populations. Before discussing the results in Section \ref{disc} and finally presenting our conclusions in Section \ref{concl}. 

\section{Numerical Method}\label{methods}

\subsection{Combining BPASS and FIRE}\label{BPASSandFIRE}

To generate a synthetic galactic population requires combining the results of the predictions for the evolution of interacting binary systems and a model of a Milky-way like galaxy. The former determines for a single burst of star formation, when and how many of the various binary systems with a compact remnant component are produced. The latter determines how many stars are formed, when they are formed, and at what metallicity. Combining these allows for the prediction of the desired binary star populations. In order to incorporate \textsc{BPASS} with FIRE, we build on the work of \citet{tang2024predictinggravitationalwavesignals}. This method has proved effective for generating theoretical galactic populations of potential sources for the future Laser Interferometer Space Antenna (LISA) gravitational wave observatory \citep{2017arXiv170200786A}.

In depth information and analysis of \textsc{BPASS} can be found in \citet{2017PASA...34...58E} and \citet{se2018}. Here we use the fiducial models over the full range of metallicity mass fractions from $Z=10^{-5}$ to $0.040$. The initial mass function (IMF) and initial binary parameter distribution used are a Kroupa IMF \citep{KroupaIMF} with a maximum stellar mass of 300~M$_{\odot}$, with the initial binary parameters taken from \citet{Ps&Qs}. The neutron-star kick employed in the population synthesis is based on \citet{2005MNRAS.360..974H}. Mass-transfer stability is determined from how the donor's radius responds to the onset of Roche-lobe overflow (RLOF). Common-envelope evolution is triggered if the radius of the donor exceeds the separation of the binary and is modelled with a $\gamma$-like common-envelope prescription \citep[see supplementary information of][]{Stevance_2023}. \textsc{BPASS} does not inherently classify compact remnants. Thus, if the model's remnant mass is between $1.39~\rm{M_{\odot}}$ and $3~\rm{M_{\odot}}$ then we assume the compact object is a NS and any mass larger than $3~\rm{M_{\odot}}$ is assumed to be a BH. For the models with a neutron star, the accretion is limited by the Eddington luminosity whereas black hole accretion can exceed the Eddington limit. We note this work builds on that of \citet{2025MNRAS.542.2087B} where we predicted the X-ray emission from \textsc{BPASS} populations for comparison to unresolved stellar populations. In this study, we are now considering the individual sources that might exist within our Galaxy. However, we do not consider modelling the emission of the accretion given the complexity of selection effects from observational surveys and also the uncertain duty cycle of these accreting compact objects. Instead, we are more concerned with the quiescent binary population.

% \textcolor{blue}{Put the two blue paragraphs from into here and reword the below to remove duplication} 
In essence, \textsc{FIRE} is a list of ``star'' particles each with an initial mass, age, metallicity, and location within this synthetic-galaxy. These particles have a minimum mass resolution of $7070\,\rm{M}_\odot$ and a spatial resolution of $1\,\rm{pc}$ \citep{Sanderson_2020}. Specifically, we make use of the \textsc{m12i} Milky Way-like galaxy model from `Latte' Simulation \citep{2023ApJS..265...44W} in ``FIRE-2'' \citep{2018MNRAS.480..800H}. This version was developed with the GIZMO\footnote{\url{http://www.tapir.caltech.edu/ phopkins/Site/GIZMO.html}} code \citep{2015MNRAS.450...53H}. GIZMO is used to solve the hydrodynamics equations using the mesh-free Lagrangian Godunov `MFM' method. For our simulation, each of these particles is interpreted to be star clusters, or simple stellar populations of a single metallicity and age. From this assumption, we use the particle's mass, age and metallicity to identify which stellar binary models exist in that cluster. We use the metallicity to select which set of \textsc{BPASS} models to use by selecting the \textsc{BPASS} metallicity set which is closest to that from the \textsc{FIRE} results. We then predict the numbers of each binary type that exist in the star particles by using the age and the mass of the star particle. Here we use \textsc{BPASS} model weights that are calculated assuming the fiducial \textsc{BPASS} initial IMF and initial binary parameter distribution as detailed above. The number of expected systems in each time bin (which are spaced in bins of $\log(T/{\rm yrs})$ of 0.1~dex size) varies from approximately $10^{-20}$ to $10^{-4}$ per M$_{\odot}$ of star formation. We use \textsc{BPASS} timebins that are 0.1~dex in size that match the age of the star particle, we count each model in that time bin and then multiply this number by the star particle mass. If an integer value of stars is not reached, we continue to sum the weight to a running total until an integer value is reached. At which point, the model that is selected is included. Any remaining weight left over is kept as a remainder carried over to the next model. This is done in order of the star particle age such that the excess values are carried over to particles of similar age. We then go through and record specific objects by linking them to stellar models. It should be noted that we only consider the \textsc{BPASS} secondary stellar models as these are those with a compact remnant companion. These are models that continue the evolution of the system after the death of the first star. While \textsc{BPASS} does calculate when RLOF occurs, the stellar evolution models do not provide identification of the binary types and so we do this as listed below in post-processing.

\subsection{Identification of the compact remnant binary subtypes}\label{binaryspeciesmethod}

An important element of this work is how to determine which models are the different binary ``species'' that we are interested in. It is useful to think of these systems in the chronological order in which we expect a binary system to experience these phenomena following the first supernova in the systems that were not unbound by the supernova kick. The binary species we consider here are: quiescent compact remnant binaries where the objects have yet to interact, wind-fed accreting XRBs, Be star XRBs (a subtype of wind-fed XRBs), RLOF accreting XRBs, compact remnant binaries experiencing a common-envelope evolution (CEE) phase, and post-interaction quiescent binaries - these include systems where the luminous companion has evolved to become a white dwarf as well as other systems which are evolving towards core-collapse. 

To identify the compact remnant binary types we use the following criteria:
\begin{enumerate}
    \item Pre-interaction: We select models that have not interacted and are not currently interacting as defined by any of the following categories. We do allow wind-fed accretion so long as the rates are low enough so that the effective temperature of the accretion is $<10^6~\rm{K}$, for details of how this temperature is calculated see the next binary type. This category practically includes all binary systems with a compact remnant that do not belong to one of the following categories. 
    
    \item Wind-fed: These are systems where the surface conditions are such that strong stellar winds occur. If the winds of these systems have suitable parameters in density and velocity then mass transfer onto the compact object would be sufficient for accretion. This is the first of two main methods of mass transfer considered. We adopt the Bondi-Hoyle-Lyttleton model \citep{1944MNRAS.104..273B} outlined in \citet{2002MNRAS.329..897H} to determine if the winds from the donor are sufficient. We make use of the mass-loss rate from the \textsc{BPASS} stellar models for this calculation \citep[see][for details]{2017PASA...34...58E}. For the wind velocities, we use the parameters from \citet{2002MNRAS.329..897H} for all stars except Wolf-Rayet stars where we use the wind velocities from \citet{2000A&A...360..227N}. With the wind parameters from the stellar models, we calculate the mean accretion rate: 
\begin{equation}\label{BH1944}
\langle\dot{M}_{2A}\rangle=\frac{-1}{\sqrt{1-e^{2}}}\left[\frac{\textrm{G}M_{2}}{v^{2}_{W}} \right]^{2}\frac{\alpha_W}{2a^{2}}\frac{1}{(1+v^{2})^{3/2}}\dot{M}_{1W}
\end{equation}
where
\begin{equation}\label{vsq}
    v^{2}=\frac{v^{2}_{\rm orb}}{v^{2}_{W}}
\end{equation}
\begin{equation}\label{v2orb}
    v^{2}_{\rm orb}=\frac{\textrm{G}M_{\rm{tot}}}{a}
\end{equation}
\begin{equation}\label{v2wind}
    v^{2}_{W}=2\beta_{W}\frac{\textrm{GG}M_{1}}{R_{1}}
\end{equation}
Where $a$ is the separation of the binary. \(\beta_{W}\) is dependent on the stars spectral type and we determine this based on the effective temp of the star in BPASS. \(\alpha_{W}\) is set to $\frac{3}{2}$ in agreement with \citet{1988A&A...205..155B} and the lower limit provided within. It should be noted that when mass leaves the donor, we assume that none is lost and all of it is accreted onto the CO. Following the work of \citet{2025MNRAS.542.2087B}, we use this accretion rate to calculate the effective temperature of the accretion, following:
\begin{equation}\label{Teffeq}
    T_{\rm eff}^4= \frac{3\textrm{G}M_{2}\langle\dot{M}_{2A}\rangle}{8\pi R_{2}^3}
\end{equation}

Provided that $T_{\rm eff}\geq10^{6}K$ and the system has an initial period $<10^4$ days, we consider the system to be a wind-fed X-ray binary. The temperature cut-off allows us to filter out systems with inadequate accretion rates as these are unlikely to be detectable in X-rays. We note that we only consider systems to be wind-fed if they have yet to fill their Roche-lobe. Finally, we do not include any mass cut-off to allow for all donor masses. It should be noted that we don't model wind-fed accretion beyond this, meaning we do not model direct impact accretion and assume that the material forms some sort of accretion disk.

    \item Be XRBs: Be XRBs are unique and well studied subset of wind-fed XRBs. Therefore we separate these out using the same classification scheme as in \citet{2025MNRAS.542.2087B}. We select models with the following parameters:
    \begin{itemize}
        \item The donor has accreted more than 5\% of it's initial mass from the primary star that formed the compact remnant.
        \item The mass is between $6\rm{M_{\odot}}$ and $\le30.1\rm{M_{\odot}}$. 
        \item The donor star is on the main sequence, determined by having no helium core.
        \item The donor star is not a WR star, i.e.  the star has a surface hydrogen mass fraction $>0.4$.
        \item The surface temperature is within the following range $4.52 \ge\log (T_{\rm eff}/{\rm K}) \ge 4.04$.
        \item The orbital period is between 6 and 1000 days.
        \item The compact object's mass must be $>1.39\rm{M_\odot}$. 
\end{itemize}

    \item RLOF: These are the most common type of interacting compact remnant binary we find in our models, and they make up the majority of XRBs observed in the Galaxy. Any system where the stellar object has filled its Roche-lobe and the donor is accreting onto the compact remnant is considered to be a RLOF system unless the star has entered a CEE phase (see next type). We also separate these systems into X-ray bright and X-ray faint binaries. This is done by determining the effective temperature of the accretion disk in the models in a similar fashion to the wind-fed systems. We do this by taking Equation \ref{Teffeq} and use the mass transfer from RLOF in the stellar models. For NSs this mass transfer rate is limited by the Eddington luminosity, while for BHs no limit is imposed and all mass is accreted. If the disk $T_{\rm{eff}}$ is $>10^6K$ than we consider it to be X-ray bright, otherwise it is flagged as X-ray faint. 
    
    \item CEE: We define CEE binaries as systems where the radius of the stellar object is greater than or equal to the separation of the masses of the two objects in the binary. Effectively the compact remnant in these systems is engulfed by the donor star and we expect such sources to be difficult to observe. Modelling of such systems is difficult and inherently a dynamical process \citep[e.g. see recent hydrodynamical models by][]{2025A&A...697A..68G}. The \textsc{BPASS} CEE model limits the mass loss rate for numerical stability and so the length of the CEE phase is overestimated by a few orders of magnitude. Therefore the number of these systems will also be overestimated.
    
    \item Post-interaction: We look for systems that have previously been flagged as one of the previous interacting types in the past but are currently not interacting. We split the results of this into systems where a white dwarf is now present alongside a NS or BH versus systems that still have a stellar companion supported by nuclear fusion.

\end{enumerate}

\subsection{Observational Data}\label{obs}

It is important to be able to compare our synthetic population to observational data to validate our predictions. Our models should predict the systems known, however, any mismatch is also useful as it reveals where further refinements will be needed in the \textsc{BPASS} stellar models. We list the observations we use as follows. 

First, we consider the quiescent binaries, which are a relatively small sample size to date given they are difficult to observe. This arises due to their lack of X-ray emission and can only be definitively identified by careful study of the stellar companion's orbit. The first observations we include are the quiescent binaries already found in extant Gaia data releases. These notably include Gaia-BH1 \citep{El_Badry2023-1}, Gaia BH-2 \citep{El_Badry2023}, and Gaia-BH3 \citep{BH3_2024}. The detection of these sources was the motivation for this study. We supplement these sources with the NS quiescent binary candidates that were also discovered in Gaia observations from \citet{elbadry2024populationneutronstarcandidates}. While they are not included, it should be noted that \citet{2026arXiv260806453E} reports on over $200$ new compact object binary sources in Gaia DR3. 

Furthermore, we also include individual binaries taken from \citet{Paduano_2021}, which includes three sources located in the star cluster NGC 3201. They are ACS ID \#5132 \cite[first reported in][]{Giesers2019}, ACS ID \#12560 \citep{Giesers2018}, and ACS ID \#21859 \citep{Giesers2019}. These are all binary systems with strong evidence that they house BHs. Interestingly, \citet{Giesers2019} reported ACS ID \#21859 to have partially filled H$\alpha$ lines without X-ray or radio counterparts. \citet{Paduano_2021} assumed this source to be a system where any accretion takes the form of wind caught by the black hole with low accretion rates. Assuming similar properties for the other two systems, we compare these systems to our quiescent binary population.

The observations we use to compare to our synthetic X-ray binary populations, either from wind-fed XRBs or Roche-lobe overflow XRBs, are from the database provided by the Binary rEvolution\footnote{\url{https://github.com/Binary-rEvolution}} project. The goal of the project is to study the current population of XRBs in the Galaxy as a tool in understanding their role as progenitors of double compact mergers. It consists of two databases, one of high-mass XRBs (HMXRBs) and the other low-mass XRBs (LMXRBs). In addition to the masses of the objects and their periods, the databases also provides a classification of what species of system if known (e.g. a BeXRB label for some within the HMXRB catalogue). The HMXRB catalogue holds 100+ entries and the LMXRB catalogue holds 300+ entries. For both catalogues we only use sources where both the donor and remnant masses are known.

We then supplement the above samples with the Australia Telescope National Facility's (ATNF) Pulsar Catalogue v2.4.0\footnote{\url{http://www.atnf.csiro.au/research/pulsar/psrcat}} \citep{Manchester_2005}. The ATNF Pulsar Catalogue includes all published rotation-powered pulsars, excluding accretion-powered pulsars. Since the masses of the NSs aren't provided for each system we aren't able to make meaningful predictions regarding the remnant masses from this database. In the cases where a remnant mass is missing, we assume a lower bound of $1.35\rm{M_\odot}$ as that is the NS mass assumed by ATNF when determining the minimum companion mass. The catalogue also provides classifications for the systems which allows us to separate the sample into systems with a main-sequence companion and those with a white dwarf companion. Thus, we are able to compare these binaries to the post-interaction quiescent binary population as well as the post-interaction quiescent binaries with a WD companion.

Lastly, we note that we need to take account of the fact that the \textsc{BPASS} models only have circular orbits. Therefore we recalculate the periods of the observations so that the system has a period in the circular orbit with the same separation as the semi-latus rectrum of the eccentric orbit with the following equation:
\begin{equation}\label{circularise}
    P_\text{circ}=P_\text{ecc}(1-e^{2})^{\frac{3}{2}},
\end{equation}
Where \textit{e} is the observed eccentricity of a system. We only do this if an eccentricity value is available for a given system. The assumption of circular orbits in the \textsc{BPASS} models arises from \citet{2002MNRAS.329..897H} and the assertion that orbits generally circularise before the onset of Roche-lobe overflow. That is to say, orbits with the same semi-latus rectra evolve down a similar path. \textsc{BPASS} also assumes baryonic masses for NSs, so we convert the observed NS masses to baryonic where applicable. This is done with Equation \ref{baryonicgravMass} from \citet{2020FrPhy..1524603G}:
\begin{equation}\label{baryonicgravMass}
    {M_{\rm b}=M_{\rm g}+ {\rm{A}} \times{M_{\rm g}^2}},
\end{equation}
where A is a constant of $0.075$ and $M_g$ and $M_b$ are in solar masses. 

\subsection{Simulated Space Telescope Surveys}\label{spacetelesim}

The methods outlined in Section \ref{BPASSandFIRE} yield a synthetic Milky Way-like galaxy where the stellar population is linked directly to \textsc{BPASS} secondary stellar models. This means we are able to retrieve any feature or parameter from the \textsc{BPASS} model associated with each binary system. This enables us to not just predict the compact remnant binary populations and compare to known systems, but also to determine, for example, their magnitudes. Thus, we can make predictions on what we can expect these surveys to detect by undertaking a mock survey in our synthetic galactic population. In this work we will consider both Roman's Galactic Bulge Time Domain Survey \citep[][GBTDS]{committee2025romanobservationstimeallocation} as well as the expected final results from the Gaia's astrometry mission, DR5 \citep{2023A&A...674A...1G}. As stated in Section \ref{obs}, Gaia has already found several quiescent binary systems with a compact object \citep{El_Badry2023-1, El_Badry2023, elbadry2024populationneutronstarcandidates, BH3_2024}. Additional quiescent binaries are expected to be found in DR3 with the promise of many more being found in DR5 \citep{Chawla_2022, 2024NewAR..9801694E, 2025arXiv250821805C, 2025PASP..137d4202N}. While no in depth study has been done to predict a similar population from the GBTDS, it has been theorised that the survey has potential to find many such systems \citep{RomanWP, 2024arXiv240614767K}.

\subsubsection{Roman's Galactic Bulge Time Domain Survey}\label{mockGBTDS}

Here we outline how we perform mock surveys upon the synthetic populations. First, we consider Roman's Galactic Bulge Time Domain Survey (GBTDS) to be undertaken by the Roman. The survey will consist of six high-cadence seasons $(\sim12.1\ \rm{mins})$, four low-cadence seasons $(\sim3\ \rm{days})$, multiband photometry snapshots, and spectroscopic snapshots. The primary science directive of the GBTDS is the study cold exoplanets and their demographics \citep{committee2025romanobservationstimeallocation}. However, the nature of this survey and Roman's sub-milliarcsecond precision will make the GBTDS a transformative dataset for Galactic time-domain astronomy. Therefore it is useful to determine how many, if any, compact remnant binaries it will detect.

Our synthesis returns a dataset where stars are given three-dimensional Cartesian coordinates on a kiloparsec (kpc) scale with the Galactic Centre at the origin. We assume the Earth is at ($8.5,0,0$~kpc), from which we can convert the population into Galactic coordinates. Following the final recommendation from the Roman Time Allocation Committee \citep{committee2025romanobservationstimeallocation} and imaged in Figure \ref{fig:GBTDSField}, we mask off any sources not within the line of sight for the survey. Additionally, any sources with periods greater than the initial mission time (five years) are not considered. This is a very conservative condition as the wobble may be significant enough for binaries with orbits longer than the mission time to still be detected. Precedent for this already exits as Gaia BH3 was found in preliminary Gaia DR4 data ($\sim66~\rm{months}$ of observations) with an orbit of $11.6~\rm{years}$ \citep{BH3_2024}. 

A key detail we use to determine observability is the apparent magnitude of the nuclear-fusing star in the binary. The \textsc{BPASS} models contain absolute magnitudes, thus we use those and distance between the Earth and the source star to calculate the apparent magnitude. Then, given that we are observing through the Galactic plane, we must account for dust and extinction. To do so we use the \texttt{mwdust} python package and apply the \texttt{combined19} dustmap \citep{2016ApJ...818..130B}. The \texttt{combined19} map provides extinction in all relevant filters (J,H,K,V,I) and has been used in similar Gaia simulations \citep{Chawla_2022, 2025arXiv250821805C}. The map returns an extinction value that depends on sky location and distance from the Earth so we can factor extinction into the apparent magnitude. 

The GBTDS will primarily be carried out in Roman's $\rm{F146}$ filter. After accounting for extinction and distance we follow the work of  \citet{2023ApJS..269....5W} to estimate $\rm{F146}$ magnitudes from \textit{JHK} magnitudes. The work in \citet{2023ApJS..269....5W} uses the AB magnitude system which is not included as part of the \textsc{BPASS} model standard magnitudes. Thus we convert the \textsc{BPASS} Vega mags to AB mags.

Our criteria for detectability requires that the``wobble'' of the binary system to be greater than the astrometric positional error as a function of magnitude. For the Roman GBTDS, we use the results of \citet{2026PASP..138d4507M}\footnote{\url{https://github.com/KevinMcK95/gaia_roman_astrometry}} and apply their model for a best case scenario of a $0.01\%$ pixel-floor post calibration for Roman. It is important to note that the result is given in the Gaia G band and the authors assume zero colour pairwise across the different filters, which we factor into our work. Additionally, at magnitudes $>22$ sources will become noise dominated \citep{2024ApJ...965..138F, 2026PASP..138d4507M} which is reflected in Figure \ref{fig:DR3DR5}. It should also be noted that we do not account for any complications from proper motion or parallax in our simulation and assume that these parameters can be adequately resolved. \citet{2026PASP..138d4507M} reports that for the GBTDS the parallax and proper motion errors should be an order of magnitude below that of the positional error so we feel that this assumption is acceptable. We also do not account for any crowding phenomena in either survey and assume that any system in our model can be observed without confusion from another source. 

The next step is to take all the systems within the the GBTDS field for Roman, and calculate their astrometric ''wobble" which we compare to the astrometric precision $\rm{\sigma_{ast}}$ computed for each source. As stated in Section \ref{obs}, \textsc{BPASS} models exclusively have circular orbits. Thus, the wobble we compare the precision to is the semi-latus rectum of the modelled system. If the wobble is greater than the error we consider it ``resolvable". Additionally, we present a separate population where we require the wobble to be $>3\sigma_{\rm{ast}}$ for a more ``pessimistic'' scenario. \citet{2022ApJ...931..107C} performed similar analysis for resolvable systems in Gaia DR3. 

\subsubsection{Gaia Astrometric Observations}\label{mockGaia}

We approach creating a mock Gaia survey result through a similar process. We start with the calculation of magnitudes in Gaia's G-band with the \textit{V} and \textit{I} magnitudes from the \textsc{BPASS} models. As Gaia is an all sky survey, we allow any system within our population to be assessed. After calculating apparent magnitude and accounting for extinction we use the results of Table 3 from \citet{2010A&A...523A..48J} to calculate the stars apparent G-band magnitude. We then calculate the binary system's wobble the same way we do for Roman so that we can compare it to Gaia's $\rm{\sigma_{ast}}$ for each system. Gaia DR3 presented data from $34$ months of data and Gaia DR4 represents $66$ months of data. In the future, DR5 will include the full $126$ months of data. Like our GBTDS simulation, we only consider systems where the orbital periods are less than time span of DR3 and DR5 in their respective simulations. As outlined above, this is a more conservative condition as Gaia BH3 was found to have an orbit of $11.6~\rm{years}$ \citep{BH3_2024}. It is also believed that with the extended mission of Gaia, DR5's observational baseline may allow for orbits of up to 20 years to be constrained \citep{2024OJAp....7E..38E}, or potentially even longer should methods like those in \citet{2023ApJ...946..111A} be implemented. 

\begin{figure}
    \centering
    \includegraphics[width=\columnwidth]{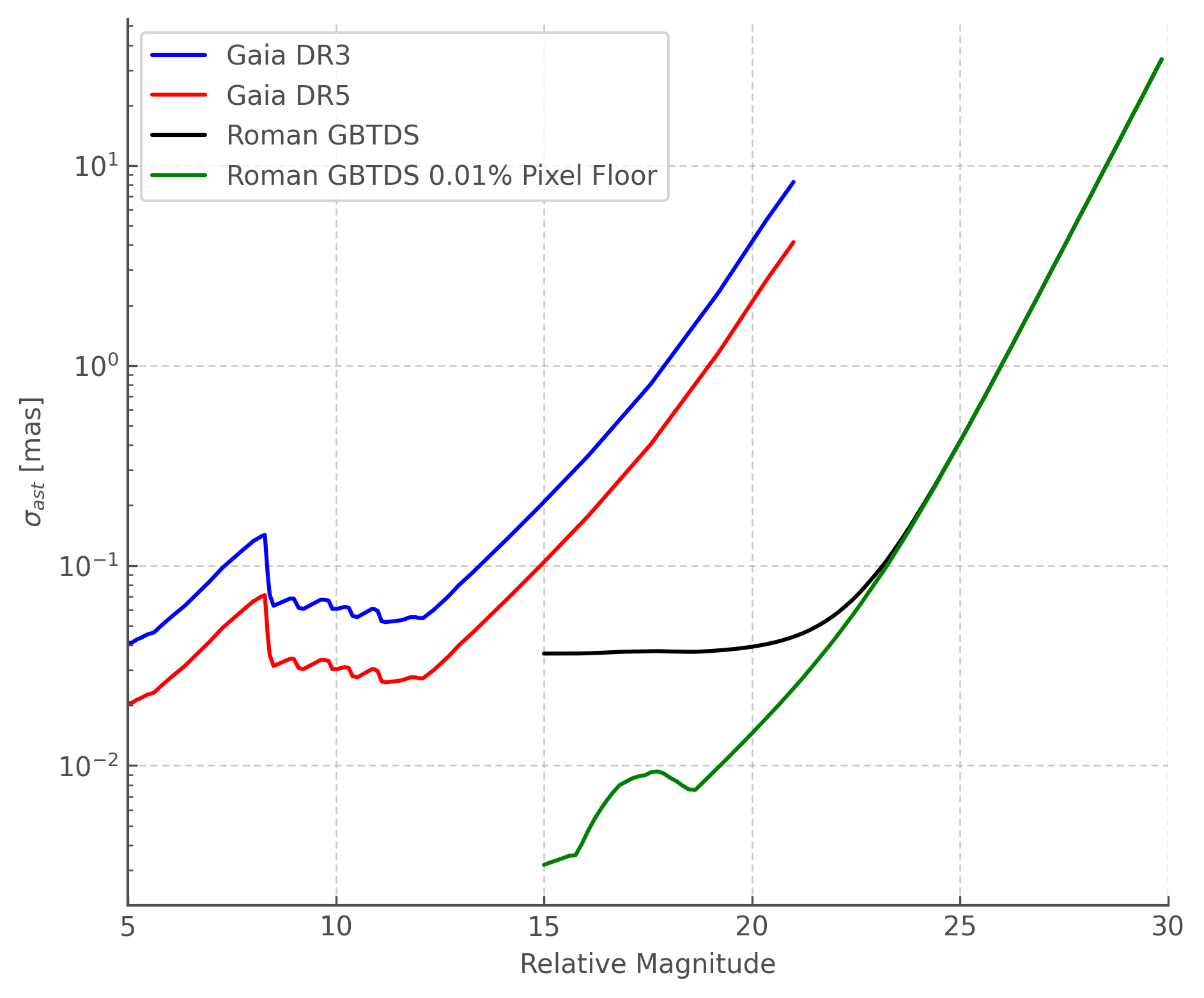}
    \caption{Simulated astrometric precisions used. The Gaia DR3 data from \protect\citet{2023JCAP...07..037C} is represented by the blue line. The extrapolated DR5 function is represented by the red line reflecting the $\sim2$ times improvement over DR3 \protect\citep{2023A&A...674A...1G}. We also include results from \protect\citet{2026PASP..138d4507M} for the Roman GBTDS (black line) as well as the $0.01\%$ pixel floor (green line).}
    \label{fig:DR3DR5}
\end{figure}

It is expected that the astrometric precision should improve with mission time at a rate of $\sqrt{t}$ \citep{2023A&A...674A...1G}. By assuming each data release contains twice the amount of observational data as the previous release, we assume a best case scenario where the precision of DR5 should reduce by a factor of $\sim2$ over DR3. Using the precision data for Gaia DR3 presented in \citet{2023JCAP...07..037C}\footnote{\url{https://github.com/mkongsore/BlipFinder/blob/main/PaperPlots/error_plot.ipynb}}, we extrapolate a theoretical Gaia DR5 precision curve. This can be seen in Figure \ref{fig:DR3DR5}. 

As for the GBTDS, but now for all sources in the entire Galaxy, we calculate their astrometric wobble and compare to the astrometric precision computed for each source. If the wobble is greater than the error we consider it ``resolvable" as well as a similar ``pessimistic" population. 

We repeat the above process at 16 different positions around our synthetic galaxy while maintaining a radius of $8.5~\rm{kpc}$ from the Galactic Centre to account for the spatial asymmetry within our model galaxy. The population counts presented in Table \ref{tab:GaiaRomanNumsTable} contain the mean and standard deviation of this dataset. 

\subsubsection{Mock Space Telescope Observations Summary}

To summarise, below are the conditions we use for determining which sources in our mock observations would be detected by the surveys:
\begin{enumerate}
    \item There must be a compact object as well as a nuclear fusing star in the binary.
    \item Specifically for GBTDS, the source must be within the line of sight of Roman. For Gaia it must be in the Galaxy.
    \item The period of the binary must be less than or equal to the mission length so a full orbit can be resolved (5 years for Roman, 10.5 years for Gaia DR5)
    \item Magnitudes must be within the limits of the astrometric precision functions found in Figure \ref{fig:DR3DR5}.
    \item The angular spacing of the orbital wobble on sky must be greater than the precision calculated for the target to be resolvable.

\end{enumerate}

we note that these criteria should be interpreted as optimistic detectability thresholds rather than a complete inclusion of the observational selection function where other factors such as cadance, crowding and parallax degeneracies would need to be included. 

\section{Results - Synthetic Galactic Populations}\label{results}

\begin{table*}
    \centering
    \label{tab:enwcts}
    \begin{tabular}{cccccc}
         \hline
         Binary Type&  Total Counts&  HM-BH&  LM-BH&  HM-NS& LM-NS\\
         \hline
         Pre-interaction& 1567784&  841& 5433&  7516& 1521405\\
         Be& 2641&  23&  2&  638& 1978\\
         Wind-Fed& 51816&  79&  34228&  73& 17395\\
         X-ray Bright RLOF& 147498&  25&  48934&  201& 98297\\
         X-ray Faint RLOF& 709544&  16&  392930&  64& 315473\\
         CEE& 11047&  0&  13&  644& 10336\\
         Post-interaction& 352693&  8&  82220&  29& 268079\\
         WD& 5938253&  0&  2212942&  0& 3609858\\
         \hline
    \end{tabular}
    \vspace{0.15cm} 
    \caption{Total binary populations from the \textsc{BPASS} models in our MW-like galaxy. We impose a cut-off of $10$M$_\odot$ for high mass and low mass donors. Additionally, if a system has a remnant mass $\ge3$M$_\odot$, than we consider its remnant to be a black hole. If the remnant mass is between $1.39$M$_\odot$ and $3$M$_\odot$ then we consider the remnant to be a neutron star.}
\end{table*}

Here we present the \textsc{BPASS} theoretical predictions for the synthetic populations of the different compact object binary systems as specified in Section \ref{binaryspeciesmethod}. We also compare these predictions to observational samples described in Section \ref{obs} to evaluate the \textsc{BPASS} predictions. These results are presented in ``chronological'' order of a system's lifetime from pre-interaction, to mass transfer stages, to post-interaction. The populations are presented as density distributions with bin sizes of 0.1 dex in the donor and remnant masses and 0.2 dex in period. For the spatial distribution, the bins are 1-kpc wide. 

It is important to note that the observational samples are heterogenous with ill defined selection effects. Thus, in our comparison we do not expect to exactly reproduce observed numbers but validate that our models fill a similar parameter space to the observations. Additionally, in the catalogues some observed systems may have mass estimates without period data, or vice versa. Therefore, we plot observations in the panels where they have the relevant data and do not plot them if we do not have that data. However, the exception to this is for systems with neutron stars, where its mass is unknown, where we select a lower mass of $1.39\rm{M_{\odot}}$. 

Finally, we present the full number of objects we predict in each population in Table \ref{tab:enwcts}. As we discuss each population below we will first discuss the total number of systems we predict before discussing their distributions. 

\subsection{Pre-interaction}\label{preint}
\begin{figure*} 
\includegraphics[width=2\columnwidth]{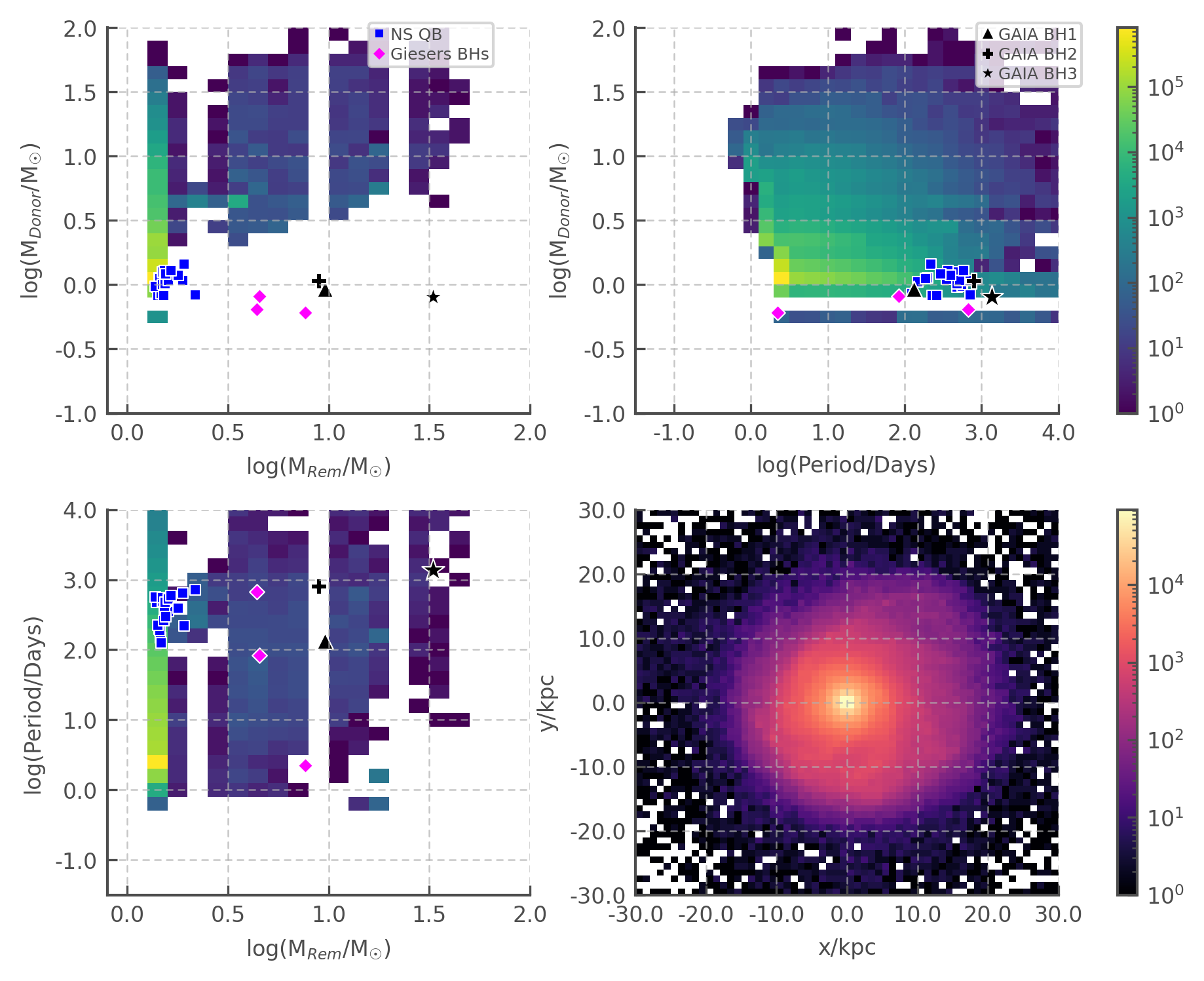}
\caption{Pre-interaction quiescent binaries synthetic populations compared to observed systems. The Gaia BH systems are represented by the three black points and the Gaia NS systems reported in \protect\citet{2024ApJ...973...75C} are the blue points. The NGC 3201 sources with data from \protect\citet{Paduano_2021} are the magenta points.}
\label{fig:quin0}
\end{figure*}

We start with the pre-interaction systems post-primary supernova (SN). Table \ref{tab:enwcts} indicates that the pre-interaction quiescent compact object binary systems are the second most populous within the different populations, with over 1.5 million objects. The majority of systems are binaries with low mass stars and a NS. The number of objects with a black hole however is still significant at about 6200 systems in the model galaxy. The lower right panel of Figure \ref{fig:quin0} shows that the density of these objects peaks in the Galactic centre while the density of objects near the Sun is of the order of several hundred per square kpc. With maybe a few to several black hole systems expected per square kpc. 

In this figure we can also see the distribution of the parameters. Overall we see a somewhat positive correlation between the model population and the Gaia NS observations \citep{elbadry2024populationneutronstarcandidates}. However, the NS masses are slightly higher than those in our model populations. Additionally, we do not reproduce systems similar to the Gaia BH systems. While we do predict binaries with similar donor masses and periods, we have no black holes with masses below approximately 3~M$_{\odot}$. This is an artifact of the \textsc{BPASS} initial parameters grid as the minimum initial mass ratio (i.e., $q=M_2/M_1$) modelled in the population is $0.1$. When this is combined with the fact that the minimum initial progenitor mass for a NS varies from 6 to $8\rm{M_{\odot}}$, dependent on metallicity and the minimum initial mass for a BH $\approx 20~\rm{M_{\odot}}$. Our models have a minimum companion star mass of 0.6 to 0.8~M$_{\odot}$ for a NS and 2~M$_{\odot}$ for a BH. This minimum companion mass increases for more massive black holes as they require more massive progenitor stars. 

We have created a small number of bespoke models in this mass range to see whether the \textsc{BPASS} stellar evolution code can create models with parameters analogous to the Gaia BH systems. While other studies have suggested that isolated binary evolution cannot create some of the Gaia-BH systems \citep{2023MNRAS.526..740R, 2024MNRAS.527.4031T}, we can do this within the \textsc{BPASS} code because our inefficient, $\gamma$-like common-envelope prescription leads to more mass being lost without significant orbital shrinkage. We find an initial primary mass of $\sim60$M$_{\odot}$, with an initial period of a few 100~days and a companion of 0.93~M$_{\odot}$ leads to a system with the same parameters as Gaia-BH1. However, assuming a flat mass ratio distribution, only 1.6\% of stars will have a companion mass equal to, or below the required mass. These objects are likely to be rare and a dynamical formation pathway is more likely for these systems. We note that Gaia-BH2 and Gaia-BH3 are much wider systems with periods of 3.5 and 11.57 years and if not dynamically formed are more likely to be the result of effectively single-star evolution with very low supernova kicks given the mass of the black holes formed. Such low kicks are suggested through observations of VFTS 243 when compared to \textsc{BPASS} models \citep{Stevance2023MNRAS.520.4740S} and other similar studies \citep{1999A&A...352L..87N, 2005ApJ...625..324W,2009ApJ...697.1057F}. 

All observations except for one, have orbital periods greater than 100 days. They are also longer than where the peak in the synthetic population period is located  (one to 10 days). We note however that the detection method for many of the objects relies on the Gaia data releases which will be biased to higher period systems that will have a greater astrometric wobble. Other survey strategies may be needed to detect these shorter period binaries. 

Additionally, \textsc{BPASS} also predicts a sizeable number of systems with a range of masses and periods above a thousand days, as seen in the upper-right and lower-left plots. Since \textsc{BPASS} circularises the orbits, they could have even longer periods if appropriate eccentricities are considered. Whether such systems exist remains an open observational question. And if they do, would they be able to be detected with currently available observatories? Understandably, they are challenging to detect as these post-SN systems have yet to go through mass transfer and would be reliant on methods similar to those employed in the discovery of the Gaia BH systems to find them. Any of these systems with periods greater than $10^{3.2}\rm{days}$ would be beyond Roman's currently planned mission lifespan, while periods greater than $10^{3.5}\rm{days}$ would even be beyond Gaia DR5's timeline. While difficult, it may not be impossible to detect such long period systems. \citet{2023ApJ...946..111A} explored the possibility of resolving the partial orbits of long period systems with Gaia data alone with promising results.

\subsection{Wind-fed \& Be XRBs}\label{Wind-fedandBe}

As discussed in Section \ref{binaryspeciesmethod}, wind-fed systems are taken to be any where in the models wind-fed accretion is significant enough to drive X-ray emission. We treat the wind-fed accretion in Be-XRBs slightly different and discuss these in Section \ref{BeXRBResults}. 

\subsubsection{Wind-fed XRBs}\label{windfedresults}
\begin{figure*}
\includegraphics[width=2\columnwidth]{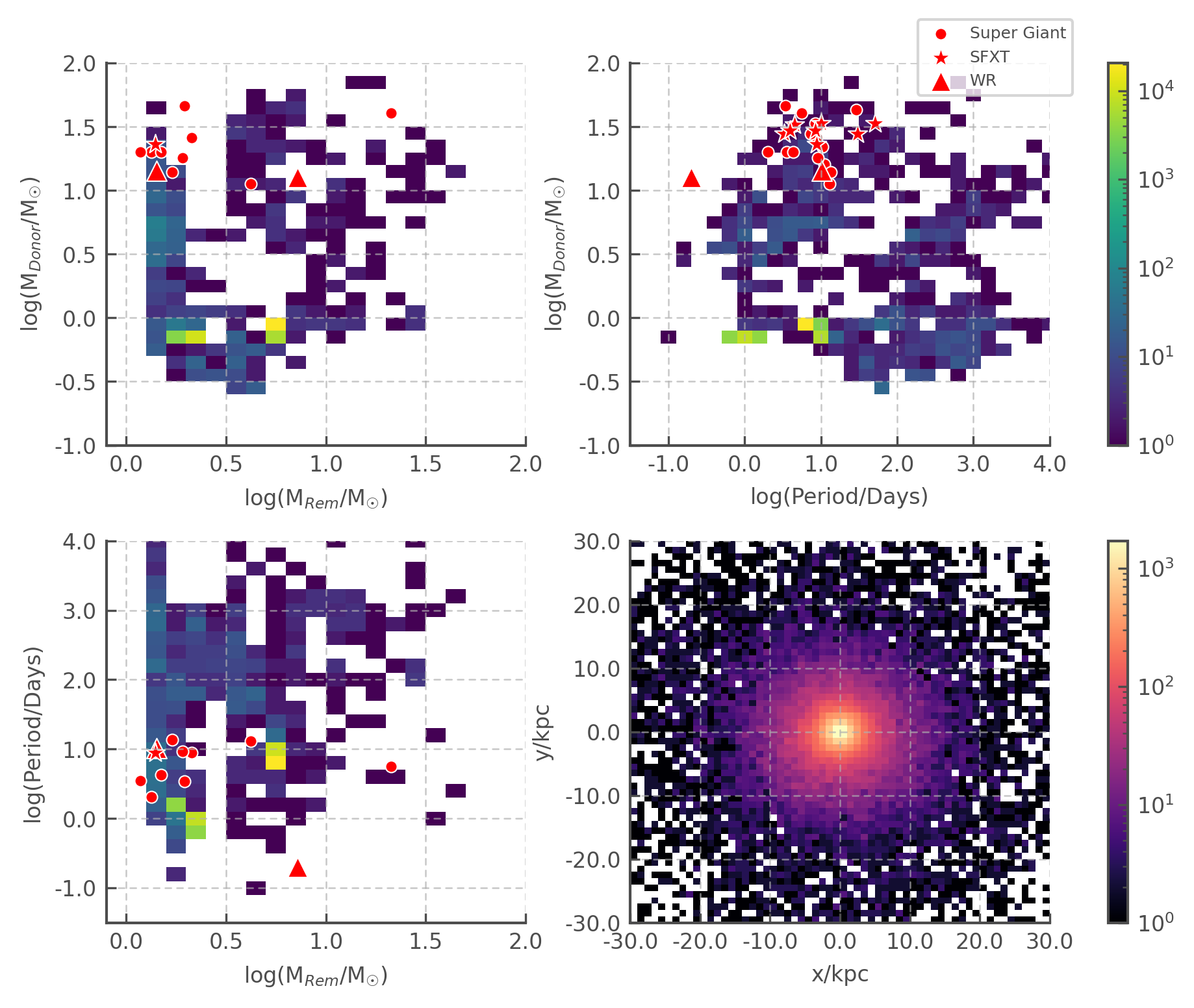}
\caption{Wind-fed XRBs synthetic populations compared to observated systems. Each point represents a system from the Binary rEvolution HMXRB catalogue. The data points have different shapes based on the tags in the Binary rEvolution HMXRB catalogue. These include super giant companion (circle), super-fast x-ray transients (SFXT, star), as well as Wolf-Rayet star companion (triangle). The BeXRBs from the catalogue are not on this plot and are instead found in Figure \ref{fig:Be}.}
\label{fig:WindFed}
\end{figure*}

For wind-fed systems we do not constrain the donor star mass. An interesting consequence of this is that we predict a large number of systems where the donor is below $10\rm{M_{\odot}}$, with a peak around a Solar mass. However, it should be noted that the observational samples we compare to only include the Binary rEvolution HMXRB catalogue. We also assume all of the systems within that catalogue are wind-driven XRBs. 

We will discuss the majority of the wind-fed XRBs first, followed by discussion of the wind-fed sub-type of BeXRBs.

Figure \ref{fig:WindFed} shows our synthetic populations for wind-fed systems. Overall, we see a good overlap between observed systems and the \textsc{BPASS} predictions, especially at high primary masses. The observed systems with donor masses $>10~\rm{M_{\odot}}$ all sit in a place that matches our synthetic population. 

Beyond this, Figure \ref{fig:WindFed}'s mass distribution plot indicates that there could be a significant population where the donor mass is at or below $10~\rm{M_{\odot}}$, which would be wind-fed LMXRBs. We have been unable to find an observational data set that includes such sources, though the predicted system parameters do overlap with those for RLOF fed LMXRBs discussed below. In the low-mass systems strong winds arise in the subgiant and giant evolution stages of Solar-like stars.

Figure \ref{fig:WindFed} and Table \ref{tab:enwcts} indicate that the majority of synthetic population systems contain a BH accretor, with the ratio of BH to NS accretors of approximately 2:1. While there is some spread to the BH systems, there is a concentration of systems with solar mass donors and $\sim10$ day periods. 

We must ask whether or not these lower mass donor XRB systems exist. The closest analogue to these wind-driven LMXRBs would be the symbiotic X-ray binary (SyXRB). Generally, these are a subclass of XRB that are theorised to feature a NS accretor and a late-life giant star \citep{2012MNRAS.424.2265L,2019MNRAS.485..851Y}, although it is also theorised in rare cases that they may instead house a BH accretor \citep{2024ApJ...977...95D}. Beyond this, SyXRBs have been known to have relatively low X-ray luminosity. The first SyXRB was discovered in 1977 \citep{1977ApJ...211..866D} and a handful have been discovered since. Some of these are found within the Binary rEvolution catalogue, however they do not have enough information to plot those systems. \citet{2019MNRAS.485..851Y} theorised there should be about 40 SyXRBs per $10^{10}~\rm{M_{\odot}}$, making them rare while their low X-ray luminosity makes them difficult to observe. While we do not classify these systems within BPASS, it is understandable that some of our models may represent these rare systems. 

\subsubsection{Be-XRBs}\label{BeXRBResults}

Here we consider Be-XRBs as a unique subtype of the wind-fed XRBs. Our criteria for the classification of Be-XRBs is described in Section \ref{methods}. Traditionally, these are assumed to be high-mass X-ray binary systems consisting of a Be star with a neutron star in a wide orbit. The unique accretion of these systems is thought to be due to an equatorial decretion disk of material thrown off the rapidly rotating Be star. The neutron star passes close to, or through the disk leading to accretion onto the NS. We include all systems labelled as Be in the Binary rEvolution catalogue. In the case where a system is classified as a Be-XRB but has no estimate for a remnant mass, we assume a lower bound of $1.39\rm{M_{\odot}}$. 

It should be noted that there is no known mechanism that would prevent a Be-XRB from having a black hole instead of a neutron star companion \citep{BePablo2011Ap&SS.332....1R}, so we also include any modelled systems where that is the case. It should also be noted that \citet{BePablo2011Ap&SS.332....1R} suggests the possibility of Be-WD systems. However, as we are only concerned with BHs and NSs they remain outside the scope of this paper. The Binary rEvolution Catalogue presents two Be-BH candidates, MWC 656 \citep{MWCXB2014ApJ...786L..11M} and HD 96670 \citep{HDisXB2021ApJ...913...48G}. However, recent studies question whether or not they do in fact house black hole accretors \citep{MWCnot2023A&A...677L...9J, HDnotBH2025A&A...696A..84N}. 

\begin{figure*}
\includegraphics[width=2\columnwidth]{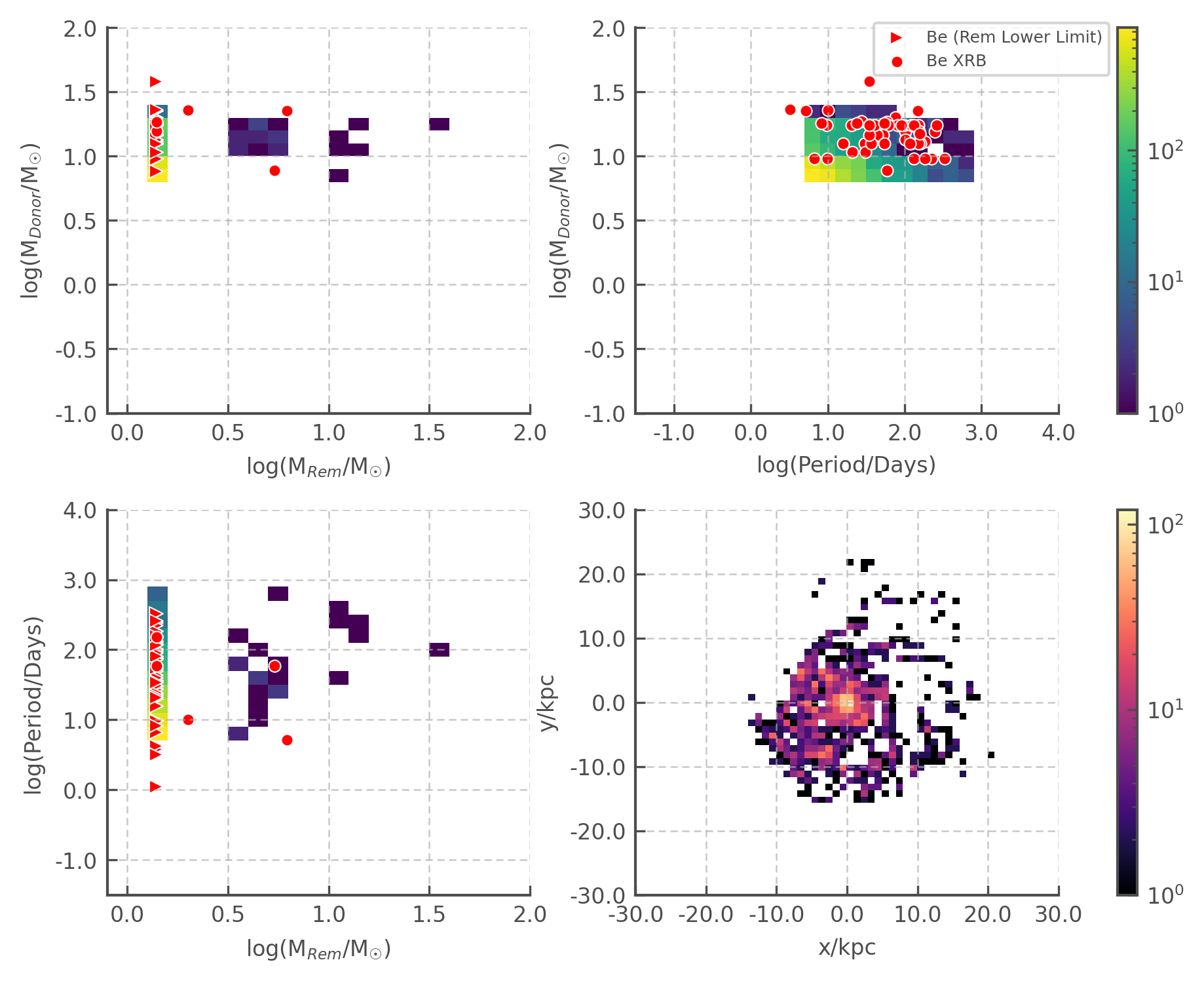}
\caption{BeXRB synthetic population compared to observed systems. The criteria used for classifying our models as BeXRBs in Section \ref{binaryspeciesmethod}. All observations here are flagged as ``Be" in the Binary rEvolution catalogue. Not all of these have remnant masses reported, so we assume a minimum mass of $1.39~\rm{M}_\odot$ for such cases and plot them as triangles.}
\label{fig:Be}
\end{figure*}

Figure \ref{fig:Be} shows a very strong correlation to the observed systems. This is unsurprising given the narrow window of parameters that give rise to the Be XRB phenomenon. As expected, the majority of the systems contain an NS as the accretor. However, while rare, we do indeed see systems where a black hole accretor should exist. These appear at a ratio of $\sim109:1$. These Be-BH systems do not seem dependent on period or donor mass as they have very similar distributions to the NS systems. Additionally we have systems analogous to MWC 656 and HD 96670 in terms of mass and period. While this should not be interpreted as evidence for a black hole accretor in those systems, it does give credence to the possibility of Be-BH XRBs. 

\subsection{Roche-lobe Overflow XRBs}\label{RLOFResults}

As stated in Section \ref{methods}, we split the results of RLOF systems into those we consider to be X-ray bright and X-ray faint based on the estimated effective disk temperatures for the model binaries. The rationale for this is that, in the models, we find a wide variety behaviour of RLOF. Additionally, at low mass transfer rates it is unlikely that every system will be luminous in the X-ray domain. In such cases, disk emission will primarily lie in the UV, optical, and infrared domain where it would be incorrect to identify them as XRBs. Although we note that we only consider the disk and have not considered possible X-ray emission from the accretion stream impacting the NSs.

\subsubsection{X-ray Bright}\label{xraybrightresults}
\begin{figure*}
\includegraphics[width=2\columnwidth]{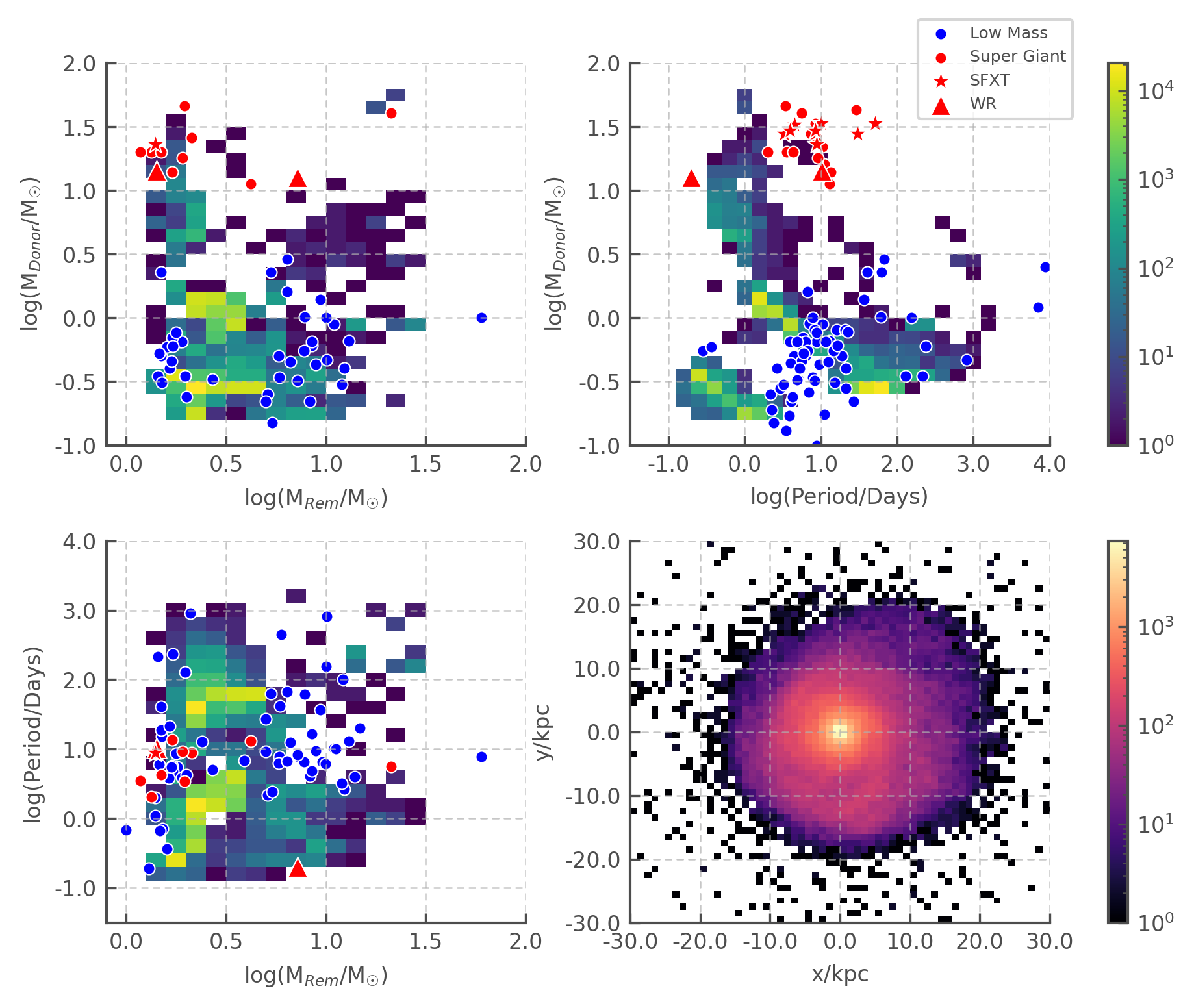}
\caption{X-ray bright RLOF synthetic populations compared to observed systems. These are \textsc{BPASS} systems where the effective temperature of the disk is $>10^6~\rm{K}$ and are expected to be X-ray luminous. We plot the entire Binary rEvolution LMXRB catalogue as blue points. Additionally, we include the HMXRB catalogue without the BeXRB systems in the same fashion as Figure \ref{fig:WindFed}.}
\label{fig:RLOF}
\end{figure*}

Figure \ref{fig:RLOF} displays the results for the x-ray bright systems. For simplicity, we assume all the sources in the LMXRB catalogue from Binary rEvolution to be X-ray bright RLOF systems, although as in Section \ref{windfedresults}, some of these could be wind-fed systems. In addition, we include the HMXRB wind-fed systems on the Figure to show that some of these could be RLOF systems. Given that we predict no RLOF systems with these same donor mass -- orbital period values, many of them are more likely to be wind-fed.

For the LMXRBs, the masses of the donors and remnants appear to correlate well with the current library of observed systems. It should be noted that some observed systems end up below the minimum donor mass on the Figure. This is due to their donor masses falling below $0.1M_{\odot}$. A reason that we have few models with such low donor masses is due to the same reason why we don't have low mass donors in the quiescent binaries. Such systems, on the borderline of being nuclear fusing stars, are beyond what we have studied with our models currently.

We see a trend in the model systems where the higher-mass donors have periods of about a day while the lower mass systems have a more diverse range of periods. In terms of remnant masses, we see a rather even distribution for the systems with NSs. Once we get to systems with BHs, outside of some outliers, they trend toward shorter periods as the mass of the BH increases. 

A key feature to note however in this, and subsequent Figures, is in the upper-right panel. Here, we see a gap in the model population for sub-solar donors between $1$ and $30$ days. While there are observed systems within this gap, our methods do not yield \textsc{BPASS} models in that range. We note that systems in the \textsc{BPASS} models do exist in this region, but they evolve through that area quickly so such systems are not within our synthetic population. This is perhaps the first sign that the physics included in the binary evolution models for stars in this mass and period range should be revised. We discuss the possible reasons for this in Section \ref{results}.

\subsubsection{X-ray Faint}\label{xrayfaintresults}
\begin{figure*}
\includegraphics[width=2\columnwidth]{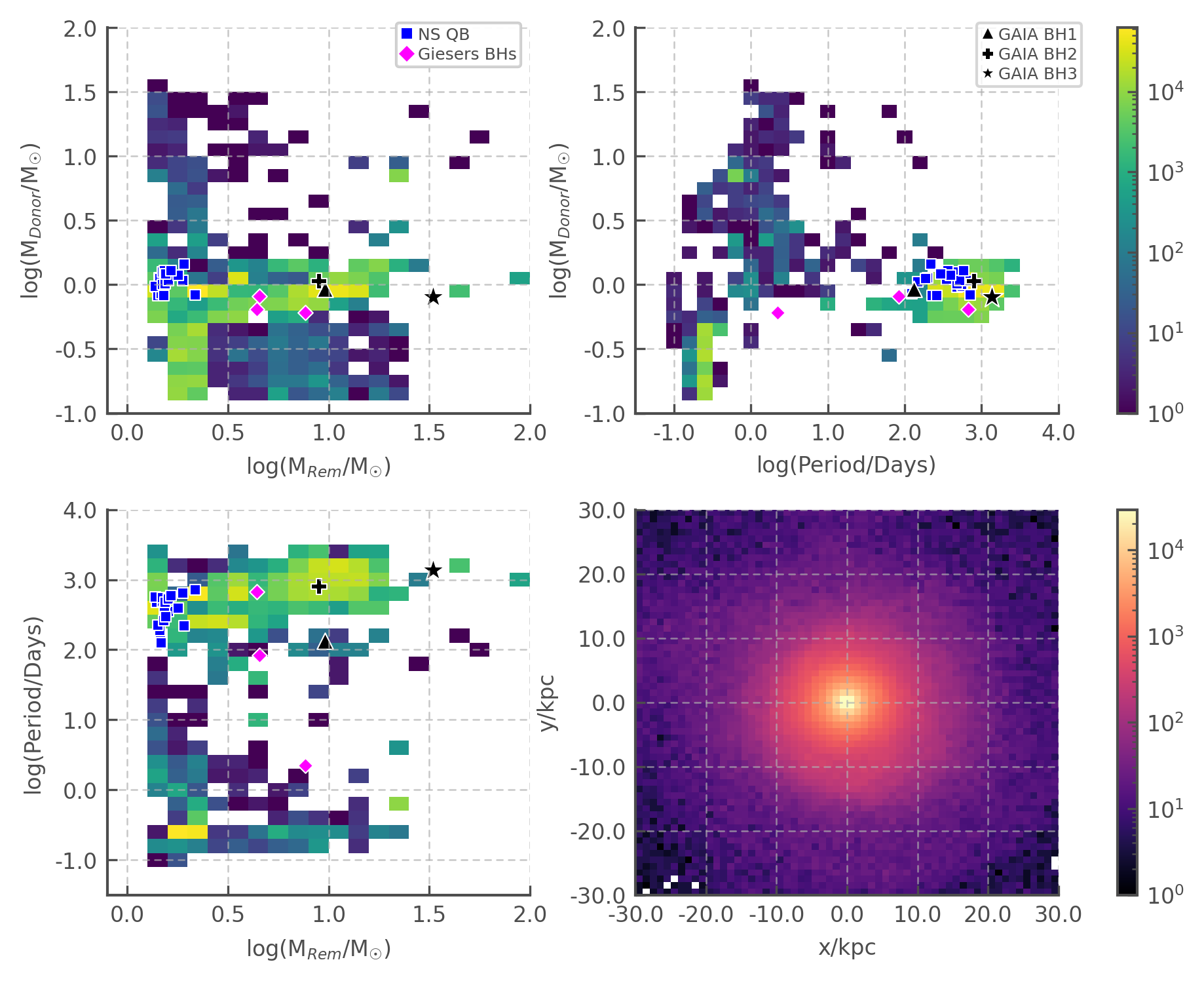}
\caption{X-ray faint RLOF XRB synthetic population compared to observed systems. \textsc{BPASS} systems where the effective temperature of the disk is $\rm{\le10^6K}$ and therefore emitting primarily in optical and IR wavelengths. We include the same observational data found in Figure \ref{fig:quin0}.}
\label{fig:quinrlof}
\end{figure*}

While observations indicate that systems found on Figure \ref{fig:quin0} are not likely to be engaging in mass transfer, we still find it pertinent to study them in the context of these X-ray faint systems. The X-ray faint predictions are shown in Figure \ref{fig:quinrlof}. These faint systems tend to have longer orbital periods than the bright ones. The donor masses here are dominated by donors of around $1{\rm M_{\odot}}$ where their low mass transfer rates can be explained by their post-main sequence evolution. Here, the core is supported by electron degeneracy pressure and hydrogen shell burning slowly increases core mass until the helium flash occurs. Therefore, when these stars fill their Roche-lobes on the giant branch mass transfer is slow and stable giving rise to the peak of $1{\rm M_{\odot}}$ donor masses and periods from 100 to 1000 days. Eventually when the helium flash does occur and the stars shrink in radius, this phase of evolution will end. Another confirmation of this is shown by Figure \ref{fig:bpassage}, the age of these systems tend to be on the older side with a peak in the age distribution at 10~Gyrs.

Using the predictions in Table \ref{tab:enwcts}, the X-ray faint population is predicted to outnumber the X-ray bright population by a factor of $\sim5$ to $1$. Our results also predict faint systems are more likely to house black holes rather than neutron stars. This is to be expected as such systems typically have longer periods, lower mass transfer rates, weaker irradiation, and larger disks leading to cooler temperatures. Additionally, the innermost stable orbit around these BHs is greater than that of their NS counterparts, again decreasing the accretion disk temperature. Furthermore, systems with a NS are more likely to have a more massive donor star meaning mass transfer is more likely to be unstable and more of these systems would enter a CEE phase.

It is interesting to consider what the observational analogues may be and whether any have been observed. A possible example is the peculiar ACS \#21859 system \citep{Giesers2019}. This is a system with partially filled H$\alpha$ lines and no X-ray or radio counterpart. Given the systems mass ratio, it is similar to that of Gaia BH1 \citep{El_Badry2023-1} and BH2 \citep{El_Badry2023}, yet has a much shorter period. While it does not line up directly with a period bin, there are systems in \textsc{BPASS} that are still close analogues to this system. Whether it is actually engaging in mass transfer or not has yet to be determined, its position on these plots provides possible explanations as to its nature. These results could be viewed as a prediction of what the Gaia-BH systems will look like in future. \citet{2024ApJ...973...75C} used Chandra X-ray Observatory observations to investigate the possibility of accretion onto Gaia BH3 but found no evidence of it. Our results could instead be interpreted to mean that such systems may eventually evolve into systems that may engage in mass transfer. Additionally, the sheer number of these systems implies the existence of a large population of growing compact objects that are more or less invisible to X-ray observation. 

\subsubsection{CEE}

Common-envelope evolution is a phase of evolution when the binary system has entered a phase of unstable mass transfer and the compact remnant orbits within the hydrogen envelope of the donor. Whether such systems will be detectable or not in X-rays is uncertain. For example, \citet{2007MNRAS.375.1000B} suggest that the CEE phase might be avoided completely in some cases. When the mass transfer rate is high, irradiation effects from high X-ray luminosity on the donor's envelope may change the nature of CEE or prevent it altogether. Finally, we note that in \textsc{BPASS}, the CEE phase occurs on a thermal rather than a dynamical timescale to retain numerical stability in the stellar models. Therefore we are over predicting the number of these systems.

Figure \ref{fig:CEE} shows the predicted population of CEE systems. We find \textsc{BPASS} predicts a large number of these systems to have sub-day periods and so may evolve in to ultra-compact XRBs, a quasi-star (a star with a BH at its centre), or Thorne–Żytkow objects (a star with a NS at its centre); stars with either a black hole or neutron star at their centre. It should be noted that the \textsc{BPASS} CEE prescription is weak in the sense that the orbital period does not decrease as much as might be expected from the standard $\alpha$-prescription CEE \citep[see supplementary information in ][for discussion on \textsc{BPASS} CEE]{2023NatAs...7..444S}. However, including these systems is important for completeness of the evolutionary stages we see within the \textsc{BPASS} stellar models.

\subsection{Post-interaction}
\begin{figure*}
\includegraphics[width=2\columnwidth]{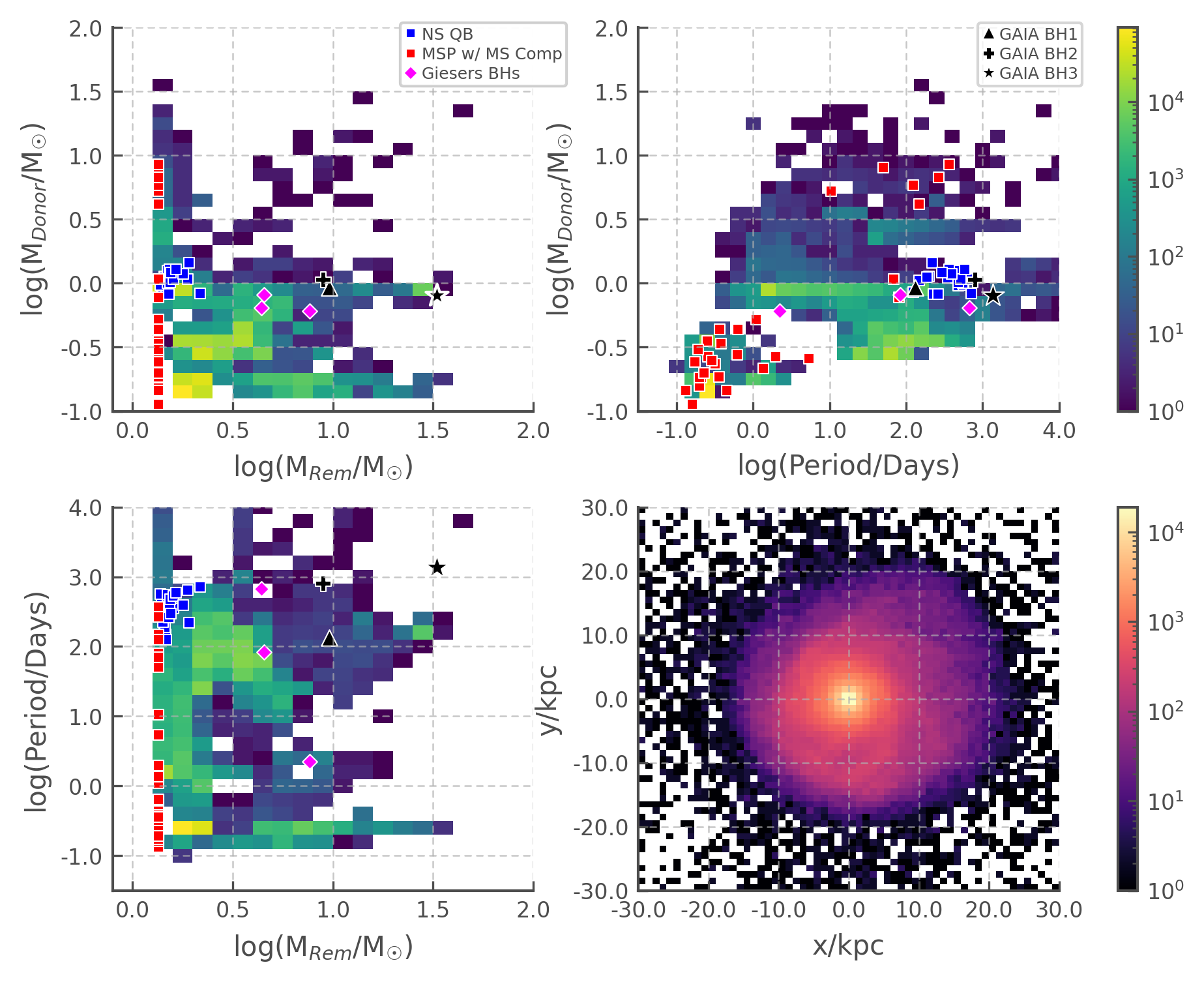}
\caption{Post-interaction Quiescent Binaries synthetic population compared to observed systems. Systems that have previously been flagged as one of the other previous interacting systems but are not currently engaging in mass transfer. We plot the observational dataset from Figure \ref{fig:quin0}. Additionally, we include millisecond pulsars from the ANTF catalogue that have main sequence donors. These are plotted as the red squares. As mentioned in Section \ref{obs}, a remnant mass of $\rm{1.35M_{\odot}}$ is assumed.}
\label{fig:quin1obs}
\end{figure*}

\begin{figure*}
\includegraphics[width=2\columnwidth]{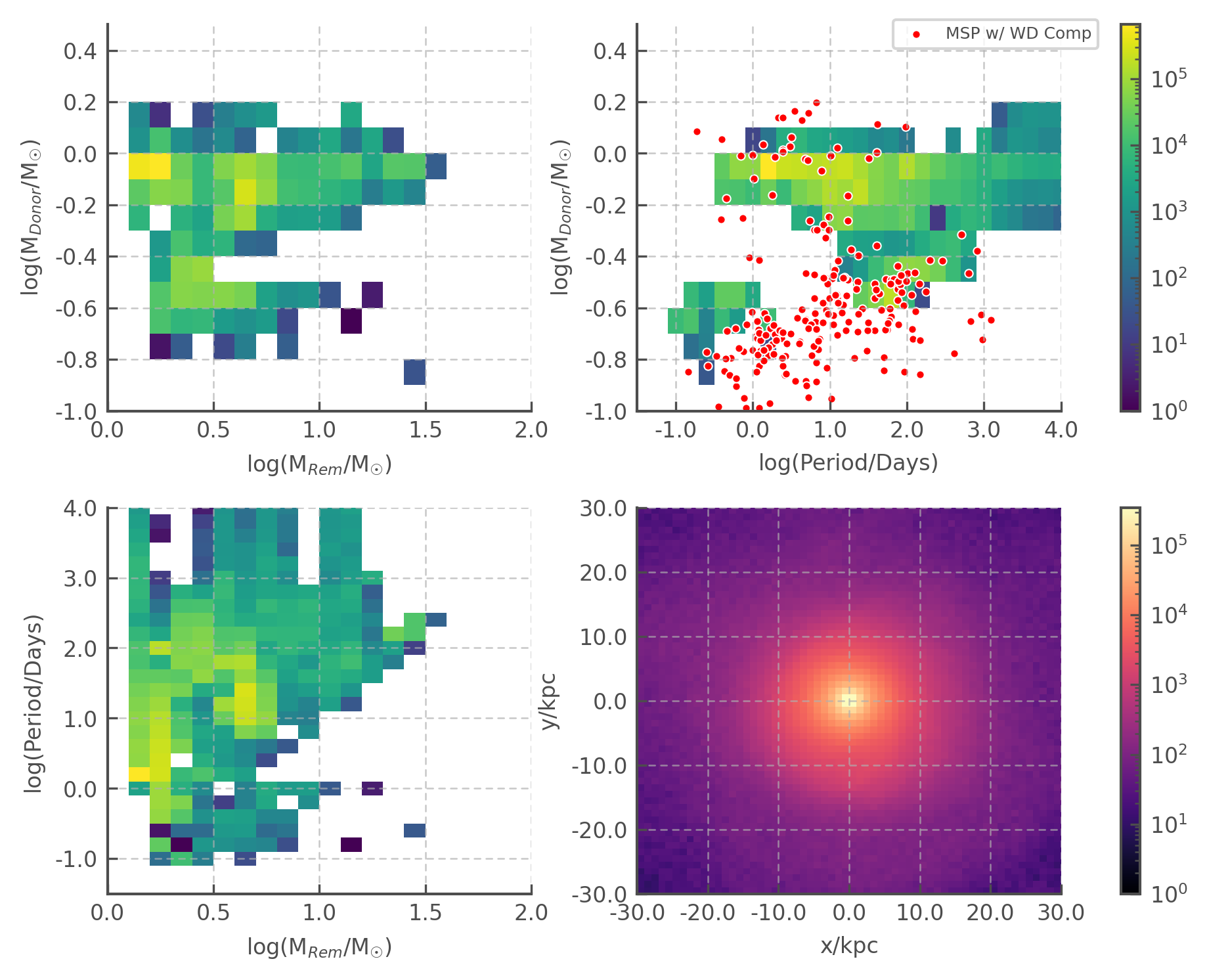}
\caption{Post-interaction Quiescent Binaries with a white dwarf companion synthetic population compared to observed systems. We compare our theoretical data of non-interacting systems with a white dwarf donor to observations in the ATNF catalogue of millisecond pulsars that have a WD companion.}
\label{fig:wd}
\end{figure*}

The final phase in the life of a compact object binary is post-interaction evolution. We present our results for this stage in Figure \ref{fig:quin1obs} and Figure \ref{fig:wd}. The latter includes the  systems where the donor has evolved into a white dwarf (WD), while Figure \ref{fig:quin1obs} has donor stars which still have nuclear fusion in the interior. Although we have included the observed samples of quiescent systems here, we note that these stars are all likely to be systems that have yet to interact. However, it is useful to compare them to the systems here. The majority of systems below a~few~${\rm M_{\odot}}$ are likely to be those that evolve into a white dwarf in future, while those that are more massive will experience a supernova and become unbound or become a double compact object binary \citep[again see][for predictions concerning these systems]{tang2024predictinggravitationalwavesignals}. Table \ref{tab:enwcts} indicates that the nuclear fusing quiescent systems are the smallest population of quiescent binaries we predict, with there being approximately five times more pre-interaction systems and 20 times more white dwarf systems.

In Figure \ref{fig:quin1obs}, we have also included the sample of millisecond pulsars with main-sequence companions. The reasoning for this is that these millisecond pulsars are likely to have arisen due to some form of mass accretion. Therefore, we investigate whether or not the observed sample matches this synthetic population. The fact that so many overly at both high and low periods and donor matches suggest this match is possible. 

The white dwarf binaries (Figure \ref{fig:wd}) are the only post-interaction population where we can compare to observations understood to be the same species. Here we use the results from the ANTF catalogue where the pulsar is known to have a white dwarf. The ANTF catalogue doesn't have values for the remnant masses, so when estimating a mass for the white dwarf donor they assume a minimum remnant mass of $1.35~\rm{M}_\odot$. This also means we can only make a comparison in the upper-right panel of donor mass against period. In the upper-right panel, we see a gap in the predicted population centred at a period of about a few days and a donor mass of $\sim0.1{\rm M_{\odot}}$. Here we see more observed stars filling the lower end of this gap. Given that such donors may have come from lower mass companions that have convective surfaces, we can hypothesize that such stars may be heavily influenced by the process of magnetic braking during their mass transfer phase which may explain the source of this gap. Additionally, this would be related to that seen in the RLOF synthetic population \citep{2000A&A...363..657E, 2025ApJ...995...99Y}. Furthermore, here we have also not considered the evolution of these systems as Spider systems that may also explain why some of the predicted masses are higher than those in the observed system \citep{1988Natur.333..832P,2017MNRAS.464..237S}.

\subsection{Comparisons across the Synthetic Populations}\label{overallstats}

Figures \ref{fig:quin0} to \ref{fig:wd} show the predicted populations for the different compact object binaries from the \textsc{BPASS} stellar evolution models. It is important to also compare the parameters of these populations and check whether the evolution of the donor mass, remnant mass, and orbital period agree with the picture that these quiescent binaries become wind-fed or RLOF XRBs, before becoming post-interaction quiescent binaries. Keeping in mind that the widest quiescent binaries will remain so for their whole evolution.

Figure \ref{fig:BPASSperiod} shows the period distribution, Figure \ref{fig:BPASSdonor} shows the donor mass distribution and Figure \ref{fig:BPASSdem} shows the remnant mass distribution. Here, we see that the period distributions of interacting systems tend more towards shorter period systems than the pre-interaction quiescent binaries, while the CEE systems have the shortest periods. One interesting detail to note here is that the post-interaction WDs have a large population with very long periods. This is expected given that the widest binaries that form white dwarfs will lose most of their initial mass during their AGB phase provided they don't interact and thus only become a wider binary. 

As for the donor mass, again the pre-interaction systems are distributed more towards higher masses than the other distributions with the post-interaction quiescent binaries having the distribution most skewed towards the lowest masses. We note that the CEE systems have a minimum mass just below a Solar mass as less massive donors won't have the time to evolve and interact within the age of the Universe (or in this case, the age of the synthetic \textsc{FIRE} galaxy). For the wind-fed systems, the strong peak at a Solar mass is due to the increase in stellar wind mass-loss rates for these stars as they ascend the giant branch.

For the remnant mass we should expect that the pre-interaction distribution is weighted towards lower masses than those of interacting systems which is indeed what we see. We note that the highest black-hole masses are obtained from X-ray faint RLOF systems. Additionally, the post-interaction systems are skewed to higher remnant masses.

Finally, we can also consider the age and metallicity distributions of our synthetic population as shown in Figure \ref{fig:bpassage} and Figure \ref{fig:bpassz}. We see that most of the compact object binaries are from systems with ages beyond 300~Myr. The most noticeable peak is for the wind-fed systems at 10~Gyrs. Again, this is due to the increase of mass-loss rates for Solar mass stars as they ascend the red-giant branch. The trend of increasing age of the population will be directly related to the age distribution of the \textsc{FIRE} simulation which is shown in Figure \ref{fig:fireage}.

Interestingly, the metallicity distribution of the population shows much greater variety than the relatively flat distribution of FIRE's metallicity shown in Figure \ref{fig:firez}. From Figure \ref{fig:bpassz}, we see that CEE is constrained to higher metallicity systems, indicating that CEE is likely to only occur for higher metallicity systems. Also,  if this phase is only linked to stars younger than 300~Myrs, this reflects the fact that recent star formation is more metal rich. We also see that wind-fed systems are more likely at higher metallicities, reflecting the fact that stellar winds are stronger at higher metallicities. An interesting result of comparing the RLOF bright and faint populations is that the bright systems are more likely at higher metallicity with the faint systems coming from low metallicity systems. Perhaps indicative of how the RLOF evolution is sensitive to how the star reacts to mass loss with the surface boundary condition being dependent on surface opacity which in turn is directly related to the metal content of the binary star. 

The comparison between populations indicates that our predictions of how the binaries evolve at least make sense. If more complete and homogeneous observational datasets could be obtained, it would be possible to see if observations indicate that these trends in the synthetic populations are also seen in nature.

\section{Results - Space Telescopes}\label{SpaceTeleResults}

\begin{figure}
\includegraphics[width=0.9\columnwidth]{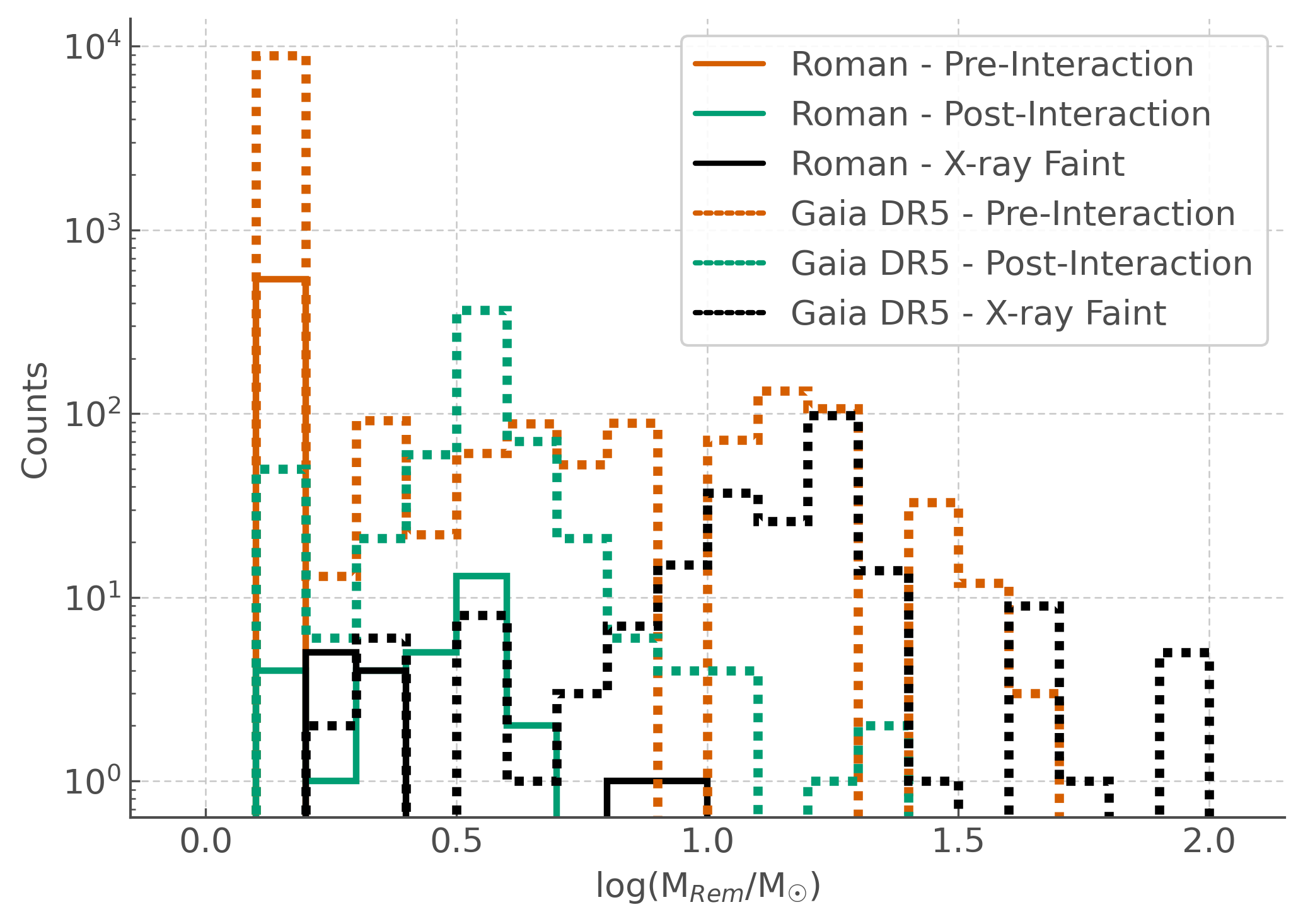}
\caption{Roman GBTDS and Gaia DR5 detectable remnant mass distribution. The solid lines represent the Roman GBTDS while the dashed lines are for Gaia DR5.}
\label{fig:teleremM}
%\end{figure}

%\begin{figure}
\includegraphics[width=0.9\columnwidth]{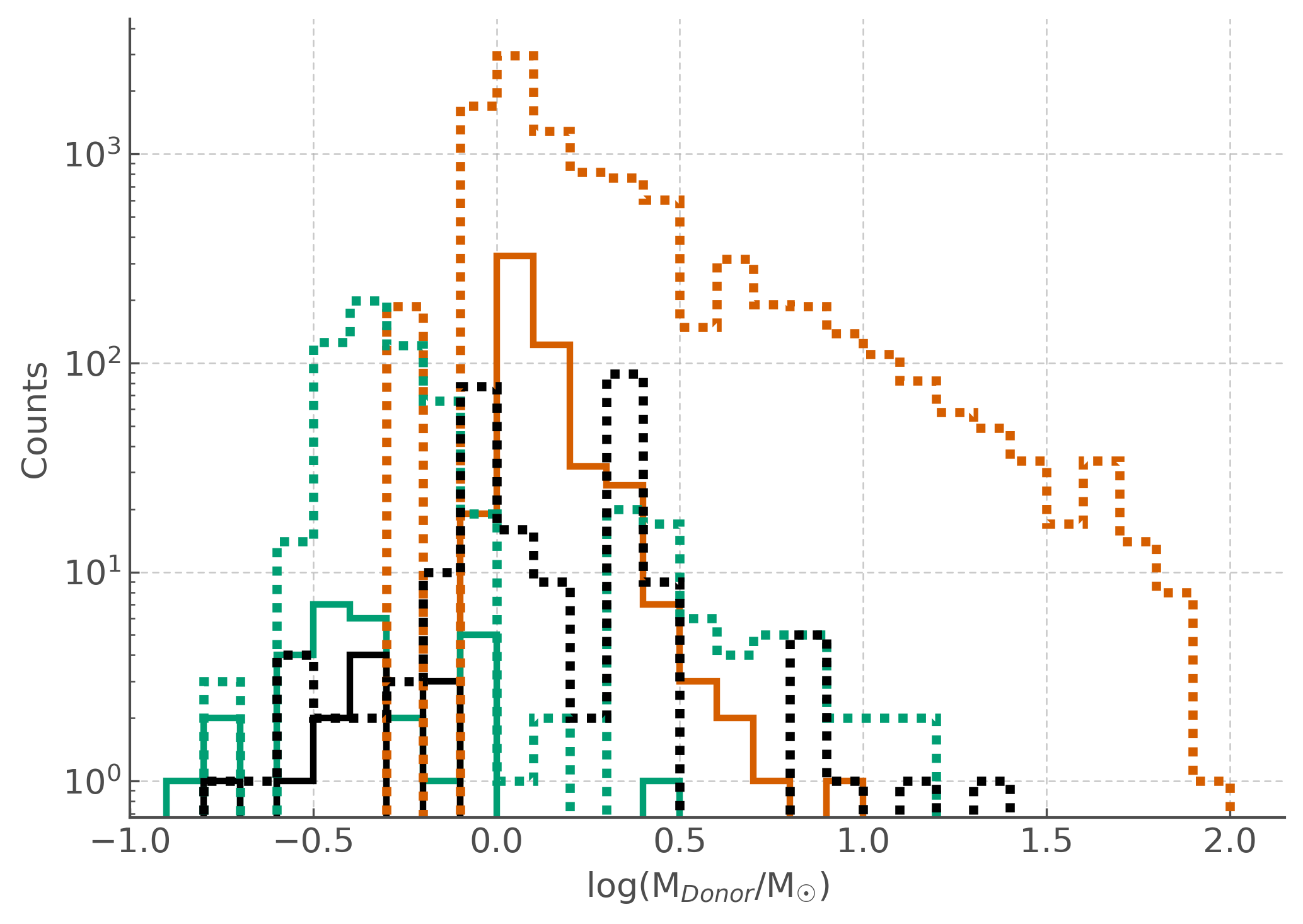}
\caption{Roman GBTDS and Gaia DR5 detectable donor mass distribution. Legend is the same as that in Figure \ref{fig:teleremM}.}
\label{fig:teledonorM}
%\end{figure}

%\begin{figure}
\includegraphics[width=0.9\columnwidth]{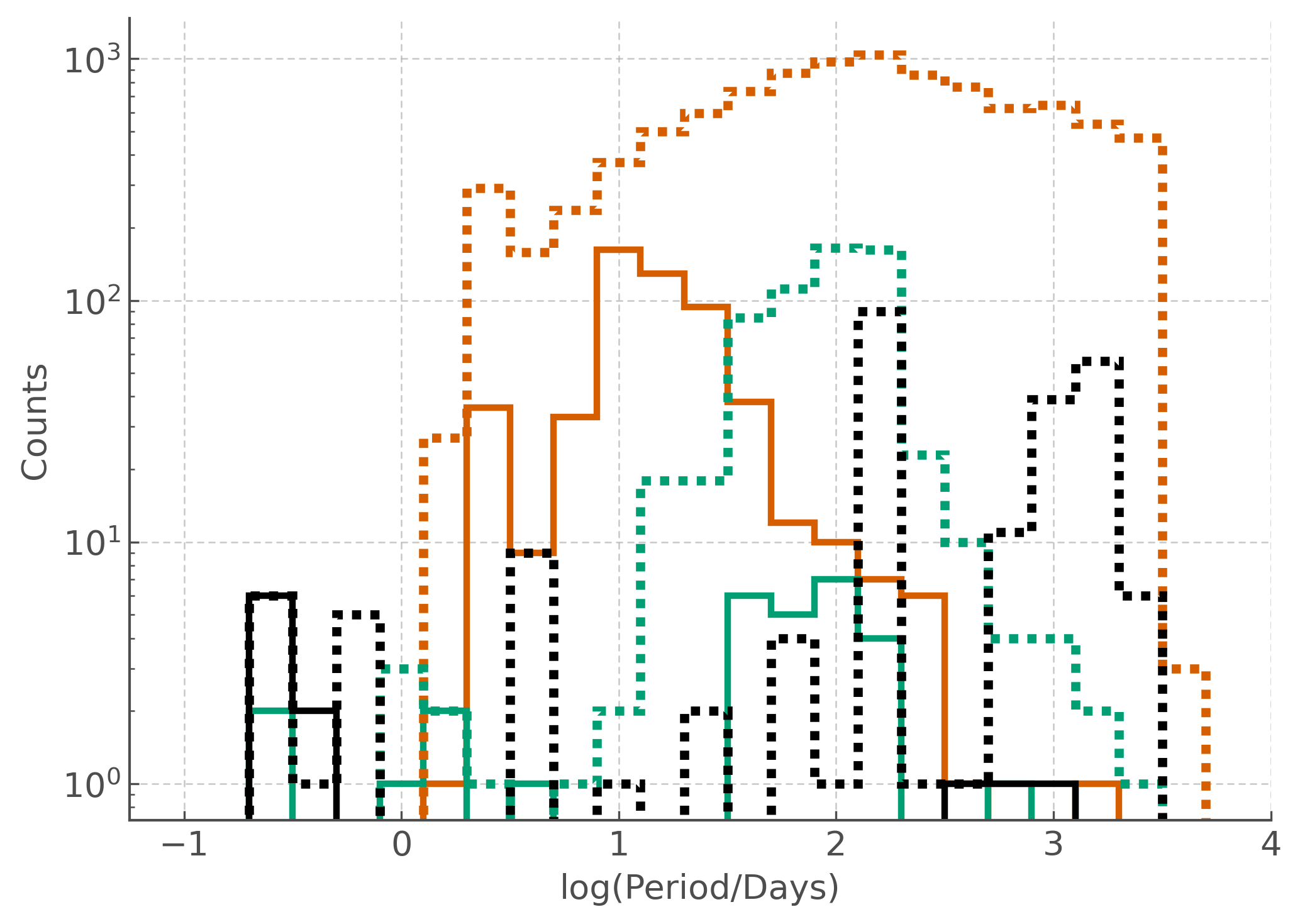}
\caption{Roman GBTDS and Gaia DR5 detectable system period distribution. Legend is the same as that in Figure \ref{fig:teleremM}.}
\label{fig:teleP}
\end{figure}

\begin{figure}
\includegraphics[width=0.9\columnwidth]{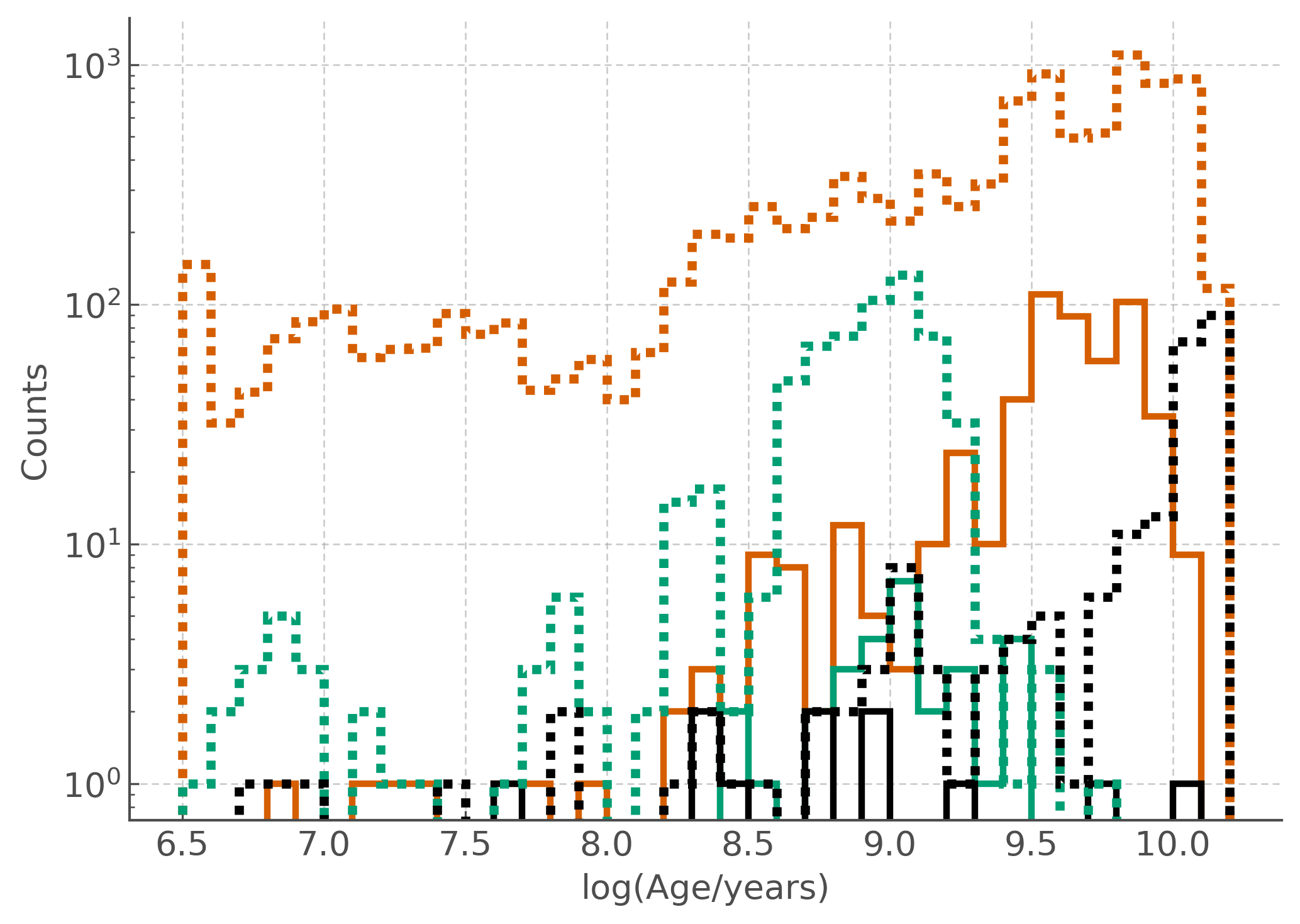}
\caption{Roman GBTDS and Gaia DR5 detectable system age distribution. Legend is the same as that in Figure \ref{fig:teleremM}.}
\label{fig:teleAGE}
%\end{figure}

%\begin{figure}
\includegraphics[width=0.9\columnwidth]{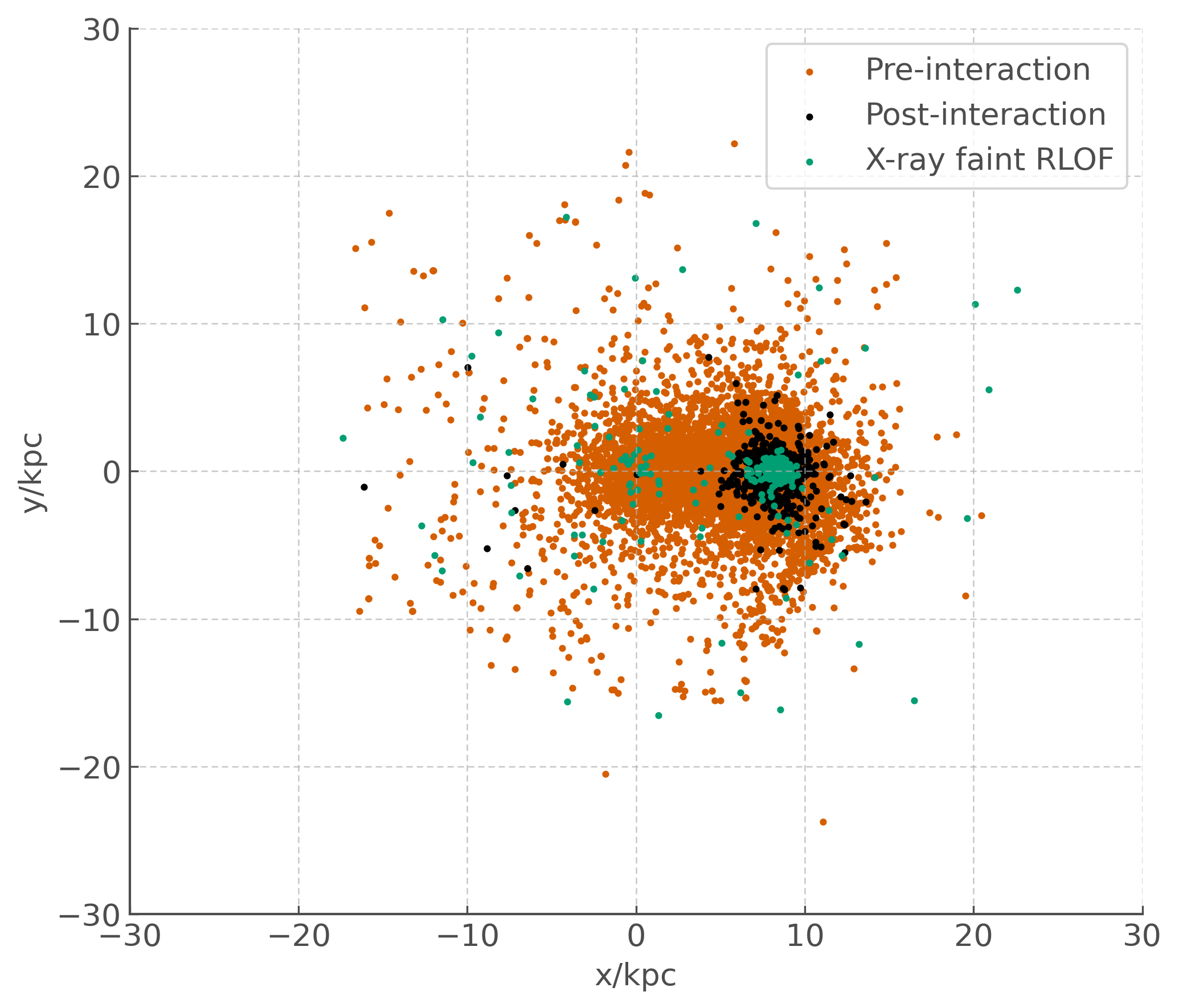}
\caption{DR5: spatial distribution of detectable systems when the Earth is placed at $\rm{(8.5,0,0)~kpc}$.} 
\label{fig:gaiaSPACE}
%\end{figure}

%\begin{figure}
\includegraphics[width=0.9\columnwidth]{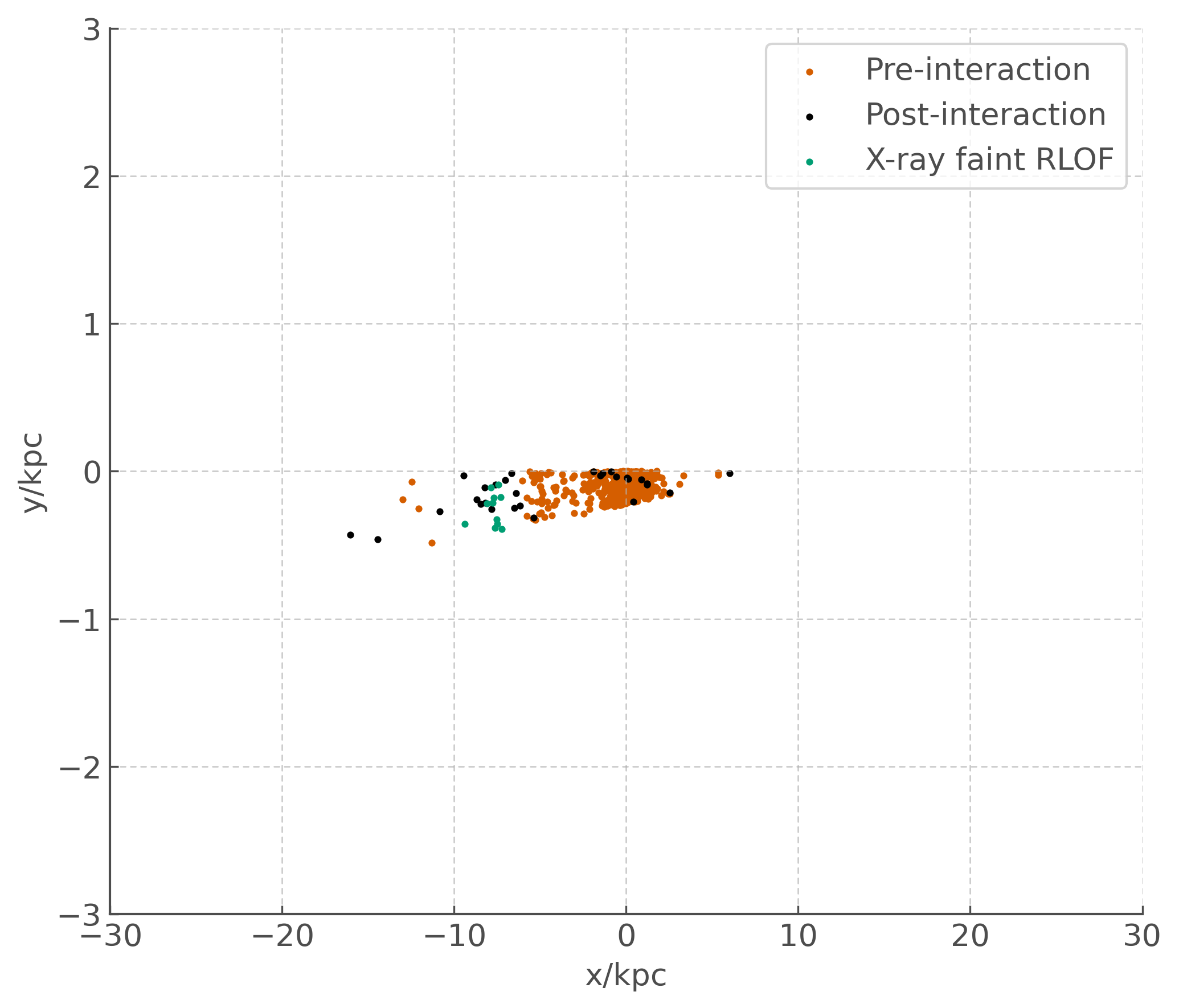}
\caption{GBTDS spatial distribution of detectable systems when the Earth is placed at $\rm{(8.5,0,0)~kpc}$.}
\label{fig:GBTDSSPACE}
\end{figure}

As discussed above, it is possible to make simple estimates of the populations that future space telescope surveys may find from their observations. From the synthetic population and the method outlined in Section \ref{spacetelesim}, we predict what populations will be found in Gaia DR3, Gaia DR5 as well as Roman's GBTDS. For each survey, we provide an ``Optimistic'' and ``Pessimistic'' prediction in a similar fashion to \citet{2025arXiv250821805C}. The former is a population where the wobble is greater than the calculated $\sigma_{ast}$ for said system, while the latter is three times greater. We present results for how many systems are detectable in Table \ref{tab:GaiaRomanNumsTable}. We break the populations down further in Table \ref{tab:appendixgaiaromantable} to the number of different types of systems that would be detectable in each survey. In addition, we provide distributions for the remnant mass, donor mass, and period in Figures \ref{fig:teleremM}, \ref{fig:teledonorM}, \ref{fig:teleP}, and \ref{fig:teleAGE}. Additionally, we visualise spatial distributions for Gaia DR5 and Roman's GBTDS in Figure \ref{fig:gaiaSPACE} and Figure \ref{fig:GBTDSSPACE} respectively. 

Comparing the different distributions in Figures \ref{fig:teleremM}, \ref{fig:teledonorM}, \ref{fig:teleP}, and \ref{fig:teleAGE}, we find the following. First, Gaia is more sensitive to higher mass remnants and black holes, as well as detecting a large number of NS binaries. This is also true of the donor masses as Gaia observed the entire Galaxy which includes regions of recent star formation. Second, Gaia extends to the youngest ages while GBTDS will mainly find systems older than 300~Myrs. 

\begin{table*}
    \centering

    \begin{tabular}{c|c|c|c|c|c|c|c}
        \hline
        \multicolumn{1}{c|}{Remnant} & \multicolumn{1}{c|}{$ P_{\rm orbit}<10.5~\rm{yr}$} & \multicolumn{3}{c|}{Optimistic} & \multicolumn{3}{c}{Pessimistic} \\
        \hline
         &  & DR3 & DR5 & GBTDS & DR3 & DR5 & GBTDS\\
         \hline
        BH & $476414\pm8921$ & $647\pm101$ & $1262\pm135$ & $36\pm6$ & $203\pm47$ & $536\pm78$ & $21\pm5$\\
        NS & $2110713\pm29461$ & $3122\pm414$ & $8338\pm806$ & $603\pm50$ & $592\pm93$ & $2112\pm262$ & $192\pm22$\\
        \hline
        
    \end{tabular}
    \vspace{0.15cm} 
    \caption{Results for the number of detectable quiescent binary systems from each survey. The population is the total number of pre-interaction, post-interaction, and X-ray faint RLOF systems that have a period less than $10.5$ years. For the surveys themselves, the results are the mean of across the 16 positions we surveyed in our simulation. The optimistic population where the wobble is greater than $\sigma_{ast}$ and the pessimistic population is where the wobble is greater than $3\times\sigma_{ast}$.}
    \label{tab:GaiaRomanNumsTable} 
\end{table*}

\subsection{Roman GBTDS}\label{gbtdsresultsection}
In our results, we find tens of BH systems and hundreds of NS systems that meet the criteria of being resolvable. 

Compared to Gaia's all-sky survey, Roman's GBTDS will be observing a very small patch of the sky, so it is expected that fewer systems will be found within the actual and mock surveys. The survey will observe a $\sim\rm{1.7deg^2}$ patch of sky, any results from this or any simulation will be dependent on the spatial distribution of systems within the line of sight. For example, in our simulation only about $\sim500$ total BH quiescent binary systems lie in the line of sight for our simulated telescope, but not all can be detected. 

Furthermore, the Galactic Bulge is a complex, crowded, and very dense region of the sky which also suffers from high differential reddening \citep{2013A&A...550A..35P, 2014A&A...566A.120S, 2016ApJ...818..130B, 2016PASA...33...24N, 2025AJ....170..328Z}. At magnitudes $>22$, sources are expected to be noise dominated \citep{2024ApJ...965..138F, 2026PASP..138d4507M} which will be further complicated with extinction. Additionally, as Roman's expected mission time is half of Gaia's, longer period systems may not be easily resolved. In the latest reporting from NASA, Roman may have its mission extended by $5$ years should launch and commissioning go smoothly. This may provide an opportunity to observe those longer period systems. However, in our simulation's line of sight , only $\sim100$ systems have periods greater than five years, though this will vary between galactic simulations. Should the survey be extended with more seasons, another benefit would be the possible positional error improvements for fainter systems that are not noise dominated. 

\subsection{Gaia}\label{gaiaresultsection}

We predict that Gaia will find a few hundred BH candidate systems and possibly thousands of NS candidate systems. The greater number of objects arises because of the larger survey volume, that should overlap with the GBTDS.

Comparing DR3 to DR5, our population suggests almost twice as many binaries would be found. This is due in part to increasing the upper limit of periods to 10.5 years to reflect total science time of the Gaia mission as well as improvements to the positional error. Outside of the longer period systems, these improvements may also yield shorter period systems that trace a smaller angle on sky. 

Additionally, we also expect the results to be limited by distance as apparent magnitudes grow fainter and wobbles smaller which is reflected in Figure \ref{fig:gaiaSPACE}. However, we do have detectable systems upwards of $20~\rm{kpc}$ away. At the time of writing the furthest Gaia quiescent binary is $\sim1.5~\rm{kpc}$ from Earth so it has yet to be seen if such distant systems can be resolved. 

We note here that identifying candidates in a survey is one thing, finding evidence to show that the objects are NS or BHs is another. Here we have made an estimate of how many candidates might be found and it is likely that we are overestimating the number of systems. Especially since at the time of writing we only know of 3 BH systems in Gaia and 21 NS systems. This may also indicate that elements of our models need revising as we discuss below.

\section{Discussion}\label{disc}
The synthetic populations we have created allow us follow compact object binary systems throughout their lives. While our results do match observed populations to some extent, there are areas of mismatch between the synthetic populations and observed systems. This mismatch is important as it indicates aspects of physics in the binary evolution need to be updated which we discuss below.

\subsection{Non-interacting and X-ray faint evolution}\label{quindisc}
The most novel synthetic populations we present are those for quiescent systems and systems experiencing RLOF that will not be observed in X-rays. The predicted number of such systems shows that compact object binaries spend most of their life in quiescence. Studying them at this time, both before and after mass transfer, is useful to better understand how the compact objects evolve over time. With the former providing firm constraints on the birth mass distribution, and the latter providing a constraint on the post-mass transfer mass distribution. 

Our pre-interaction predictions in Figure \ref{fig:quin0} provide an understanding of these binary systems from their birth up to interaction with the donor star, or their companion death in the widest systems. As noted in Section \ref{preint}, we do not have any systems that exactly match these Gaia BH systems due to the minimum mass ratio resolution of our models. However, we have created bespoke models with lower companion masses of a Solar mass with a massive enough primary to form the black hole that show these systems can be formed by the \textsc{BPASS} stellar evolution code. These systems do experience common-envelope evolution but do not merge due to the weakness of our CEE implementation. This is at odds with other binary evolution models that try to reproduce the Gaia BH systems. Other studies have shown that Gaia BH1 is difficult to reconcile with standard isolated binary evolution because the progenitor is expected to undergo a common-envelope phase, yet a successful common-envelope episode would generally leave a system much tighter than the observed 185-day orbit (e.g. \citet{2023MNRAS.526..740R, 2023MNRAS.522.1763G}). However, while such models are possible they are also very unlikely given the extreme mass ratio required. To this end, it is more likely that most of the Gaia BH systems have been dynamically formed. Comparatively, the neutron star systems presented in \citet{elbadry2024populationneutronstarcandidates} also have analogous systems within the \textsc{BPASS} synthetic population models. However, shorter period systems are again more likely. For both object types there is likely to be a selection effect against shorter period systems being detected by astrometry with Gaia.

We can see that the population of the compact remnant binaries shifts post-interaction in Figure \ref{fig:quin1obs}, where we show post-interaction systems before any evolve into WDs. Comparing the synthetic population to the observations plotted over, we can start to get a picture of what these systems may evolve into over time. In general, after mass transfer, we have systems with shorter periods, more massive remnants and less massive donors. 

Most noticeable in Figure \ref{fig:quin1obs} but present to a degree in every plot, there is an ``island'' of systems in the lower left of the period versus donor mass plots. This ``island'' is filled with sub-solar donors with sub-day periods. As stated previously, this population is a representation of a current MW-like galaxy. This island population matches with observed short period millisecond pulsars, suggesting the \textsc{BPASS} models are reproducing these systems. As for the gap, the \textsc{BPASS} models have time steps where models are in that area, however they do not remain there for long. This is an indication that we need to reassess the physics implemented in the synthesis of these models.

When comparing Figure \ref{fig:quin0} to Figure \ref{fig:wd}, we can see the end-state of MSPs and how the bifurcation period may affect their progenitors \citep{1988A&A...191...57P,1989A&A...208...52P,2009ApJ...691.1611M}. Those with periods longer than a day are expected to have their period diverge into a MSP with a HeWD. If we look at the systems with neutron stars, those with stellar mass donors and periods $\lesssim1\rm{day}$ at the onset of RLOF are expected to converge into UCXRBs \citep{2002ApJ...565.1107P, 1988A&A...191...57P,1989A&A...208...52P}. These are expected to consist of systems in continuous RLOF in a tight orbit \citep{2002ApJ...565.1107P}, or systems with a NS and HeWD with an orbital period less than nine hours that contact within a Hubble time due to gravitational wave emission. Given the approximations in Section \ref{methods} and the current physics in \textsc{BPASS}, some systems plotted in Figure \ref{fig:wd} may contain UCXRBs. While we do not include magnetic braking within the \textsc{BPASS} models, this is a likely explanation for the increase in subsolar donors with sub-day periods in Figure \ref{fig:quin1obs} and Figure \ref{fig:wd}. 

\subsection{Interacting Systems}\label{xraydisc}

Interacting systems have been widely studied in previous population studies \citep[e.g.][]{Corral_Santana_2016,2024MNRAS.527.5023L,2023ApJ...954..212S,2003ApJ...597.1036P,2005MNRAS.356..401R}. These have typically focused on binaries in either low-mass or high-mass groups. We have not made this division here and instead look at the full range of objects to see if such a division occurs within the synthetic populations.

While we utilised the Bondi-Hoyle-Lyttleton model for most wind-fed XRBs, we matched our Be-XRBs models to the known characteristics of this observed population \citep{2024MNRAS.527.5023L,2025MNRAS.542.2087B}. In Figure \ref{fig:Be} we can see that we match the Be-XRB population rather well. An interesting result is that while most systems have a neutron star, our synthetic population yields a handful of black hole-Be systems at a ratio of $\sim109_{\rm{NS}}:1_{\rm{BH}}$. This is a larger ratio than found by \citet{LackofBeBH2009ApJ...707..870B} of $30_{\rm{NS}}:1_{\rm{BH}}$. Assuming that MWC 656 \citep{MWCXB2014ApJ...786L..11M} and HD 96670 \citep{HDisXB2021ApJ...913...48G} do contain black holes, which suggests an observed ratio of $79_{\rm{NS}}:1_{\rm{BH}}$ (assuming that all unidentified systems in the Binary rEvolution catalogue are NSs), then our results are closer to what is found in observations. However, further observational work is needed to verify this. This may prove difficult as Be-BH systems are predicted to be transient sources that spend a significant amount  of time in quiescence \citep{BePablo2011Ap&SS.332....1R}. One key formation factor that \citet{LackofBeBH2009ApJ...707..870B} presents is that the common envelope phase may lead to systems where more comparable masses have a better chance of surviving. \citet{LackofBeBH2009ApJ...707..870B} also asserts that the total galactic population of Be-XRBs should be approximately one or two orders of magnitude larger than the current observed population, which aligns with our predicted population.

When we consider the rest of the wind-driven systems in our population, we again see strong correlation with the observational data for HMXRBs. Comparing Figures \ref{fig:WindFed} and \ref{fig:RLOF}, we see that the observed HMXRBs have a greater overlap with the wind-fed models. While the RLOF binaries are in that parameter space, our models suggest they are more likely to be wind-fed as the observations suggest.

Additionally, in our synthetic populations we see a significant number of wind-fed systems that have donors below $3M_{\odot}$, the majority of which are in shorter orbits. Our wind-driven population is of higher metallicity on average as stronger winds in short orbits make it more likely for wind-fed accretion. We do anticipate these lower mass systems will be hard to identify. According to theories such as those presented in  \citet{1944MNRAS.104..273B} and \citet{1976ApJ...204..555S}, these systems would form an accretion disk like structure without the donor filling their Roche-lobe. Understandably, these systems may appear as a RLOF systems rather than wind driven systems. Additionally, as discussed in Section \ref{windfedresults}, these could also be SyXRBs where there is an NS accretor with a late-life giant star \citep{2012MNRAS.424.2265L,2019MNRAS.485..851Y}. Nonetheless, our work shows there should be a large number of these smaller wind driven donors and this still presents an interesting focus of study. 

When we consider our two RLOF categories, in Figures \ref{fig:RLOF} and \ref{fig:quinrlof}, we see that there are many more X-ray faint systems than the bright ones. Additionally, the vast majority of the faint systems contain black holes rather than neutron stars. This is to be expected as in Equation \ref{Teffeq}, $\rm{T_{eff}}$ is dependent on the mass transfer rate and inversely dependent on the radius inner edge of the accretion disk of the compact object. This leads to a situation where, in our calculations for equal mass loss rates, $\rm{T_{eff}}$ will be higher for a NS than a BH. Thus, the majority of our systems with NS are X-ray bright compared to the BH systems. Physically speaking, a BH lacks the boundary layer possessed by a weakly magnetised NS. Thus, the amount of energy that infalling matter can release is limited. Without exotic conditions such as very high mass transfer rates and rapidly rotating BHs, the accretion disk of a BH will have weaker irradiation compared to that of a NS, leading to a lower effective temperature. Hence, it is understandable that these systems primarily contain a black hole. In particular, these BH systems are expected to be soft X-ray transients (SXTs) due to disk instability events rather than continuous X-ray emission. Conversely, NS systems make up the majority of the X-ray bright systems. The boundary layer of a weakly magnetised NS can reach temperatures exceeding $10^8\rm{K}$ and emits a large amount of X-rays. \citep{2001ApJ...547..355P} Overall, these systems will have a higher $\rm{T_{eff}}$ than their equivalent BH system. Finally, the majority of the faint RLOF systems have long periods with Solar mass donors. These stars will be filling their Roche lobe as red giants. The evolution of such stars occurs on a nuclear timescale as the helium core grows through hydrogen shell burning before the helium flash. This means the mass transfer rates are low given that nuclear evolution is driving the RLOF, again making these systems faint.

%%(Pfahl predicts anywhere from 100 to 100,000(ish) luminous low and intermediate sources. Remarks that these numbers are sensitive to the "parameters in the analytic formula describing the CEE.)

\subsection{Space Telescope Surveys}\label{STSec}

Our predicted results for the future space telescope surveys are not attempts to reproduce the exact observations that will be undertaken, but rather an estimate based on calculations of extinction, astrometric wobble, and the astrometric precision achievable by the telescopes. Additionally, our simulations assume that we view a binary ``face on'' so that the magnitude of the luminous companion is constant throughout the orbit and that the compact object does not eclipse the luminous companion. Changes in magnitude over an orbit (eclipsing binary) may serve as another test to determine the nature of these quiescent systems, however, such calculations are outside the scope of this paper. 

While we have considered the telescopes separately here, it is important to note that Roman and Gaia will provide important results when analysed together. \citet{2026PASP..138d4507M} reported on how the two can work together to improve upon their position errors, parallax, and proper motion measurements. Part of their work included the development of a new tool\footnote{\url{https://github.com/KevinMcK95/gaia_roman_astrometry}} to simulate combined precision from the two telescopes. They found that in some surveys, Roman will deliver Gaia DR3-quality proper motions down to $G = 29.0$. In terms of the GBTDS, Roman will deliver PMs and Parallaxes greater than Gaia DR5 alone. Their methods of utilising both telescopes yields even greater results for the faintest systems. 

Simulations of the core Roman surveys is a relatively new area of study. Previous studies have explored exoplanet yields \citep{2023ApJS..269....5W} as well as the possibility of finding primordial black holes with the GBTDS \citet{2024ApJ...965..138F}. While a few white papers have explored the use of microlensing for isolated black holes \citep{2023arXiv230612514L} as well as potential self-microlensing binary populations in the Galactic Bulge \citep{2023arXiv230616479G}, no in-depth study of potential populations such as this exists yet. 

In this, we predict that the GBTDS holds promise for the future of quiescent binary study with tens of black holes and hundreds of neutron stars that meet our criteria. While fewer in number, the high-cadence, high-precision nature of the survey should lead to very well constrained systems. In the case of Roman BHs, we predict them to primarily be post-interaction systems. Should this be the case these systems would provide valuable insight into what systems look like after engaging in mass-transfer.

In addition to the astrometric observations, Roman is set to collect grism spectra for sources in the field. The final report from the time allocation committee \citep{committee2025romanobservationstimeallocation} explains that this data will be used to measure stellar temperatures, metallicity, and radial velocities. The spectra snapshots are confusion-limited at magnitudes $\rm{K\lesssim16}$ \citep{committee2025romanobservationstimeallocation}. Although our analysis shows that the vast majority of our theoretical companions are within this limit, our analysis only looks at binary stars so other bright objects in the field will no doubt complicate this. $30$ epochs of spectroscopic snapshots will be taken over the survey. Comparing these grism spectral snapshots with colour information will no doubt prove useful in studying and classifying any potential observed systems.

As for Gaia, several studies for quiescent binary populations within the data releases exist \citep{RevealingBHwGaia, 2019ApJ...886...68A, 2025PASP..137d4202N, 2025arXiv250821805C}. Several of these studies share methodology with ours such as the use of the same dustmap and \textsc{FIRE} simulation \citep{2025arXiv250821805C}.

In general, our results for Gaia contain hundreds of BHs and thousands of NSs. While our numbers are similar to those in \citet{RevealingBHwGaia}, they are roughly twice the number presented in \citet{2025arXiv250821805C}. Several factors contribute to this. \citet{2025arXiv250821805C} uses the same dustmap and \textsc{FIRE} model as this study but do not consider X-ray faint systems in their population. Additionally, we make use of a theoretical DR5 error function where other studies apply the DR3 function to longer period systems. Our larger population is, in part, due the CEE prescription in \texttt{BPASS} being ``weak". That is to say, in our code, we have fewer mergers than other codes and therefore have more systems in our population. 

Taking a closer look at Figure \ref{fig:teleP}, we see that the Gaia systems skew toward longer periods that trace larger angles on the sky. Additionally, in Figure \ref{fig:gaiaSPACE} we can see how the detectability of a system is related to distance. The post-interaction systems are the most constrained by distance. The donors are generally about $\sim0.5\rm{M_{\odot}}$ and have previously gone through some sort of mass transfer. At this point, they will be relatively dim and difficult to detect at larger distances. Some pre-interaction systems in our population are detectable upwards of tens of kiloparsecs away in DR3 and even further in DR5. Whether or not these are actually able to be detected has yet to be seen. Of the observed quiescent binaries, Gaia BH2 is the furthest BH at $\rm{1.16~kpc}$ away \citep{El_Badry2023} while the furthest Gaia NS is $\rm{1.5~kpc}$ away \citep{elbadry2024populationneutronstarcandidates}. 

When talking about these telescopes, we must also consider any observational complications. While we include X-ray faint systems in our population, these will be challenging to observe and constrain using astrometric measurements. When computing the magnitude of the X-ray faint systems, we rely solely on the magnitude of the donor star and do not model any contribution from the accretion disk. That is to say, we assume the disk and accretion stream is relatively dim or invisible in a respective passband. \citet{2025MNRAS.542.2087B} documents the start of integrating X-ray emission into the \textsc{BPASS} suite. While the authors report that the disk emission does not dominate the donor emission in their models, disk emission should still be considered. Cooler accretion disks ($\rm{T\lesssim10^{6}K}$), are expected to primarily radiate in the IR, optical, and UV. A relatively bright disk can cause confusion when analysing observations. Systems such as $\rm{V1408\ Aquilae (4U\ 1957+115)}$ and its subsequent studies \citep{2011ApJ...730...43B, 2014MNRAS.444.3802H, 2015ApJ...809....9G} provide an example on the confusion between disk and stellar emission. Additionally, given temperature and mass distributions, it is not unreasonable to expect some of our X-ray faint systems to be soft X-ray transients \citep[SXTs]{1996ApJ...457..821N, 1997ApJ...489..865E, 1999MNRAS.303..139D, 2001A&A...373..251D}. This confusing or ``veiling" effect has been observed in both optical \citep{2015ApJ...806...92W,2016ApJ...825...46W,2022ApJ...925...83Z} as well as IR \citep{2007MNRAS.374..657R}.

As mentioned previously, the Galactic Bulge is not easily studied with differential extinction being one of the biggest factors to consider \citep{2013A&A...550A..35P, 2014A&A...566A.120S, 2016ApJ...818..130B, 2016PASA...33...24N, 2025AJ....170..328Z}. While these effects are more noticeable in optical wavelengths, it is still a concern in IR and was considered when selecting the GBTDS field \citep{committee2025romanobservationstimeallocation}. Beyond extinction, dense stellar regions like the bulge and plane have been known to complicate observations resulting in inaccurate astrometric solutions \citep{2023A&A...677A.185L}. While the above were not considered in this work, they are caveats that astronomers will need to keep in mind. 

\section{Conclusions}\label{concl}

In this study, we have combined the \textsc{BPASS} stellar models with a synthetic \textsc{FIRE} Milky Way analogue to create synthetic populations of binary systems of stars with compact objects across all their evolutionary stages. We are able to predict total numbers of systems as well as mass distributions and periods of said systems. From this, we are able to also evaluate our models for their strengths and weaknesses. The new pipeline developed for this project will also be used to evaluate future developments of the \textsc{BPASS} project.

Our conclusions are as follows,
\begin{enumerate}
    \item We predict that the Galaxy contains 1567784 quiescent pre-interaction binaries, which have periods peaking at a few days out to a maximum of 10,000 days. The compact remnants in these binaries are predominately neutron stars but black hole masses of up to $30{\rm M_{\odot}}$ are possible. Donor masses range from $0.8{\rm M_{\odot}}$ up to $100{\rm M_{\odot}}$. Lower mass companions are possible but this is through the low-mass stars evolving and losing mass without interacting with the compact remnant.
    \item We do not find exact matches to the Gaia-BH systems, however this is due to the minimum mass ratio of \textsc{BPASS} models. Bespoke models indicate that such models are possible with the \textsc{BPASS} stellar evolution code but will be rare due to the low mass ratios required. We find reasonable matches for the observed Gaia NS candidates found in \citet{elbadry2024populationneutronstarcandidates}.
    \item For interacting systems we find good agreement between our predicted HMXRB wind-fed population and that observed in the Galaxy. However we predict there should be a population of lower mass wind-fed systems formed as Solar mass stars ascend the RGB.
    \item Our synthetic Be-XRB population matches the observed population. This validates the scheme used to identify such systems as outlined \citet{2025MNRAS.542.2087B}.
    \item Our RLOF predictions match the observed systems reasonably well, although we note there is a gap in our synthetic populations in donor mass--period space. While our models do evolve through this region, they move through it rapidly such that we predict few systems in that space. As there are observations in that region, this indicates that revisions to the mass transfer physics of low-mass donors may be required to improve the \textsc{BPASS} stellar evolution model (i.e. magnetic wind braking).
    \item Our RLOF models also predict a large number of X-ray faint systems that outnumber the X-ray bright RLOF systems. These systems are expected to primarily radiate in IR, optical, and UV. Some of these may fall into the category of soft X-ray transients. We can expect that the known Gaia-BHs may evolve into such systems.
    \item The post-interaction quiescent binaries agree with the observed distribution of MSPs with main-sequence companions. Quiescent binaries where the donor star has evolved into a white dwarf match some of our models, but as discussed above we see a gap for low-mass systems. Again, indicating missing physics in the binary evolution for low-mass donors.
    \item We predict that Roman's GBTDS should yield tens of well constrained BH quiescent binaries and hundreds containing NSs and is a promising survey for time-domain astronomy. 
    \item Gaia-DR5 should roughly double the number of resolvable quiescent binaries over DR3. We also  predict far more systems with periods less than 10.5 years compared to comparable studies (see Table \ref{tab:GaiaRomanNumsTable}). 
    \item Our model galaxy can be used to make predictions for future observational missions as well as help refine the \textsc{BPASS} stellar models. 
\end{enumerate}

Taking the above conclusions together, we can have some confidence that the predicted synthetic populations match the observations within an order of magnitude. While exact matches are not possible for every subtype we do predict systems close to those we have observed in the Galaxy. However, there are strong signs that we are not correctly reproducing compact remnant binaries with low-mass donors. This indicates that we are missing important physics, most likely including magnetic braking and potentially enhanced angular momentum losses such as gravitational-wave radiation.. Furthermore, using the numerical architecture created it will be possible to vary the physics within the \textsc{BPASS} model to determine how the synthetic populations predicted depend on these uncertain processes as well as others such as the supernova kicks and remnant mass distributions.
\section*{Acknowledgements}\label{ack}
EE, JJE, and JCB acknowledge the support provided by the University of Auckland and funding from the Royal Society Te Ap$\bar{\textrm{a}}$rangi of New Zealand Marsden Grant Scheme. Additionally, this work made use of the eResearch Infrastructure Platform hosted by the Crown company, Research and Education Advanced Network New Zealand (REANNZ) Ltd., and funded by the Ministry of Business, Innovation \& Employment\footnote{\url{https://www.reannz.co.nz}}.

This work includes data from the catalogues maintained by the Binary rEvolution\footnote{\url{https://github.com/Binary-rEvolution}} team as well as the ANTF pulsar catalogue\footnote{\url{http://www.atnf.csiro.au/research/pulsar/psrcat}}. \citep{Manchester_2005}
\section*{Data Availability}\label{data_avail}

\textsc{BPASS} stellar models are available from the \textsc{BPASS} websites at \url{http://bpass.auckland.ac.nz} and \url{https://warwick.ac/uk/bpass}. Additional data is available upon request from the authors.
% \end{thebibliography}
\bibliographystyle{paslike}
\bibliography{bib}

@ARTICLE{Giesers2018,
       author = {{Giesers}, Benjamin and {Dreizler}, Stefan and {Husser}, Tim-Oliver and {Kamann}, Sebastian and {Anglada Escud{\'e}}, Guillem and {Brinchmann}, Jarle and {Carollo}, C. Marcella and {Roth}, Martin M. and {Weilbacher}, Peter M. and {Wisotzki}, Lutz},
        title = "{A detached stellar-mass black hole candidate in the globular cluster NGC 3201}",
      journal = {MNRAS},
         year = 2018,
        month = mar,
       volume = {475},
       number = {1},
        pages = {L15-L19},
          doi = {10.1093/mnrasl/slx203},
archivePrefix = {arXiv},
       eprint = {1801.05642},
 primaryClass = {astro-ph.SR},
       adsurl = {https://ui.adsabs.harvard.edu/abs/2018MNRAS.475L..15G}
}

@ARTICLE{Giesers2019,
       author = {{Giesers}, Benjamin and {Kamann}, Sebastian and {Dreizler}, Stefan and {Husser}, Tim-Oliver and {Askar}, Abbas and {G{\"o}ttgens}, Fabian and {Brinchmann}, Jarle and {Latour}, Marilyn and {Weilbacher}, Peter M. and {Wendt}, Martin and {Roth}, Martin M.},
        title = "{A stellar census in globular clusters with MUSE: Binaries in NGC 3201}",
      journal = {\aap},
         year = 2019,
        month = dec,
       volume = {632},
          eid = {A3},
        pages = {A3},
          doi = {10.1051/0004-6361/201936203},
archivePrefix = {arXiv},
       eprint = {1909.04050},
 primaryClass = {astro-ph.SR},
       adsurl = {https://ui.adsabs.harvard.edu/abs/2019A&A...632A...3G}
}

@article{El_Badry2023,
   title={A red giant orbiting a black hole},
   volume={521},
   ISSN={1365-2966},
   url={http://dx.doi.org/10.1093/mnras/stad799},
   DOI={10.1093/mnras/stad799},
   number={3},
   journal={Monthly Notices of the Royal Astronomical Society},
   publisher={Oxford University Press (OUP)},
   author={El-Badry, Kareem and Rix, Hans-Walter and Cendes, Yvette and Rodriguez, Antonio C and Conroy, Charlie and Quataert, Eliot and Hawkins, Keith and Zari, Eleonora and Hobson, Melissa and Breivik, Katelyn and Rau, Arne and Berger, Edo and Shahaf, Sahar and Seeburger, Rhys and Burdge, Kevin B and Latham, David W and Buchhave, Lars A and Bieryla, Allyson and Bashi, Dolev and Mazeh, Tsevi and Faigler, Simchon},
   year={2023},
   month=mar, pages={4323–4348} }

@article{se2018,
    author = {Stanway, E R and Eldridge, J J},
    title = "{Re-evaluating old stellar populations}",
    journal = {Monthly Notices of the Royal Astronomical Society},
    volume = {479},
    number = {1},
    pages = {75-93},
    year = {2018},
    month = {05},
    issn = {0035-8711},
    doi = {10.1093/mnras/sty1353},
    url = {https://doi.org/10.1093/mnras/sty1353},
    eprint = {https://academic.oup.com/mnras/article-pdf/479/1/75/25111406/sty1353.pdf},
}

@ARTICLE{2023ApJS..265...44W,
       author = {{Wetzel}, Andrew and {Hayward}, Christopher C. and {Sanderson}, Robyn E. and {Ma}, Xiangcheng and {Angl{\'e}s-Alc{\'a}zar}, Daniel and {Feldmann}, Robert and {Chan}, T.~K. and {El-Badry}, Kareem and {Wheeler}, Coral and {Garrison-Kimmel}, Shea and {Nikakhtar}, Farnik and {Panithanpaisal}, Nondh and {Arora}, Arpit and {Gurvich}, Alexander B. and {Samuel}, Jenna and {Sameie}, Omid and {Pandya}, Viraj and {Hafen}, Zachary and {Hummels}, Cameron and {Loebman}, Sarah and {Boylan-Kolchin}, Michael and {Bullock}, James S. and {Faucher-Gigu{\`e}re}, Claude-Andr{\'e} and {Kere{\v{s}}}, Du{\v{s}}an and {Quataert}, Eliot and {Hopkins}, Philip F.},
        title = "{Public Data Release of the FIRE-2 Cosmological Zoom-in Simulations of Galaxy Formation}",
      journal = {\apjs},
         year = 2023,
        month = apr,
       volume = {265},
       number = {2},
          eid = {44},
        pages = {44},
          doi = {10.3847/1538-4365/acb99a},
archivePrefix = {arXiv},
       eprint = {2202.06969},
 primaryClass = {astro-ph.GA},
       adsurl = {https://ui.adsabs.harvard.edu/abs/2023ApJS..265...44W}
}

@ARTICLE{2014MNRAS.445..581H,
       author = {{Hopkins}, Philip F. and {Kere{\v{s}}}, Du{\v{s}}an and {O{\~n}orbe}, Jos{\'e} and {Faucher-Gigu{\`e}re}, Claude-Andr{\'e} and {Quataert}, Eliot and {Murray}, Norman and {Bullock}, James S.},
        title = "{Galaxies on FIRE (Feedback In Realistic Environments): stellar feedback explains cosmologically inefficient star formation}",
      journal = {MNRAS},
         year = 2014,
        month = nov,
       volume = {445},
       number = {1},
        pages = {581-603},
          doi = {10.1093/mnras/stu1738},
archivePrefix = {arXiv},
       eprint = {1311.2073},
 primaryClass = {astro-ph.CO},
       adsurl = {https://ui.adsabs.harvard.edu/abs/2014MNRAS.445..581H}
}

@ARTICLE{2018MNRAS.480..800H,
       author = {{Hopkins}, Philip F. and {Wetzel}, Andrew and {Kere{\v{s}}}, Du{\v{s}}an and {Faucher-Gigu{\`e}re}, Claude-Andr{\'e} and {Quataert}, Eliot and {Boylan-Kolchin}, Michael and {Murray}, Norman and {Hayward}, Christopher C. and {Garrison-Kimmel}, Shea and {Hummels}, Cameron and {Feldmann}, Robert and {Torrey}, Paul and {Ma}, Xiangcheng and {Angl{\'e}s-Alc{\'a}zar}, Daniel and {Su}, Kung-Yi and {Orr}, Matthew and {Schmitz}, Denise and {Escala}, Ivanna and {Sanderson}, Robyn and {Grudi{\'c}}, Michael Y. and {Hafen}, Zachary and {Kim}, Ji-Hoon and {Fitts}, Alex and {Bullock}, James S. and {Wheeler}, Coral and {Chan}, T.~K. and {Elbert}, Oliver D. and {Narayanan}, Desika},
        title = "{FIRE-2 simulations: physics versus numerics in galaxy formation}",
      journal = {MNRAS},
         year = 2018,
        month = oct,
       volume = {480},
       number = {1},
        pages = {800-863},
          doi = {10.1093/mnras/sty1690},
archivePrefix = {arXiv},
       eprint = {1702.06148},
 primaryClass = {astro-ph.GA},
       adsurl = {https://ui.adsabs.harvard.edu/abs/2018MNRAS.480..800H}
}

@ARTICLE{2015MNRAS.450...53H,
       author = {{Hopkins}, Philip F.},
        title = "{A new class of accurate, mesh-free hydrodynamic simulation methods}",
      journal = {MNRAS},
         year = 2015,
        month = jun,
       volume = {450},
       number = {1},
        pages = {53-110},
          doi = {10.1093/mnras/stv195},
archivePrefix = {arXiv},
       eprint = {1409.7395},
 primaryClass = {astro-ph.CO},
       adsurl = {https://ui.adsabs.harvard.edu/abs/2015MNRAS.450...53H}
}

@article{El_Badry2023-1,
    author = {El-Badry, Kareem and Rix, Hans-Walter and Quataert, Eliot and Howard, Andrew W and Isaacson, Howard and Fuller, Jim and Hawkins, Keith and Breivik, Katelyn and Wong, Kaze W K and Rodriguez, Antonio C and Conroy, Charlie and Shahaf, Sahar and Mazeh, Tsevi and Arenou, Frédéric and Burdge, Kevin B and Bashi, Dolev and Faigler, Simchon and Weisz, Daniel R and Seeburger, Rhys and Almada Monter, Silvia and Wojno, Jennifer},
    title = "{A Sun-like star orbiting a black hole}",
    journal = {Monthly Notices of the Royal Astronomical Society},
    volume = {518},
    number = {1},
    pages = {1057-1085},
    year = {2022},
    month = {11},
    issn = {0035-8711},
    doi = {10.1093/mnras/stac3140},
    url = {https://doi.org/10.1093/mnras/stac3140},
    eprint = {https://academic.oup.com/mnras/article-pdf/518/1/1057/47139767/stac3140.pdf},
}

@article{Manchester_2005,
   title={The Australia Telescope National Facility Pulsar Catalogue},
   volume={129},
   ISSN={1538-3881},
   url={http://dx.doi.org/10.1086/428488},
   DOI={10.1086/428488},
   number={4},
   journal={The Astronomical Journal},
   publisher={American Astronomical Society},
   author={Manchester, R. N. and Hobbs, G. B. and Teoh, A. and Hobbs, M.},
   year={2005},
   month=apr, pages={1993–2006} }

@misc{elbadry2024populationneutronstarcandidates,
      title={A population of neutron star candidates in wide orbits from Gaia astrometry}, 
      author={Kareem El-Badry and Hans-Walter Rix and David W. Latham and Sahar Shahaf and Tsevi Mazeh and Allyson Bieryla and Lars A. Buchhave and René Andrae and Natsuko Yamaguchi and Howard Isaacson and Andrew W. Howard and Alessandro Savino and Ilya V. Ilyin},
      year={2024},
      eprint={2405.00089},
      archivePrefix={arXiv},
      primaryClass={astro-ph.SR},
      doi={https://doi.org/10.33232/001c.121261},
      url={https://arxiv.org/abs/2405.00089}, 
}

@misc{tang2024predictinggravitationalwavesignals,
      title={Predicting gravitational wave signals from BPASS White Dwarf Binary and Black Hole Binary populations of a Milky Way-like galaxy model for LISA}, 
      author={Petra Tang and Jan Eldridge and Renate Meyer and Astrid Lamberts and Guillaume Boileau and Wouter van Zeist},
      year={2024},
      eprint={2405.20484},
      archivePrefix={arXiv},
      primaryClass={astro-ph.GA},
      url={https://arxiv.org/abs/2405.20484}, 
}

@article{Eldridge_2008,
   title={The effect of massive binaries on stellar populations and supernova progenitors},
   volume={384},
   ISSN={1365-2966},
   url={http://dx.doi.org/10.1111/j.1365-2966.2007.12738.x},
   DOI={10.1111/j.1365-2966.2007.12738.x},
   number={3},
   journal={Monthly Notices of the Royal Astronomical Society},
   publisher={Oxford University Press (OUP)},
   author={Eldridge, J.J. and Izzard, Robert G. and Tout, Christopher A.},
   year={2008},
   month=mar, pages={1109–1118} }

@ARTICLE{Sana2012,
       author = {{Sana}, H. and {de Mink}, S.~E. and {de Koter}, A. and {Langer}, N. and {Evans}, C.~J. and {Gieles}, M. and {Gosset}, E. and {Izzard}, R.~G. and {Le Bouquin}, J. -B. and {Schneider}, F.~R.~N.},
        title = "{Binary Interaction Dominates the Evolution of Massive Stars}",
      journal = {Science},
         year = 2012,
        month = jul,
       volume = {337},
       number = {6093},
        pages = {444},
          doi = {10.1126/science.1223344},
archivePrefix = {arXiv},
       eprint = {1207.6397},
 primaryClass = {astro-ph.SR},
       adsurl = {https://ui.adsabs.harvard.edu/abs/2012Sci...337..444S}
}

@article{Corral_Santana_2016,
   title={BlackCAT: A catalogue of stellar-mass black holes in X-ray transients},
   volume={587},
   ISSN={1432-0746},
   url={http://dx.doi.org/10.1051/0004-6361/201527130},
   DOI={10.1051/0004-6361/201527130},
   journal={Astronomy \& Astrophysics},
   publisher={EDP Sciences},
   author={Corral-Santana, J. M. and Casares, J. and Muñoz-Darias, T. and Bauer, F. E. and Martínez-Pais, I. G. and Russell, D. M.},
   year={2016},
   month=feb, pages={A61} }

@article{Kobayashi_2023,
   title={Can Neutron Star Mergers Alone Explain the r-process Enrichment of the Milky Way?},
   volume={943},
   ISSN={2041-8213},
   url={http://dx.doi.org/10.3847/2041-8213/acad82},
   DOI={10.3847/2041-8213/acad82},
   number={2},
   journal={The Astrophysical Journal Letters},
   publisher={American Astronomical Society},
   author={Kobayashi, Chiaki and Mandel, Ilya and Belczynski, Krzysztof and Goriely, Stephane and Janka, Thomas H. and Just, Oliver and Ruiter, Ashley J. and Vanbeveren, Dany and Kruckow, Matthias U. and Briel, Max M. and Eldridge, Jan J. and Stanway, Elizabeth},
   year={2023},
   month=jan, pages={L12} }

@article{BH3_2024,
   title={Discovery of a dormant 33 solar-mass black hole in pre-release Gaia astrometry},
   volume={686},
   ISSN={1432-0746},
   url={http://dx.doi.org/10.1051/0004-6361/202449763},
   DOI={10.1051/0004-6361/202449763},
   journal={Astronomy \& Astrophysics},
   publisher={EDP Sciences},
   author={Panuzzo, P. and Mazeh, T. and Arenou, F. and Holl, B. and Caffau, E. and Jorissen, A. and Babusiaux, C. and Gavras, P. and Sahlmann, J. and Bastian, U. and Wyrzykowski, Ł. and Eyer, L. and Leclerc, N. and Bauchet, N. and Bombrun, A. and Mowlavi, N. and Seabroke, G. M. and Teyssier, D. and Balbinot, E. and Helmi, A. and Brown, A. G. A. and Vallenari, A. and Prusti, T. and de Bruijne, J. H. J. and Barbier, A. and Biermann, M. and Creevey, O. L. and Ducourant, C. and Evans, D. W. and Guerra, R. and Hutton, A. and Jordi, C. and Klioner, S. A. and Lammers, U. and Lindegren, L. and Luri, X. and Mignard, F. and Nicolas, C. and Randich, S. and Sartoretti, P. and Smiljanic, R. and Tanga, P. and Walton, N. A. and Aerts, C. and Bailer-Jones, C. A. L. and Cropper, M. and Drimmel, R. and Jansen, F. and Katz, D. and Lattanzi, M. G. and Soubiran, C. and Thévenin, F. and van Leeuwen, F. and Andrae, R. and Audard, M. and Bakker, J. and Blomme, R. and Castañeda, J. and De Angeli, F. and Fabricius, C. and Fouesneau, M. and Frémat, Y. and Galluccio, L. and Guerrier, A. and Heiter, U. and Masana, E. and Messineo, R. and Nienartowicz, K. and Pailler, F. and Riclet, F. and Roux, W. and Sordo, R. and Gracia-Abril, G. and Portell, J. and Altmann, M. and Benson, K. and Berthier, J. and Burgess, P. W. and Busonero, D. and Busso, G. and Cacciari, C. and Cánovas, H. and Carrasco, J. M. and Carry, B. and Cellino, A. and Cheek, N. and Clementini, G. and Damerdji, Y. and Davidson, M. and de Teodoro, P. and Delchambre, L. and Dell’Oro, A. and Fraile Garcia, E. and Garabato, D. and García-Lario, P. and Haigron, R. and Hambly, N. C. and Harrison, D. L. and Hatzidimitriou, D. and Hernández, J. and Hestroffer, D. and Hodgkin, S. T. and Jamal, S. and Jevardat de Fombelle, G. and Jordan, S. and Krone-Martins, A. and Lanzafame, A. C. and Löffler, W. and Lorca, A. and Marchal, O. and Marrese, P. M. and Moitinho, A. and Muinonen, K. and Nuñez Campos, M. and Oreshina-Slezak, I. and Osborne, P. and Pancino, E. and Pauwels, T. and Recio-Blanco, A. and Riello, M. and Rimoldini, L. and Robin, A. C. and Roegiers, T. and Sarro, L. M. and Schultheis, M. and Smith, M. and Sozzetti, A. and Utrilla, E. and van Leeuwen, M. and Weingrill, K. and Abbas, U. and Ábrahám, P. and Abreu Aramburu, A. and Ahmed, S. and Altavilla, G. and Álvarez, M. A. and Anders, F. and Anderson, R. I. and Anglada Varela, E. and Antoja, T. and Baig, S. and Baines, D. and Baker, S. G. and Balaguer-Núñez, L. and Balog, Z. and Barache, C. and Barros, M. and Barstow, M. A. and Bartolomé, S. and Bashi, D. and Bassilana, J.-L. and Baudeau, N. and Becciani, U. and Bedin, L. R. and Bellas-Velidis, I. and Bellazzini, M. and Beordo, W. and Bernet, M. and Bertolotto, C. and Bertone, S. and Bianchi, L. and Binnenfeld, A. and Blanco-Cuaresma, S. and Bland-Hawthorn, J. and Blazere, A. and Boch, T. and Bossini, D. and Bouquillon, S. and Bragaglia, A. and Braine, J. and Bratsolis, E. and Breedt, E. and Bressan, A. and Brouillet, N. and Brugaletta, E. and Bucciarelli, B. and Butkevich, A. G. and Buzzi, R. and Camut, A. and Cancelliere, R. and Cantat-Gaudin, T. and Capilla Guilarte, D. and Carballo, R. and Carlucci, T. and Carnerero, M. I. and Carretero, J. and Carton, S. and Casamiquela, L. and Casey, A. and Castellani, M. and Castro-Ginard, A. and Ceraj, L. and Cesare, V. and Charlot, P. and Chaudet, C. and Chemin, L. and Chiavassa, A. and Chornay, N. and Chosson, D. and Cooper, W. J. and Cornez, T. and Cowell, S. and Crosta, M. and Crowley, C. and Cruz Reyes, M. and Dafonte, C. and Dal Ponte, M. and David, M. and de Laverny, P. and De Luise, F. and De March, R. and de Torres, A. and del Peloso, E. F. and Delbo, M. and Delgado, A. and Delisle, J.-B. and Demouchy, C. and Denis, E. and Dharmawardena, T. E. and Di Giacomo, F. and Diener, C. and Distefano, E. and Dolding, C. and Dsilva, K. and Enke, H. and Fabre, C. and Fabrizio, M. and Faigler, S. and Fatović, M. and Fedorets, G. and Fernández-Hernández, J. and Fernique, P. and Figueras, F. and Fouron, C. and Fragkoudi, F. and Gai, M. and Galinier, M. and Garcia-Serrano, A. and García-Torres, M. and Garofalo, A. and Gerlach, E. and Geyer, R. and Giacobbe, P. and Gilmore, G. and Girona, S. and Giuffrida, G. and Gomboc, A. and Gomez, A. and González-Santamaría, I. and Gosset, E. and Granvik, M. and Gregori Barrera, V. and Gutiérrez-Sánchez, R. and Haywood, M. and Helmer, A. and Hidalgo, S. L. and Hilger, T. and Hobbs, D. and Hottier, C. and Huckle, H. E. and Jiménez-Arranz, Ó. and Juaristi Campillo, J. and Kaczmarek, Z. and Kervella, P. and Khanna, S. and Kontizas, M. and Kordopatis, G. and Korn, A. J. and Kóspál, Á and Kostrzewa-Rutkowska, Z. and Kruszyńska, K. and Kun, M. and Lambert, S. and Lanza, A. F. and Lebreton, Y. and Lebzelter, T. and Leccia, S. and Lecoutre, G. and Liao, S. and Liberato, L. and Licata, E. and Livanou, E. and Lobel, A. and López-Miralles, J. and Loup, C. and Madarász, M. and Mahy, L. and Mann, R. G. and Manteiga, M. and Marcellino, C. P. and Marchant, J. M. and Marconi, M. and Marín Pina, D. and Marinoni, S. and Marshall, D. J. and Martín Lozano, J. and Martin Polo, L. and Martín-Fleitas, J. M. and Marton, G. and Mascarenhas, D. and Masip, A. and Mastrobuono-Battisti, A. and McMillan, P. J. and Meichsner, J. and Merc, J. and Messina, S. and Millar, N. R. and Mints, A. and Mohamed, D. and Molina, D. and Molinaro, R. and Molnár, L. and Monguió, M. and Montegriffo, P. and Monti, L. and Mora, A. and Morbidelli, R. and Morris, D. and Mudimadugula, R. and Muraveva, T. and Musella, I. and Nagy, Z. and Nardetto, N. and Navarrete, C. and Oh, S. and Ordenovic, C. and Orenstein, O. and Pagani, C. and Pagano, I. and Palaversa, L. and Palicio, P. A. and Pallas-Quintela, L. and Pawlak, M. and Penttilä, A. and Pesciullesi, P. and Pinamonti, M. and Plachy, E. and Planquart, L. and Plum, G. and Poggio, E. and Pourbaix, D. and Price-Whelan, A. M. and Pulone, L. and Rabin, V. and Rainer, M. and Raiteri, C. M. and Ramos, P. and Ramos-Lerate, M. and Ratajczak, M. and Re Fiorentin, P. and Regibo, S. and Reylé, C. and Ripepi, V. and Riva, A. and Rix, H.-W. and Rixon, G. and Robert, G. and Robichon, N. and Robin, C. and Romero-Gómez, M. and Rowell, N. and Ruz Mieres, D. and Rybicki, K. A. and Sadowski, G. and Sagristà Sellés, A. and Sanna, N. and Santoveña, R. and Sarasso, M. and Sarmiento, M. H. and Sarrate Riera, C. and Sciacca, E. and Ségransan, D. and Semczuk, M. and Shahaf, S. and Siebert, A. and Slezak, E. and Smart, R. L. and Snaith, O. N. and Solano, E. and Solitro, F. and Souami, D. and Souchay, J. and Spitoni, E. and Spoto, F. and Squillante, L. A. and Steele, I. A. and Steidelmüller, H. and Surdej, J. and Szabados, L. and Taris, F. and Taylor, M. B. and Teixeira, R. and Tepper-Garcia, T. and Thuillot, W. and Tolomei, L. and Tonello, N. and Torra, F. and Torralba Elipe, G. and Trabucchi, M. and Trentin, E. and Tsantaki, M. and Turon, C. and Ulla, A. and Unger, N. and Valtchanov, I. and Vanel, O. and Vecchiato, A. and Vicente, D. and Villar, E. and Weiler, M. and Zhao, H. and Zorec, J. and Zucker, S. and Župić, A. and Zwitter, T.},
   year={2024},
   month=may, pages={L2} }

@article{Paduano_2021,
    author = {Paduano, Alessandro and Bahramian, Arash and Miller-Jones, James C A and Kawka, Adela and Göttgens, Fabian and Strader, Jay and Chomiuk, Laura and Kamann, Sebastian and Dreizler, Stefan and Heinke, Craig O and Husser, Tim-Oliver and Maccarone, Thomas J and Tremou, Evangelia and Zhao, Yue},
    title = "{The MAVERIC Survey: The first radio and X-ray limits on the detached black holes in NGC 3201}",
    journal = {Monthly Notices of the Royal Astronomical Society},
    volume = {510},
    number = {3},
    pages = {3658-3673},
    year = {2021},
    month = {12},
    issn = {0035-8711},
    doi = {10.1093/mnras/stab3743},
    url = {https://doi.org/10.1093/mnras/stab3743},
    eprint = {https://academic.oup.com/mnras/article-pdf/510/3/3658/42151802/stab3743.pdf},
}

@ARTICLE{2002MNRAS.329..897H,
       author = {{Hurley}, Jarrod R. and {Tout}, Christopher A. and {Pols}, Onno R.},
        title = "{Evolution of binary stars and the effect of tides on binary populations}",
      journal = {MNRAS},
         year = 2002,
        month = feb,
       volume = {329},
       number = {4},
        pages = {897-928},
          doi = {10.1046/j.1365-8711.2002.05038.x},
archivePrefix = {arXiv},
       eprint = {astro-ph/0201220},
 primaryClass = {astro-ph},
       adsurl = {https://ui.adsabs.harvard.edu/abs/2002MNRAS.329..897H}
}

@ARTICLE{2024ApJ...973...75C,
       author = {{Cappelluti}, Nico and {Pacucci}, Fabio and {Hasinger}, G{\"u}nther},
        title = "{Constraining Wind-driven Accretion onto Gaia BH3 with Chandra}",
      journal = {ApJ},
         year = 2024,
        month = oct,
       volume = {973},
       number = {2},
          eid = {75},
        pages = {75},
          doi = {10.3847/1538-4357/ad6f96},
archivePrefix = {arXiv},
       eprint = {2406.07602},
 primaryClass = {astro-ph.HE},
       adsurl = {https://ui.adsabs.harvard.edu/abs/2024ApJ...973...75C}
}

@article{Sanderson_2020,
   title={Synthetic Gaia Surveys from the FIRE Cosmological Simulations of Milky Way-mass Galaxies},
   volume={246},
   ISSN={1538-4365},
   url={http://dx.doi.org/10.3847/1538-4365/ab5b9d},
   DOI={10.3847/1538-4365/ab5b9d},
   number={1},
   journal={The Astrophysical Journal Supplement Series},
   publisher={American Astronomical Society},
   author={Sanderson, Robyn E. and Wetzel, Andrew and Loebman, Sarah and Sharma, Sanjib and Hopkins, Philip F. and Garrison-Kimmel, Shea and Faucher-Giguère, Claude-André and Kereš, Dušan and Quataert, Eliot},
   year={2020},
   month=jan, pages={6} }

@ARTICLE{1944MNRAS.104..273B,
       author = {{Bondi}, H. and {Hoyle}, F.},
        title = "{On the mechanism of accretion by stars}",
      journal = {MNRAS},
         year = 1944,
        month = jan,
       volume = {104},
        pages = {273},
          doi = {10.1093/mnras/104.5.273},
       adsurl = {https://ui.adsabs.harvard.edu/abs/1944MNRAS.104..273B}
}

@ARTICLE{RomanWP,
       author = {{Gandhi}, P. and {Dashwood Brown}, C. and {Zhao}, Y. and {El-Badry}, K. and {Maccarone}, T.~J. and {Knigge}, C. and {Anderson}, J. and {Middleton}, M. and {Miller-Jones}, J.~C.~A.},
        title = "{New Compact Object Binary Populations with Precision Astrometry (Roman White Paper)}",
      journal = {arXiv e-prints},
         year = 2023,
        month = jun,
          eid = {arXiv:2306.16479},
        pages = {arXiv:2306.16479},
          doi = {10.48550/arXiv.2306.16479},
archivePrefix = {arXiv},
       eprint = {2306.16479},
 primaryClass = {astro-ph.IM},
       adsurl = {https://ui.adsabs.harvard.edu/abs/2023arXiv230616479G}
}

@ARTICLE{RevealingBHwGaia,
       author = {{Breivik}, Katelyn and {Chatterjee}, Sourav and {Larson}, Shane L.},
        title = "{Revealing Black Holes with Gaia}",
      journal = {\apjl},
         year = 2017,
        month = nov,
       volume = {850},
       number = {1},
          eid = {L13},
        pages = {L13},
          doi = {10.3847/2041-8213/aa97d5},
archivePrefix = {arXiv},
       eprint = {1710.04657},
 primaryClass = {astro-ph.SR},
       adsurl = {https://ui.adsabs.harvard.edu/abs/2017ApJ...850L..13B}
}

@ARTICLE{2022ApJ...931..107C,
       author = {{Chawla}, Chirag and {Chatterjee}, Sourav and {Breivik}, Katelyn and {Moorthy}, Chaithanya Krishna and {Andrews}, Jeff J. and {Sanderson}, Robyn E.},
        title = "{Gaia May Detect Hundreds of Well-characterized Stellar Black Holes}",
      journal = {ApJ},
         year = 2022,
        month = jun,
       volume = {931},
       number = {2},
          eid = {107},
        pages = {107},
          doi = {10.3847/1538-4357/ac60a5},
archivePrefix = {arXiv},
       eprint = {2110.05979},
 primaryClass = {astro-ph.GA},
       adsurl = {https://ui.adsabs.harvard.edu/abs/2022ApJ...931..107C}
}

@article{Chawla_2022,
doi = {10.3847/1538-4357/ac60a5},
url = {https://dx.doi.org/10.3847/1538-4357/ac60a5},
year = {2022},
month = {may},
publisher = {The American Astronomical Society},
volume = {931},
number = {2},
pages = {107},
author = {Chawla, Chirag and Chatterjee, Sourav and Breivik, Katelyn and Moorthy, Chaithanya Krishna and Andrews, Jeff J. and Sanderson, Robyn E.},
title = {Gaia May Detect Hundreds of Well-characterized Stellar Black Holes},
journal = {The Astrophysical Journal}
}

@article{Stevance_2023,
   title={End-to-end study of the host galaxy and genealogy of the first binary neutron star merger},
   volume={7},
   ISSN={2397-3366},
   url={http://dx.doi.org/10.1038/s41550-022-01873-y},
   DOI={10.1038/s41550-022-01873-y},
   number={4},
   journal={Nature Astronomy},
   publisher={Springer Science and Business Media LLC},
   author={Stevance, Heloise F. and Eldridge, Jan J. and Stanway, Elizabeth R. and Lyman, Joe and McLeod, Anna F. and Levan, Andrew J.},
   year={2023},
   month=jan, pages={444–450} }

@ARTICLE{2017PASA...34...58E,
       author = {{Eldridge}, J.~J. and {Stanway}, E.~R. and {Xiao}, L. and {McClelland}, L.~A.~S. and {Taylor}, G. and {Ng}, M. and {Greis}, S.~M.~L. and {Bray}, J.~C.},
        title = "{Binary Population and Spectral Synthesis Version 2.1: Construction, Observational Verification, and New Results}",
      journal = {\pasa},
         year = 2017,
        month = nov,
       volume = {34},
          eid = {e058},
        pages = {e058},
          doi = {10.1017/pasa.2017.51},
archivePrefix = {arXiv},
       eprint = {1710.02154},
 primaryClass = {astro-ph.SR},
       adsurl = {https://ui.adsabs.harvard.edu/abs/2017PASA...34...58E}
}

@ARTICLE{2022EldridgeStanway,
       author = {{Eldridge}, Jan J. and {Stanway}, Elizabeth R.},
        title = "{New Insights into the Evolution of Massive Stars and Their Effects on Our Understanding of Early Galaxies}",
      journal = {\araa},
         year = 2022,
        month = aug,
       volume = {60},
        pages = {455-494},
          doi = {10.1146/annurev-astro-052920-100646},
archivePrefix = {arXiv},
       eprint = {2202.01413},
 primaryClass = {astro-ph.GA},
       adsurl = {https://ui.adsabs.harvard.edu/abs/2022ARA&A..60..455E}
}

@ARTICLE{KroupaIMF,
       author = {{Kroupa}, Pavel},
        title = "{On the variation of the initial mass function}",
      journal = {MNRAS},
         year = 2001,
        month = apr,
       volume = {322},
       number = {2},
        pages = {231-246},
          doi = {10.1046/j.1365-8711.2001.04022.x},
archivePrefix = {arXiv},
       eprint = {astro-ph/0009005},
 primaryClass = {astro-ph},
       adsurl = {https://ui.adsabs.harvard.edu/abs/2001MNRAS.322..231K}
}

@ARTICLE{Stevance2023MNRAS.520.4740S,
       author = {{Stevance}, H.~F. and {Ghodla}, S. and {Richards}, S. and {Eldridge}, J.~J. and {Briel}, M.~M. and {Tang}, P.},
        title = "{VFTS 243 as predicted by the BPASS fiducial models}",
      journal = {MNRAS},
         year = 2023,
        month = apr,
       volume = {520},
       number = {3},
        pages = {4740-4746},
          doi = {10.1093/mnras/stad362},
archivePrefix = {arXiv},
       eprint = {2208.02258},
 primaryClass = {astro-ph.SR},
       adsurl = {https://ui.adsabs.harvard.edu/abs/2023MNRAS.520.4740S}
}

@ARTICLE{BePablo2011Ap&SS.332....1R,
       author = {{Reig}, Pablo},
        title = "{Be/X-ray binaries}",
      journal = {\apss},
         year = 2011,
        month = mar,
       volume = {332},
       number = {1},
        pages = {1-29},
          doi = {10.1007/s10509-010-0575-8},
archivePrefix = {arXiv},
       eprint = {1101.5036},
 primaryClass = {astro-ph.HE},
       adsurl = {https://ui.adsabs.harvard.edu/abs/2011Ap&SS.332....1R}
}

@ARTICLE{HDnotBH2025A&A...696A..84N,
       author = {{Naz{\'e}}, Ya{\"e}l and {Rauw}, Gregor},
        title = "{Another one (BH+OB pair) bites the dust}",
      journal = {\aap},
         year = 2025,
        month = apr,
       volume = {696},
          eid = {A84},
        pages = {A84},
          doi = {10.1051/0004-6361/202453493},
archivePrefix = {arXiv},
       eprint = {2503.08190},
 primaryClass = {astro-ph.SR},
       adsurl = {https://ui.adsabs.harvard.edu/abs/2025A&A...696A..84N}
}

@ARTICLE{HDisXB2021ApJ...913...48G,
       author = {{Gomez}, Sebastian and {Grindlay}, Jonathan E.},
        title = "{Optical Analysis and Modeling of HD96670, a New Black Hole X-Ray Binary Candidate}",
      journal = {ApJ},
         year = 2021,
        month = may,
       volume = {913},
       number = {1},
          eid = {48},
        pages = {48},
          doi = {10.3847/1538-4357/abf24c},
archivePrefix = {arXiv},
       eprint = {2211.04518},
 primaryClass = {astro-ph.HE},
       adsurl = {https://ui.adsabs.harvard.edu/abs/2021ApJ...913...48G}
}

@ARTICLE{MWCXB2014ApJ...786L..11M,
       author = {{Munar-Adrover}, P. and {Paredes}, J.~M. and {Rib{\'o}}, M. and {Iwasawa}, K. and {Zabalza}, V. and {Casares}, J.},
        title = "{Discovery of X-Ray Emission from the First Be/Black Hole System}",
      journal = {\apjl},
         year = 2014,
        month = may,
       volume = {786},
       number = {2},
          eid = {L11},
        pages = {L11},
          doi = {10.1088/2041-8205/786/2/L11},
archivePrefix = {arXiv},
       eprint = {1404.0901},
 primaryClass = {astro-ph.HE},
       adsurl = {https://ui.adsabs.harvard.edu/abs/2014ApJ...786L..11M}
}

@ARTICLE{MWCnot2023A&A...677L...9J,
       author = {{Janssens}, S. and {Shenar}, T. and {Degenaar}, N. and {Bodensteiner}, J. and {Sana}, H. and {Audenaert}, J. and {Frost}, A.~J.},
        title = "{MWC 656 is unlikely to contain a black hole}",
      journal = {\aap},
         year = 2023,
        month = sep,
       volume = {677},
          eid = {L9},
        pages = {L9},
          doi = {10.1051/0004-6361/202347318},
archivePrefix = {arXiv},
       eprint = {2308.08642},
 primaryClass = {astro-ph.SR},
       adsurl = {https://ui.adsabs.harvard.edu/abs/2023A&A...677L...9J}
}

@ARTICLE{LackofBeBH2009ApJ...707..870B,
       author = {{Belczynski}, Krzysztof and {Ziolkowski}, Janusz},
        title = "{On the Apparent Lack of Be X-Ray Binaries with Black Holes}",
      journal = {ApJ},
         year = 2009,
        month = dec,
       volume = {707},
       number = {2},
        pages = {870-877},
          doi = {10.1088/0004-637X/707/2/870},
archivePrefix = {arXiv},
       eprint = {0907.4990},
 primaryClass = {astro-ph.GA},
       adsurl = {https://ui.adsabs.harvard.edu/abs/2009ApJ...707..870B}
}

@ARTICLE{1988A&A...205..155B,
       author = {{Boffin}, H.~M.~J. and {Jorissen}, A.},
        title = "{Can a barium star be produced by wind accretion in a detached binary ?}",
      journal = {\aap},
         year = 1988,
        month = oct,
       volume = {205},
        pages = {155-163},
       adsurl = {https://ui.adsabs.harvard.edu/abs/1988A&A...205..155B}
}

@ARTICLE{2024ApJ...965..138F,
       author = {{Fardeen}, James and {McGill}, Peter and {Perkins}, Scott E. and {Dawson}, William A. and {Abrams}, Natasha S. and {Lu}, Jessica R. and {Ho}, Ming-Feng and {Bird}, Simeon},
        title = "{Astrometric Microlensing by Primordial Black Holes with the Roman Space Telescope}",
      journal = {ApJ},
         year = 2024,
        month = apr,
       volume = {965},
       number = {2},
          eid = {138},
        pages = {138},
          doi = {10.3847/1538-4357/ad3243},
archivePrefix = {arXiv},
       eprint = {2312.13249},
 primaryClass = {astro-ph.GA},
       adsurl = {https://ui.adsabs.harvard.edu/abs/2024ApJ...965..138F}
}

@ARTICLE{2019ApJ...886...68A,
       author = {{Andrews}, Jeff J. and {Breivik}, Katelyn and {Chatterjee}, Sourav},
        title = "{Weighing the Darkness: Astrometric Mass Measurement of Hidden Stellar Companions Using Gaia}",
      journal = {ApJ},
         year = 2019,
       volume = {886},
        pages = {68},
}

@ARTICLE{Ps&Qs,
       author = {{Moe}, Maxwell and {Di Stefano}, Rosanne},
        title = "{Mind Your Ps and Qs: The Interrelation between Period (P) and Mass-ratio (Q) Distributions of Binary Stars}",
      journal = {\apjs},
         year = 2017,
        month = jun,
       volume = {230},
       number = {2},
          eid = {15},
        pages = {15},
          doi = {10.3847/1538-4365/aa6fb6},
archivePrefix = {arXiv},
       eprint = {1606.05347},
 primaryClass = {astro-ph.SR},
       adsurl = {https://ui.adsabs.harvard.edu/abs/2017ApJS..230...15M}
}

@ARTICLE{2016ApJ...818..130B,
       author = {{Bovy}, Jo and {Rix}, Hans-Walter and {Green}, Gregory M. and {Schlafly}, Edward F. and {Finkbeiner}, Douglas P.},
        title = "{On Galactic Density Modeling in the Presence of Dust Extinction}",
      journal = {ApJ},
         year = 2016,
        month = feb,
       volume = {818},
       number = {2},
          eid = {130},
        pages = {130},
          doi = {10.3847/0004-637X/818/2/130},
archivePrefix = {arXiv},
       eprint = {1509.06751},
 primaryClass = {astro-ph.GA},
       adsurl = {https://ui.adsabs.harvard.edu/abs/2016ApJ...818..130B}
}

@ARTICLE{2023ApJS..269....5W,
       author = {{Wilson}, Robert F. and {Barclay}, Thomas and {Powell}, Brian P. and {Schlieder}, Joshua and {Hedges}, Christina and {Montet}, Benjamin T. and {Quintana}, Elisa and {Mcdonald}, Iain and {Penny}, Matthew T. and {Espinoza}, N{\'e}stor and {Kerins}, Eamonn},
        title = "{Transiting Exoplanet Yields for the Roman Galactic Bulge Time Domain Survey Predicted from Pixel-level Simulations}",
      journal = {\apjs},
         year = 2023,
        month = nov,
       volume = {269},
       number = {1},
          eid = {5},
        pages = {5},
          doi = {10.3847/1538-4365/acf3df},
archivePrefix = {arXiv},
       eprint = {2305.16204},
 primaryClass = {astro-ph.EP},
       adsurl = {https://ui.adsabs.harvard.edu/abs/2023ApJS..269....5W}
}

@ARTICLE{2010A&A...523A..48J,
       author = {{Jordi}, C. and {Gebran}, M. and {Carrasco}, J.~M. and {de Bruijne}, J. and {Voss}, H. and {Fabricius}, C. and {Knude}, J. and {Vallenari}, A. and {Kohley}, R. and {Mora}, A.},
        title = "{Gaia broad band photometry}",
      journal = {\aap},
         year = 2010,
        month = nov,
       volume = {523},
          eid = {A48},
        pages = {A48},
          doi = {10.1051/0004-6361/201015441},
archivePrefix = {arXiv},
       eprint = {1008.0815},
 primaryClass = {astro-ph.IM},
       adsurl = {https://ui.adsabs.harvard.edu/abs/2010A&A...523A..48J}
}

@ARTICLE{2025PASP..137d4202N,
       author = {{Nagarajan}, Pranav and {El-Badry}, Kareem and {Chawla}, Chirag and {Di Carlo}, Ugo Niccol{\`o} and {Breivik}, Katelyn and {Rodriguez}, Carl L. and {Agrawal}, Poojan and {Delfavero}, Vera and {Chatterjee}, Sourav},
        title = "{Realistic Predictions for Gaia Black Hole Discoveries: Comparison of Isolated Binary and Dynamical Formation Models}",
      journal = {PASP},
         year = 2025,
        month = apr,
       volume = {137},
       number = {4},
          eid = {044202},
        pages = {044202},
          doi = {10.1088/1538-3873/adc839},
archivePrefix = {arXiv},
       eprint = {2502.03527},
 primaryClass = {astro-ph.GA},
       adsurl = {https://ui.adsabs.harvard.edu/abs/2025PASP..137d4202N}
}

@ARTICLE{2020FrPhy..1524603G,
       author = {{Gao}, He and {Ai}, Shun-Ke and {Cao}, Zhou-Jian and {Zhang}, Bing and {Zhu}, Zhen-Yu and {Li}, Ang and {Zhang}, Nai-Bo and {Bauswein}, Andreas},
        title = "{Relation between gravitational mass and baryonic mass for non-rotating and rapidly rotating neutron stars}",
      journal = {Frontiers of Physics},
         year = 2020,
        month = jan,
       volume = {15},
       number = {2},
          eid = {24603},
        pages = {24603},
          doi = {10.1007/s11467-019-0945-9},
archivePrefix = {arXiv},
       eprint = {1905.03784},
 primaryClass = {astro-ph.HE},
       adsurl = {https://ui.adsabs.harvard.edu/abs/2020FrPhy..1524603G}
}

@ARTICLE{committee2025romanobservationstimeallocation,
       author = {{Observations Time Allocation Committee}, Roman and {Community Survey Definition Committees}, Core},
        title = "{Roman Observations Time Allocation Committee: Final Report and Recommendations}",
      journal = {arXiv e-prints},
         year = 2025,
        month = may,
          eid = {arXiv:2505.10574},
        pages = {arXiv:2505.10574},
          doi = {10.48550/arXiv.2505.10574},
archivePrefix = {arXiv},
       eprint = {2505.10574},
 primaryClass = {astro-ph.IM},
       adsurl = {https://ui.adsabs.harvard.edu/abs/2025arXiv250510574O}
}

@ARTICLE{2023JCAP...07..037C,
       author = {{Chen}, I. -Kai and {Kongsore}, Marius and {Tilburg}, Ken Van},
        title = "{Detecting dark compact objects in Gaia DR4: A data analysis pipeline for transient astrometric lensing searches}",
      journal = {\jcap},
         year = 2023,
        month = jul,
       volume = {2023},
       number = {7},
          eid = {037},
        pages = {037},
          doi = {10.1088/1475-7516/2023/07/037},
archivePrefix = {arXiv},
       eprint = {2301.00822},
 primaryClass = {astro-ph.GA},
       adsurl = {https://ui.adsabs.harvard.edu/abs/2023JCAP...07..037C}
}

@ARTICLE{2023ApJ...946..111A,
       author = {{Andrews}, Jeff J. and {Breivik}, Katelyn and {Chawla}, Chirag and {Rodriguez}, Carl L. and {Chatterjee}, Sourav},
        title = "{Weighing the Darkness. II. Astrometric Measurement of Partial Orbits with Gaia}",
      journal = {ApJ},
         year = 2023,
        month = apr,
       volume = {946},
       number = {2},
          eid = {111},
        pages = {111},
          doi = {10.3847/1538-4357/acbb5f},
archivePrefix = {arXiv},
       eprint = {2110.05549},
 primaryClass = {astro-ph.SR},
       adsurl = {https://ui.adsabs.harvard.edu/abs/2023ApJ...946..111A}
}

@ARTICLE{2024MNRAS.527.5023L,
       author = {{Liu}, Boyuan and {Sartorio}, Nina S. and {Izzard}, Robert G. and {Fialkov}, Anastasia},
        title = "{Population synthesis of Be X-ray binaries: metallicity dependence of total X-ray outputs}",
      journal = {MNRAS},
         year = 2024,
        month = jan,
       volume = {527},
       number = {3},
        pages = {5023-5048},
          doi = {10.1093/mnras/stad3475},
archivePrefix = {arXiv},
       eprint = {2308.06154},
 primaryClass = {astro-ph.HE},
       adsurl = {https://ui.adsabs.harvard.edu/abs/2024MNRAS.527.5023L}
}

@ARTICLE{2023ApJ...954..212S,
       author = {{Siegel}, Jared C. and {Kiato}, Ilia and {Kalogera}, Vicky and {Berry}, Christopher P.~L. and {Maccarone}, Thomas J. and {Breivik}, Katelyn and {Andrews}, Jeff J. and {Bavera}, Simone S. and {Dotter}, Aaron and {Fragos}, Tassos and et al.},
        title = "{Investigating the Lower Mass Gap with Low-mass X-Ray Binary Population Synthesis}",
      journal = {ApJ},
         year = 2023,
        month = sep,
       volume = {954},
       number = {2},
          eid = {212},
        pages = {212},
          doi = {10.3847/1538-4357/ace9d9},
archivePrefix = {arXiv},
       eprint = {2209.06844},
 primaryClass = {astro-ph.HE},
       adsurl = {https://ui.adsabs.harvard.edu/abs/2023ApJ...954..212S}
}

@ARTICLE{2003ApJ...597.1036P,
       author = {{Pfahl}, Eric and {Rappaport}, Saul and {Podsiadlowski}, Philipp},
        title = "{The Galactic Population of Low- and Intermediate-Mass X-Ray Binaries}",
      journal = {ApJ},
         year = 2003,
        month = nov,
       volume = {597},
       number = {2},
        pages = {1036-1048},
          doi = {10.1086/378632},
archivePrefix = {arXiv},
       eprint = {astro-ph/0303300},
 primaryClass = {astro-ph},
       adsurl = {https://ui.adsabs.harvard.edu/abs/2003ApJ...597.1036P}
}

@ARTICLE{2005MNRAS.356..401R,
       author = {{Rappaport}, S.~A. and {Podsiadlowski}, Ph. and {Pfahl}, E.},
        title = "{Stellar-mass black hole binaries as ultraluminous X-ray sources}",
      journal = {MNRAS},
         year = 2005,
        month = jan,
       volume = {356},
       number = {2},
        pages = {401-414},
          doi = {10.1111/j.1365-2966.2004.08489.x},
archivePrefix = {arXiv},
       eprint = {astro-ph/0408032},
 primaryClass = {astro-ph},
       adsurl = {https://ui.adsabs.harvard.edu/abs/2005MNRAS.356..401R}
}

@ARTICLE{2024PASP..136a4202N,
       author = {{Nagarajan}, Pranav and {El-Badry}, Kareem and {Triaud}, Amaury H.~M.~J. and {Baycroft}, Thomas A. and {Latham}, David and {Bieryla}, Allyson and {Buchhave}, Lars A. and {Rix}, Hans-Walter and {Quataert}, Eliot and {Howard}, Andrew and et al.},
        title = "{ESPRESSO Observations of Gaia BH1: High-precision Orbital Constraints and no Evidence for an Inner Binary}",
      journal = {PASP},
         year = 2024,
        month = jan,
       volume = {136},
       number = {1},
          eid = {014202},
        pages = {014202},
          doi = {10.1088/1538-3873/ad1ba7},
archivePrefix = {arXiv},
       eprint = {2312.05313},
 primaryClass = {astro-ph.SR},
       adsurl = {https://ui.adsabs.harvard.edu/abs/2024PASP..136a4202N}
}

@ARTICLE{2009ApJ...691.1611M,
       author = {{Ma}, Bo and {Li}, Xiang-Dong},
        title = "{The Bifurcation Periods in Low-Mass X-Ray Binaries: The Effect of Magnetic Braking and Mass Loss}",
      journal = {ApJ},
         year = 2009,
        month = feb,
       volume = {691},
       number = {2},
        pages = {1611-1617},
          doi = {10.1088/0004-637X/691/2/1611},
archivePrefix = {arXiv},
       eprint = {0810.2009},
 primaryClass = {astro-ph},
       adsurl = {https://ui.adsabs.harvard.edu/abs/2009ApJ...691.1611M}
}

@ARTICLE{2002ApJ...565.1107P,
       author = {{Podsiadlowski}, Ph. and {Rappaport}, S. and {Pfahl}, E.~D.},
        title = "{Evolutionary Sequences for Low- and Intermediate-Mass X-Ray Binaries}",
      journal = {ApJ},
         year = 2002,
        month = feb,
       volume = {565},
       number = {2},
        pages = {1107-1133},
          doi = {10.1086/324686},
archivePrefix = {arXiv},
       eprint = {astro-ph/0107261},
 primaryClass = {astro-ph},
       adsurl = {https://ui.adsabs.harvard.edu/abs/2002ApJ...565.1107P}
}

@ARTICLE{1988A&A...191...57P,
       author = {{Pylyser}, E. and {Savonije}, G.~J.},
        title = "{Evolution of low-mass close binary systems with a compact mass accreting component.}",
      journal = {\aap},
         year = 1988,
        month = feb,
       volume = {191},
        pages = {57-70},
       adsurl = {https://ui.adsabs.harvard.edu/abs/1988A&A...191...57P}
}

@ARTICLE{1989A&A...208...52P,
       author = {{Pylyser}, E.~H.~P. and {Savonije}, G.~J.},
        title = "{The evolution of low-mass close binary systems with a compact component. II. Systems captured by angular momentum losses.}",
      journal = {\aap},
         year = 1989,
        month = jan,
       volume = {208},
        pages = {52-62},
       adsurl = {https://ui.adsabs.harvard.edu/abs/1989A&A...208...52P}
}

@ARTICLE{2025MNRAS.542.2087B,
       author = {{Bray}, J.~C. and {Stanway}, E.~R. and {Eldridge}, J.~J.},
        title = "{X-BPASS : self-consistent modelling of stellar populations and their associated X-ray binary emission in a binary stellar evolution framework}",
      journal = {MNRAS},
         year = 2025,
        month = sep,
       volume = {542},
       number = {3},
        pages = {2087-2104},
          doi = {10.1093/mnras/staf1348},
archivePrefix = {arXiv},
       eprint = {2508.18628},
 primaryClass = {astro-ph.HE},
       adsurl = {https://ui.adsabs.harvard.edu/abs/2025MNRAS.542.2087B}
}

@MISC{2017arXiv170200786A,
       author = {{Amaro-Seoane}, Pau and {Audley}, Heather and {Babak}, Stanislav and {Baker}, John and {Barausse}, Enrico and {Bender}, Peter and {Berti}, Emanuele and {Binetruy}, Pierre and {Born}, Michael and {Bortoluzzi}, Daniele and et al.},
        title = "{Laser Interferometer Space Antenna}",
      journal = {arXiv e-prints},
         year = 2017,
        month = feb,
          eid = {arXiv:1702.00786},
        pages = {arXiv:1702.00786},
          doi = {10.48550/arXiv.1702.00786},
archivePrefix = {arXiv},
       eprint = {1702.00786},
 primaryClass = {astro-ph.IM},
       adsurl = {https://ui.adsabs.harvard.edu/abs/2017arXiv170200786A}
}

@ARTICLE{2001ApJ...547..355P,
       author = {{Popham}, Robert and {Sunyaev}, Rashid},
        title = "{Accretion Disk Boundary Layers around Neutron Stars: X-Ray Production in Low-Mass X-Ray Binaries}",
      journal = {ApJ},
         year = 2001,
        month = jan,
       volume = {547},
       number = {1},
        pages = {355-383},
          doi = {10.1086/318336},
archivePrefix = {arXiv},
       eprint = {astro-ph/0004017},
 primaryClass = {astro-ph},
       adsurl = {https://ui.adsabs.harvard.edu/abs/2001ApJ...547..355P}
}

@ARTICLE{1999MNRAS.303..139D,
       author = {{Dubus}, Guillaume and {Lasota}, Jean-Pierre and {Hameury}, Jean-Marie and {Charles}, Phil},
        title = "{X-ray irradiation in low-mass binary systems}",
      journal = {MNRAS},
         year = 1999,
        month = feb,
       volume = {303},
       number = {1},
        pages = {139-147},
          doi = {10.1046/j.1365-8711.1999.02212.x},
archivePrefix = {arXiv},
       eprint = {astro-ph/9809036},
 primaryClass = {astro-ph},
       adsurl = {https://ui.adsabs.harvard.edu/abs/1999MNRAS.303..139D}
}

@ARTICLE{1996ApJ...457..821N,
       author = {{Narayan}, Ramesh and {McClintock}, Jeffrey E. and {Yi}, Insu},
        title = "{A New Model for Black Hole Soft X-Ray Transients in Quiescence}",
      journal = {ApJ},
         year = 1996,
        month = feb,
       volume = {457},
        pages = {821},
          doi = {10.1086/176777},
archivePrefix = {arXiv},
       eprint = {astro-ph/9508014},
 primaryClass = {astro-ph},
       adsurl = {https://ui.adsabs.harvard.edu/abs/1996ApJ...457..821N}
}

@ARTICLE{1997ApJ...489..865E,
       author = {{Esin}, Ann A. and {McClintock}, Jeffrey E. and {Narayan}, Ramesh},
        title = "{Advection-Dominated Accretion and the Spectral States of Black Hole X-Ray Binaries: Application to Nova Muscae 1991}",
      journal = {ApJ},
         year = 1997,
        month = nov,
       volume = {489},
       number = {2},
        pages = {865-889},
          doi = {10.1086/304829},
archivePrefix = {arXiv},
       eprint = {astro-ph/9705237},
 primaryClass = {astro-ph},
       adsurl = {https://ui.adsabs.harvard.edu/abs/1997ApJ...489..865E}
}

@ARTICLE{2014MNRAS.444.3802H,
       author = {{Hakala}, Pasi and {Muhli}, Panu and {Charles}, Phil},
        title = "{Simultaneous optical and near-IR photometry of 4U1957+115 - a missing secondary star}",
      journal = {MNRAS},
         year = 2014,
        month = nov,
       volume = {444},
       number = {4},
        pages = {3802-3808},
          doi = {10.1093/mnras/stu1687},
archivePrefix = {arXiv},
       eprint = {1408.4025},
 primaryClass = {astro-ph.HE},
       adsurl = {https://ui.adsabs.harvard.edu/abs/2014MNRAS.444.3802H}
}

@ARTICLE{2001A&A...373..251D,
       author = {{Dubus}, G. and {Hameury}, J.-M. and {Lasota}, J.-P.},
        title = "{The disc instability model for X-ray transients: Evidence for truncation and irradiation}",
      journal = {\aap},
         year = 2001,
        month = jul,
       volume = {373},
        pages = {251-271},
          doi = {10.1051/0004-6361:20010632},
archivePrefix = {arXiv},
       eprint = {astro-ph/0102237},
 primaryClass = {astro-ph},
       adsurl = {https://ui.adsabs.harvard.edu/abs/2001A&A...373..251D}
}

@ARTICLE{2023A&A...674A...1G,
       author = {{Gaia Collaboration} and {Vallenari}, A. and {Brown}, A.~G.~A. and {Prusti}, T. and {de Bruijne}, J.~H.~J. and {Arenou}, F. and {Babusiaux}, C. and {Biermann}, M. and {Creevey}, O.~L. and {Ducourant}, C. and {Evans}, D.~W. and {Eyer}, L. and {Guerra}, R. and {Hutton}, A. and {Jordi}, C. and {Klioner}, S.~A. and {Lammers}, U.~L. and {Lindegren}, L. and {Luri}, X. and {Mignard}, F. and {Panem}, C. and {Pourbaix}, D. and {Randich}, S. and {Sartoretti}, P. and {Soubiran}, C. and {Tanga}, P. and {Walton}, N.~A. and {Bailer-Jones}, C.~A.~L. and {Bastian}, U. and {Drimmel}, R. and {Jansen}, F. and {Katz}, D. and {Lattanzi}, M.~G. and {van Leeuwen}, F. and {Bakker}, J. and {Cacciari}, C. and {Casta{\~n}eda}, J. and {De Angeli}, F. and {Fabricius}, C. and {Fouesneau}, M. and {Fr{\'e}mat}, Y. and {Galluccio}, L. and {Guerrier}, A. and {Heiter}, U. and {Masana}, E. and {Messineo}, R. and {Mowlavi}, N. and {Nicolas}, C. and {Nienartowicz}, K. and {Pailler}, F. and {Panuzzo}, P. and {Riclet}, F. and {Roux}, W. and {Seabroke}, G.~M. and {Sordo}, R. and {Th{\'e}venin}, F. and {Gracia-Abril}, G. and {Portell}, J. and {Teyssier}, D. and {Altmann}, M. and {Andrae}, R. and {Audard}, M. and {Bellas-Velidis}, I. and {Benson}, K. and {Berthier}, J. and {Blomme}, R. and {Burgess}, P.~W. and {Busonero}, D. and {Busso}, G. and {C{\'a}novas}, H. and {Carry}, B. and {Cellino}, A. and {Cheek}, N. and {Clementini}, G. and {Damerdji}, Y. and {Davidson}, M. and {de Teodoro}, P. and {Nu{\~n}ez Campos}, M. and {Delchambre}, L. and {Dell'Oro}, A. and {Esquej}, P. and {Fern{\'a}ndez-Hern{\'a}ndez}, J. and {Fraile}, E. and {Garabato}, D. and {Garc{\'\i}a-Lario}, P. and {Gosset}, E. and {Haigron}, R. and {Halbwachs}, J.-L. and {Hambly}, N.~C. and {Harrison}, D.~L. and {Hern{\'a}ndez}, J. and {Hestroffer}, D. and {Hodgkin}, S.~T. and {Holl}, B. and {Jan{\ss}en}, K. and {Jevardat de Fombelle}, G. and {Jordan}, S. and {Krone-Martins}, A. and {Lanzafame}, A.~C. and {L{\"o}ffler}, W. and {Marchal}, O. and {Marrese}, P.~M. and {Moitinho}, A. and {Muinonen}, K. and {Osborne}, P. and {Pancino}, E. and {Pauwels}, T. and {Recio-Blanco}, A. and {Reyl{\'e}}, C. and {Riello}, M. and {Rimoldini}, L. and {Roegiers}, T. and {Rybizki}, J. and {Sarro}, L.~M. and {Siopis}, C. and {Smith}, M. and {Sozzetti}, A. and {Utrilla}, E. and {van Leeuwen}, M. and {Abbas}, U. and {{\'A}brah{\'a}m}, P. and {Abreu Aramburu}, A. and {Aerts}, C. and {Aguado}, J.~J. and {Ajaj}, M. and {Aldea-Montero}, F. and {Altavilla}, G. and {{\'A}lvarez}, M.~A. and {Alves}, J. and {Anders}, F. and {Anderson}, R.~I. and {Anglada Varela}, E. and {Antoja}, T. and {Baines}, D. and {Baker}, S.~G. and {Balaguer-N{\'u}{\~n}ez}, L. and {Balbinot}, E. and {Balog}, Z. and {Barache}, C. and {Barbato}, D. and {Barros}, M. and {Barstow}, M.~A. and {Bartolom{\'e}}, S. and {Bassilana}, J.-L. and {Bauchet}, N. and {Becciani}, U. and {Bellazzini}, M. and {Berihuete}, A. and {Bernet}, M. and {Bertone}, S. and {Bianchi}, L. and {Binnenfeld}, A. and {Blanco-Cuaresma}, S. and {Blazere}, A. and {Boch}, T. and {Bombrun}, A. and {Bossini}, D. and {Bouquillon}, S. and {Bragaglia}, A. and {Bramante}, L. and {Breedt}, E. and {Bressan}, A. and {Brouillet}, N. and {Brugaletta}, E. and {Bucciarelli}, B. and {Burlacu}, A. and {Butkevich}, A.~G. and {Buzzi}, R. and {Caffau}, E. and {Cancelliere}, R. and {Cantat-Gaudin}, T. and {Carballo}, R. and {Carlucci}, T. and {Carnerero}, M.~I. and {Carrasco}, J.~M. and {Casamiquela}, L. and {Castellani}, M. and {Castro-Ginard}, A. and {Chaoul}, L. and {Charlot}, P. and {Chemin}, L. and {Chiaramida}, V. and {Chiavassa}, A. and {Chornay}, N. and {Comoretto}, G. and {Contursi}, G. and {Cooper}, W.~J. and {Cornez}, T. and {Cowell}, S. and {Crifo}, F. and {Cropper}, M. and {Crosta}, M. and {Crowley}, C. and {Dafonte}, C. and {Dapergolas}, A. and {David}, M. and {David}, P. and {de Laverny}, P. and {De Luise}, F. and {De March}, R.},
        title = "{Gaia Data Release 3. Summary of the content and survey properties}",
      journal = {\aap},
         year = 2023,
        month = jun,
       volume = {674},
          eid = {A1},
        pages = {A1},
          doi = {10.1051/0004-6361/202243940},
archivePrefix = {arXiv},
       eprint = {2208.00211},
 primaryClass = {astro-ph.GA},
       adsurl = {https://ui.adsabs.harvard.edu/abs/2023A&A...674A...1G}
}

@ARTICLE{2011ApJ...730...43B,
       author = {{Bayless}, Amanda J. and {Robinson}, Edward L. and {Mason}, Paul A. and {Robertson}, Paul},
        title = "{The Optical Orbital Light Curve of the Low-mass X-ray Binary V1408 Aquilae (= 4U 1957+115)}",
      journal = {ApJ},
         year = 2011,
        month = mar,
       volume = {730},
       number = {1},
          eid = {43},
        pages = {43},
          doi = {10.1088/0004-637X/730/1/43},
archivePrefix = {arXiv},
       eprint = {1004.4904},
 primaryClass = {astro-ph.HE},
       adsurl = {https://ui.adsabs.harvard.edu/abs/2011ApJ...730...43B}
}

@ARTICLE{2015ApJ...809....9G,
       author = {{Gomez}, Sebastian and {Mason}, Paul A. and {Robinson}, Edward L.},
        title = "{The Case for a Low Mass Black Hole in the Low Mass X-Ray Binary V1408 Aquilae (= 4U 1957+115)}",
      journal = {ApJ},
         year = 2015,
        month = aug,
       volume = {809},
       number = {1},
          eid = {9},
        pages = {9},
          doi = {10.1088/0004-637X/809/1/9},
archivePrefix = {arXiv},
       eprint = {1506.00181},
 primaryClass = {astro-ph.SR},
       adsurl = {https://ui.adsabs.harvard.edu/abs/2015ApJ...809....9G}
}

@ARTICLE{2007MNRAS.375.1000B,
       author = {{Beer}, Martin E. and {Dray}, Lynnette M. and {King}, Andrew R. and {Wynn}, Graham A.},
        title = "{An alternative to common envelope evolution}",
      journal = {MNRAS},
         year = 2007,
        month = mar,
       volume = {375},
       number = {3},
        pages = {1000-1008},
          doi = {10.1111/j.1365-2966.2006.11386.x},
archivePrefix = {arXiv},
       eprint = {astro-ph/0612251},
 primaryClass = {astro-ph},
       adsurl = {https://ui.adsabs.harvard.edu/abs/2007MNRAS.375.1000B}
}

@ARTICLE{2023NatAs...7..444S,
       author = {{Stevance}, Heloise F. and {Eldridge}, Jan J. and {Stanway}, Elizabeth R. and {Lyman}, Joe and {McLeod}, Anna F. and {Levan}, Andrew J.},
        title = "{End-to-end study of the host galaxy and genealogy of the first binary neutron star merger}",
      journal = {Nature Astronomy},
         year = 2023,
        month = apr,
       volume = {7},
        pages = {444-450},
          doi = {10.1038/s41550-022-01873-y},
archivePrefix = {arXiv},
       eprint = {2301.05236},
 primaryClass = {astro-ph.HE},
       adsurl = {https://ui.adsabs.harvard.edu/abs/2023NatAs...7..444S}
}

@ARTICLE{2025arXiv250821805C,
       author = {{Chawla}, Chirag and {Chatterjee}, Sourav and {Breivik}, Katelyn},
        title = "{Gaia's promise to detect compact-object binaries: where we stand with the third data release}",
      journal = {arXiv e-prints},
         year = 2025,
        month = aug,
          eid = {arXiv:2508.21805},
        pages = {arXiv:2508.21805},
          doi = {10.48550/arXiv.2508.21805},
archivePrefix = {arXiv},
       eprint = {2508.21805},
 primaryClass = {astro-ph.SR},
       adsurl = {https://ui.adsabs.harvard.edu/abs/2025arXiv250821805C}
}

@ARTICLE{2023arXiv230616479G,
       author = {{Gandhi}, P. and {Dashwood Brown}, C. and {Zhao}, Y. and {El-Badry}, K. and {Maccarone}, T.~J. and {Knigge}, C. and {Anderson}, J. and {Middleton}, M. and {Miller-Jones}, J.~C.~A.},
        title = "{New Compact Object Binary Populations with Precision Astrometry (Roman White Paper)}",
      journal = {arXiv e-prints},
         year = 2023,
        month = jun,
          eid = {arXiv:2306.16479},
        pages = {arXiv:2306.16479},
          doi = {10.48550/arXiv.2306.16479},
archivePrefix = {arXiv},
       eprint = {2306.16479},
 primaryClass = {astro-ph.IM},
       adsurl = {https://ui.adsabs.harvard.edu/abs/2023arXiv230616479G}
}

@ARTICLE{2024arXiv240614767K,
       author = {{Kruszy{\'n}ska}, Katarzyna and {Street}, Rachel A. and {Gough-Kelly}, Steven and {Bonito}, Rosaria and {Prisinzano}, Loredana and {Trivedi}, Oem and {Gandhi}, Poshak and {Hundertmark}, Markus and {Tsapras}, Yiannis and {Di Criscienzo}, Marcella and et al.},
        title = "{Synergies between Roman Galactic Plane Survey and other major surveys}",
      journal = {arXiv e-prints},
         year = 2024,
        month = jun,
          eid = {arXiv:2406.14767},
        pages = {arXiv:2406.14767},
          doi = {10.48550/arXiv.2406.14767},
archivePrefix = {arXiv},
       eprint = {2406.14767},
 primaryClass = {astro-ph.GA},
       adsurl = {https://ui.adsabs.harvard.edu/abs/2024arXiv240614767K}
}

@ARTICLE{2024NewAR..9801694E,
       author = {{El-Badry}, Kareem},
        title = "{Gaia's binary star renaissance}",
      journal = {\nar},
         year = 2024,
        month = jun,
       volume = {98},
          eid = {101694},
        pages = {101694},
          doi = {10.1016/j.newar.2024.101694},
archivePrefix = {arXiv},
       eprint = {2403.12146},
 primaryClass = {astro-ph.SR},
       adsurl = {https://ui.adsabs.harvard.edu/abs/2024NewAR..9801694E}
}

@ARTICLE{2026PASP..138d4507M,
       author = {{McKinnon}, Kevin A. and {van der Marel}, Roeland P.},
        title = "{Simulating Roman+Gaia Combined Astrometry, Parallaxes, and Proper Motions}",
      journal = {PASP},
         year = 2026,
        month = apr,
       volume = {138},
       number = {4},
          eid = {044507},
        pages = {044507},
          doi = {10.1088/1538-3873/ae5a73},
archivePrefix = {arXiv},
       eprint = {2602.00310},
 primaryClass = {astro-ph.IM},
       adsurl = {https://ui.adsabs.harvard.edu/abs/2026PASP..138d4507M}
}

@ARTICLE{2012MNRAS.424.2265L,
       author = {{L{\"u}}, G.-L. and {Zhu}, C.-H. and {Postnov}, K.~A. and {Yungelson}, L.~R. and {Kuranov}, A.~G. and {Wang}, N.},
        title = "{Population synthesis for symbiotic X-ray binaries}",
      journal = {MNRAS},
         year = 2012,
        month = aug,
       volume = {424},
       number = {3},
        pages = {2265-2275},
          doi = {10.1111/j.1365-2966.2012.21395.x},
archivePrefix = {arXiv},
       eprint = {1205.5696},
 primaryClass = {astro-ph.SR},
       adsurl = {https://ui.adsabs.harvard.edu/abs/2012MNRAS.424.2265L}
}

@ARTICLE{2019MNRAS.485..851Y,
       author = {{Yungelson}, Lev R. and {Kuranov}, Alexandre G. and {Postnov}, Konstantin A.},
        title = "{Wind-accreting symbiotic X-ray binaries}",
      journal = {MNRAS},
         year = 2019,
        month = may,
       volume = {485},
       number = {1},
        pages = {851-860},
          doi = {10.1093/mnras/stz467},
archivePrefix = {arXiv},
       eprint = {1902.06060},
 primaryClass = {astro-ph.HE},
       adsurl = {https://ui.adsabs.harvard.edu/abs/2019MNRAS.485..851Y}
}

@ARTICLE{1977ApJ...211..866D,
       author = {{Davidsen}, A. and {Malina}, R. and {Bowyer}, S.},
        title = "{The optical counterpart of GX 1+4: a symbiotic star.}",
      journal = {ApJ},
         year = 1977,
        month = feb,
       volume = {211},
        pages = {866-871},
          doi = {10.1086/154996},
       adsurl = {https://ui.adsabs.harvard.edu/abs/1977ApJ...211..866D}
}

@ARTICLE{2024ApJ...977...95D,
       author = {{Deng}, Zhu-Ling and {Li}, Xiang-Dong},
        title = "{Are There Black Hole Symbiotic X-Ray Binaries?}",
      journal = {ApJ},
         year = 2024,
        month = dec,
       volume = {977},
       number = {1},
          eid = {95},
        pages = {95},
          doi = {10.3847/1538-4357/ad90ab},
archivePrefix = {arXiv},
       eprint = {2411.07548},
 primaryClass = {astro-ph.HE},
       adsurl = {https://ui.adsabs.harvard.edu/abs/2024ApJ...977...95D}
}

@ARTICLE{2000A&A...363..657E,
       author = {{Ergma}, E. and {Sarna}, M.~J.},
        title = "{The eclipsing binary millisecond pulsar PSR B1744-24A - possible test for a magnetic braking mechanism}",
      journal = {\aap},
         year = 2000,
        month = nov,
       volume = {363},
        pages = {657-659},
          doi = {10.48550/arXiv.astro-ph/0010262},
archivePrefix = {arXiv},
       eprint = {astro-ph/0010262},
 primaryClass = {astro-ph},
       adsurl = {https://ui.adsabs.harvard.edu/abs/2000A&A...363..657E}
}

@ARTICLE{2025ApJ...995...99Y,
       author = {{Yang}, Xing-Peng and {Chen}, Wen-Cong},
        title = "{Formation of Binary Millisecond Pulsars with Helium White Dwarfs in a New Magnetic Braking Prescription}",
      journal = {ApJ},
         year = 2025,
        month = dec,
       volume = {995},
       number = {1},
          eid = {99},
        pages = {99},
          doi = {10.3847/1538-4357/ae1ca5},
archivePrefix = {arXiv},
       eprint = {2511.05986},
 primaryClass = {astro-ph.HE},
       adsurl = {https://ui.adsabs.harvard.edu/abs/2025ApJ...995...99Y}
}

@ARTICLE{1988Natur.333..832P,
       author = {{Phinney}, E.~S. and {Evans}, C.~R. and {Blandford}, R.~D. and {Kulkarni}, S.~R.},
        title = "{Ablating dwarf model for eclipsing millisecond pulsar 1957 + 20}",
      journal = {\nat},
         year = 1988,
        month = jun,
       volume = {333},
       number = {6176},
        pages = {832-834},
          doi = {10.1038/333832a0},
       adsurl = {https://ui.adsabs.harvard.edu/abs/1988Natur.333..832P}
}

@ARTICLE{2017MNRAS.464..237S,
       author = {{Smedley}, Sarah L. and {Tout}, Christopher A. and {Ferrario}, Lilia and {Wickramasinghe}, Dayal T.},
        title = "{The implications of a companion enhanced wind on millisecond pulsar production}",
      journal = {MNRAS},
         year = 2017,
        month = jan,
       volume = {464},
       number = {1},
        pages = {237-245},
          doi = {10.1093/mnras/stw2333},
archivePrefix = {arXiv},
       eprint = {1709.08289},
 primaryClass = {astro-ph.SR},
       adsurl = {https://ui.adsabs.harvard.edu/abs/2017MNRAS.464..237S}
}

@ARTICLE{2015ApJ...806...92W,
       author = {{Wu}, Jianfeng and {Orosz}, Jerome A. and {McClintock}, Jeffrey E. and {Steeghs}, Danny and {Longa-Pe{\~n}a}, Pen{\'e}lope and {Callanan}, Paul J. and {Gou}, Lijun and {Ho}, Luis C. and {Jonker}, Peter G. and {Reynolds}, Mark T. and et al.},
        title = "{A Dynamical Study of the Black Hole X-Ray Binary Nova Muscae 1991}",
      journal = {ApJ},
         year = 2015,
        month = jun,
       volume = {806},
       number = {1},
          eid = {92},
        pages = {92},
          doi = {10.1088/0004-637X/806/1/92},
archivePrefix = {arXiv},
       eprint = {1501.00982},
 primaryClass = {astro-ph.HE},
       adsurl = {https://ui.adsabs.harvard.edu/abs/2015ApJ...806...92W}
}

@ARTICLE{2016ApJ...825...46W,
       author = {{Wu}, Jianfeng and {Orosz}, Jerome A. and {McClintock}, Jeffrey E. and {Hasan}, Imran and {Bailyn}, Charles D. and {Gou}, Lijun and {Chen}, Zihan},
        title = "{The Mass of the Black Hole in the X-ray Binary Nova Muscae 1991}",
      journal = {ApJ},
         year = 2016,
        month = jul,
       volume = {825},
       number = {1},
          eid = {46},
        pages = {46},
          doi = {10.3847/0004-637X/825/1/46},
archivePrefix = {arXiv},
       eprint = {1601.00616},
 primaryClass = {astro-ph.HE},
       adsurl = {https://ui.adsabs.harvard.edu/abs/2016ApJ...825...46W}
}

@ARTICLE{2022ApJ...925...83Z,
       author = {{Zheng}, Wan-Min and {Wu}, Qiaoya and {Wu}, Jianfeng and {Wang}, Song and {Sun}, Mouyuan and {Guo}, Jing and {Liu}, Junhui and {Yi}, Tuan and {Zhang}, Zhi-Xiang and {Gu}, Wei-Min and et al.},
        title = "{The Disk Veiling Effect of the Black Hole Low-mass X-Ray Binary A0620-00}",
      journal = {ApJ},
         year = 2022,
        month = jan,
       volume = {925},
       number = {1},
          eid = {83},
        pages = {83},
          doi = {10.3847/1538-4357/ac4332},
archivePrefix = {arXiv},
       eprint = {2112.07842},
 primaryClass = {astro-ph.HE},
       adsurl = {https://ui.adsabs.harvard.edu/abs/2022ApJ...925...83Z}
}

@ARTICLE{2007MNRAS.374..657R,
       author = {{Reynolds}, Mark T. and {Callanan}, Paul J. and {Filippenko}, Alexei V.},
        title = "{Keck infrared observations of GRO J0422+32 in quiescence}",
      journal = {MNRAS},
         year = 2007,
        month = jan,
       volume = {374},
       number = {2},
        pages = {657-663},
          doi = {10.1111/j.1365-2966.2006.11180.x},
archivePrefix = {arXiv},
       eprint = {astro-ph/0610272},
 primaryClass = {astro-ph},
       adsurl = {https://ui.adsabs.harvard.edu/abs/2007MNRAS.374..657R}
}

@ARTICLE{2013A&A...550A..35P,
       author = {{Pottasch}, S.~R. and {Bernard-Salas}, J.},
        title = "{Dust properties in the Galactic bulge}",
      journal = {\aap},
         year = 2013,
        month = feb,
       volume = {550},
          eid = {A35},
        pages = {A35},
          doi = {10.1051/0004-6361/201219647},
archivePrefix = {arXiv},
       eprint = {1301.3732},
 primaryClass = {astro-ph.GA},
       adsurl = {https://ui.adsabs.harvard.edu/abs/2013A&A...550A..35P}
}

@ARTICLE{2025AJ....170..328Z,
       author = {{Zelakiewicz}, Aiden S. and {Johnson}, Samson A. and {Gaudi}, B. Scott and {Bryden}, Geoffrey and {Nataf}, David M. and {Shvartzvald}, Yossi},
        title = "{A Near-infrared Extinction and Reddening Map toward the Galactic Bulge Using UKIRT}",
      journal = {AJ},
         year = 2025,
        month = dec,
       volume = {170},
       number = {6},
          eid = {328},
        pages = {328},
          doi = {10.3847/1538-3881/ae0c0b},
archivePrefix = {arXiv},
       eprint = {2509.25440},
 primaryClass = {astro-ph.GA},
       adsurl = {https://ui.adsabs.harvard.edu/abs/2025AJ....170..328Z}
}

@ARTICLE{2014A&A...566A.120S,
       author = {{Schultheis}, M. and {Chen}, B.~Q. and {Jiang}, B.~W. and {Gonzalez}, O.~A. and {Enokiya}, R. and {Fukui}, Y. and {Torii}, K. and {Rejkuba}, M. and {Minniti}, D.},
        title = "{Mapping the Milky Way bulge at high resolution: the 3D dust extinction, CO, and X factor maps}",
      journal = {\aap},
         year = 2014,
        month = jun,
       volume = {566},
          eid = {A120},
        pages = {A120},
          doi = {10.1051/0004-6361/201322788},
archivePrefix = {arXiv},
       eprint = {1405.0503},
 primaryClass = {astro-ph.GA},
       adsurl = {https://ui.adsabs.harvard.edu/abs/2014A&A...566A.120S}
}

@ARTICLE{2016PASA...33...24N,
       author = {{Nataf}, David M.},
        title = "{The Interstellar Extinction Towards the Milky Way Bulge with Planetary Nebulae, Red Clump, and RR Lyrae Stars}",
      journal = {\pasa},
         year = 2016,
        month = jun,
       volume = {33},
          eid = {e024},
        pages = {e024},
          doi = {10.1017/pasa.2016.16},
archivePrefix = {arXiv},
       eprint = {1603.06951},
 primaryClass = {astro-ph.SR},
       adsurl = {https://ui.adsabs.harvard.edu/abs/2016PASA...33...24N}
}

@ARTICLE{2024OJAp....7E..38E,
       author = {{El-Badry}, Kareem},
        title = "{On the formation of a 33 solar-mass black hole in a low-metallicity binary}",
      journal = {The Open Journal of Astrophysics},
         year = 2024,
        month = may,
       volume = {7},
          eid = {38},
        pages = {38},
          doi = {10.33232/001c.117652},
archivePrefix = {arXiv},
       eprint = {2404.13047},
 primaryClass = {astro-ph.SR},
       adsurl = {https://ui.adsabs.harvard.edu/abs/2024OJAp....7E..38E}
}

@ARTICLE{2023arXiv230612514L,
       author = {{Lam}, Casey Y. and {Abrams}, Natasha and {Andrews}, Jeff and {Bachelet}, Etienne and {Bahramian}, Arash and {Bennett}, David and {Bozza}, Valerio and {Broekgaarden}, Floor and {Chakrabarti}, Sukanya and {Dawson}, William and et al.},
        title = "{Roman CCS White Paper: Characterizing the Galactic population of isolated black holes}",
      journal = {arXiv e-prints},
         year = 2023,
        month = jun,
          eid = {arXiv:2306.12514},
        pages = {arXiv:2306.12514},
          doi = {10.48550/arXiv.2306.12514},
archivePrefix = {arXiv},
       eprint = {2306.12514},
 primaryClass = {astro-ph.IM},
       adsurl = {https://ui.adsabs.harvard.edu/abs/2023arXiv230612514L}
}

@ARTICLE{2023A&A...677A.185L,
       author = {{Luna}, Alonso and {Marchetti}, Tommaso and {Rejkuba}, Marina and {Minniti}, Dante},
        title = "{Astrometry in crowded fields towards the Galactic bulge}",
      journal = {\aap},
         year = 2023,
        month = sep,
       volume = {677},
          eid = {A185},
        pages = {A185},
          doi = {10.1051/0004-6361/202346257},
archivePrefix = {arXiv},
       eprint = {2307.13719},
 primaryClass = {astro-ph.GA},
       adsurl = {https://ui.adsabs.harvard.edu/abs/2023A&A...677A.185L}
}

@ARTICLE{2005MNRAS.360..974H,
       author = {{Hobbs}, G. and {Lorimer}, D.~R. and {Lyne}, A.~G. and {Kramer}, M.},
        title = "{A statistical study of 233 pulsar proper motions}",
      journal = {\mnras},
         year = 2005,
        month = jul,
       volume = {360},
       number = {3},
        pages = {974-992},
          doi = {10.1111/j.1365-2966.2005.09087.x},
archivePrefix = {arXiv},
       eprint = {astro-ph/0504584},
 primaryClass = {astro-ph},
       adsurl = {https://ui.adsabs.harvard.edu/abs/2005MNRAS.360..974H}
}

@ARTICLE{2000A&A...360..227N,
       author = {{Nugis}, T. and {Lamers}, H.~J.~G.~L.~M.},
        title = "{Mass-loss rates of Wolf-Rayet stars as a function of stellar parameters}",
      journal = {\aap},
         year = 2000,
        month = aug,
       volume = {360},
        pages = {227-244},
       adsurl = {https://ui.adsabs.harvard.edu/abs/2000A&A...360..227N}
}

@ARTICLE{2025A&A...697A..68G,
       author = {{Gagnier}, Damien and {Pejcha}, Ond{\v{r}}ej},
        title = "{Journey to the center of the common envelope evolution: Inner dynamics of the post-dynamical inspiral}",
      journal = {\aap},
         year = 2025,
        month = may,
       volume = {697},
          eid = {A68},
        pages = {A68},
          doi = {10.1051/0004-6361/202452616},
archivePrefix = {arXiv},
       eprint = {2412.04419},
 primaryClass = {astro-ph.SR},
       adsurl = {https://ui.adsabs.harvard.edu/abs/2025A&A...697A..68G}
}

@ARTICLE{2023MNRAS.526..740R,
       author = {{Rastello}, Sara and {Iorio}, Giuliano and {Mapelli}, Michela and {Arca-Sedda}, Manuel and {Di Carlo}, Ugo N. and {Escobar}, Gast{\'o}n J. and {Shenar}, Tomer and {Torniamenti}, Stefano},
        title = "{Dynamical formation of Gaia BH1 in a young star cluster}",
      journal = {\mnras},
         year = 2023,
        month = nov,
       volume = {526},
       number = {1},
        pages = {740-749},
          doi = {10.1093/mnras/stad2757},
archivePrefix = {arXiv},
       eprint = {2306.14679},
 primaryClass = {astro-ph.SR},
       adsurl = {https://ui.adsabs.harvard.edu/abs/2023MNRAS.526..740R}
}

@ARTICLE{2024MNRAS.527.4031T,
       author = {{Tanikawa}, Ataru and {Cary}, Savannah and {Shikauchi}, Minori and {Wang}, Long and {Fujii}, Michiko S.},
        title = "{Compact binary formation in open star clusters - I. High formation efficiency of Gaia BHs and their multiplicities}",
      journal = {\mnras},
         year = 2024,
        month = jan,
       volume = {527},
       number = {2},
        pages = {4031-4039},
          doi = {10.1093/mnras/stad3294},
archivePrefix = {arXiv},
       eprint = {2303.05743},
 primaryClass = {astro-ph.GA},
       adsurl = {https://ui.adsabs.harvard.edu/abs/2024MNRAS.527.4031T}
}

@ARTICLE{2009ApJ...697.1057F,
       author = {{Fragos}, T. and {Willems}, B. and {Kalogera}, V. and {Ivanova}, N. and {Rockefeller}, G. and {Fryer}, C.~L. and {Young}, P.~A.},
        title = "{Understanding Compact Object Formation and Natal Kicks. II. The Case of XTE J1118 + 480}",
      journal = {\apj},
         year = 2009,
        month = jun,
       volume = {697},
       number = {2},
        pages = {1057-1070},
          doi = {10.1088/0004-637X/697/2/1057},
archivePrefix = {arXiv},
       eprint = {0809.1588},
 primaryClass = {astro-ph},
       adsurl = {https://ui.adsabs.harvard.edu/abs/2009ApJ...697.1057F}
}

@ARTICLE{1999A&A...352L..87N,
       author = {{Nelemans}, G. and {Tauris}, T.~M. and {van den Heuvel}, E.~P.~J.},
        title = "{Constraints on mass ejection in black hole formation derived from black hole X-ray binaries}",
      journal = {\aap},
         year = 1999,
        month = dec,
       volume = {352},
        pages = {L87-L90},
          doi = {10.48550/arXiv.astro-ph/9911054},
archivePrefix = {arXiv},
       eprint = {astro-ph/9911054},
 primaryClass = {astro-ph},
       adsurl = {https://ui.adsabs.harvard.edu/abs/1999A&A...352L..87N}
}

@ARTICLE{2005ApJ...625..324W,
       author = {{Willems}, B. and {Henninger}, M. and {Levin}, T. and {Ivanova}, N. and {Kalogera}, V. and {McGhee}, K. and {Timmes}, F.~X. and {Fryer}, C.~L.},
        title = "{Understanding Compact Object Formation and Natal Kicks. I. Calculation Methods and the Case of GRO J1655-40}",
      journal = {\apj},
         year = 2005,
        month = may,
       volume = {625},
       number = {1},
        pages = {324-346},
          doi = {10.1086/429557},
archivePrefix = {arXiv},
       eprint = {astro-ph/0411423},
 primaryClass = {astro-ph},
       adsurl = {https://ui.adsabs.harvard.edu/abs/2005ApJ...625..324W}
}

@ARTICLE{2023MNRAS.522.1763G,
       author = {{Generozov}, A. and {Perets}, H.~B.},
        title = "{Capture of stars into gaseous discs around massive black holes: alignment, circularization, and growth}",
      journal = {\mnras},
         year = 2023,
        month = jun,
       volume = {522},
       number = {2},
        pages = {1763-1778},
          doi = {10.1093/mnras/stad1016},
archivePrefix = {arXiv},
       eprint = {2212.11301},
 primaryClass = {astro-ph.GA},
       adsurl = {https://ui.adsabs.harvard.edu/abs/2023MNRAS.522.1763G}
}

@ARTICLE{1976ApJ...204..555S,
       author = {{Shapiro}, S.~L. and {Lightman}, A.~P.},
        title = "{Black holes in X-ray binaries: marginal existence and rotation reversals of accretion disks.}",
      journal = {\apj},
         year = 1976,
        month = mar,
       volume = {204},
        pages = {555-560},
          doi = {10.1086/154203},
       adsurl = {https://ui.adsabs.harvard.edu/abs/1976ApJ...204..555S}
}

@ARTICLE{2026arXiv260806453E,
       author = {{El-Badry}, Kareem and {Latham}, David W. and {Rix}, Hans-Walter and {Bieryla}, Allyson and {Buchhave}, Lars A. and {Shahaf}, Sahar and {Mazeh}, Tsevi and {M{\"u}ller-Horn}, Johanna and {Nagarajan}, Pranav and {Yamaguchi}, Natsuko and et al.},
        title = "{Spectroscopic follow-up of compact object binary candidates from Gaia DR3: White dwarfs, neutron stars, black holes, and the parallax zeropoint}",
      journal = {arXiv e-prints},
         year = 2026,
        month = aug,
          eid = {arXiv:2608.06453},
        pages = {arXiv:2608.06453},
          doi = {10.48550/arXiv.2608.06453},
archivePrefix = {arXiv},
       eprint = {2608.06453},
 primaryClass = {astro-ph.SR},
       adsurl = {https://ui.adsabs.harvard.edu/abs/2026arXiv260806453E}
}

\clearpage
\onecolumn
\begin{appendix}

% Override the class's appendix numbering completely
\makeatletter
\renewcommand{\thesection}{\Alph{section}}
\setcounter{section}{1}  % Set to 1 so \thesection becomes "A"
\renewcommand{\thefigure}{\thesection\arabic{figure}}
\renewcommand{\thetable}{\thesection\arabic{table}}
\setcounter{figure}{0}
\setcounter{table}{0}
\makeatother

% Use section* to hide the "A" in the title, but keep the counter
\section*{Appendix}
\addcontentsline{toc}{section}{Appendix}
% \noindent Supplementary tables and figures are presented below.

% test
% \clearpage
\begin{table}[H]
    \centering 

    \begin{tabular}{c|c|c|c|c|c|c}
        \hline
        \multicolumn{1}{c|}{Remnant} &  \multicolumn{3}{c|}{Optimistic} & \multicolumn{3}{c}{Pessimistic} \\
        \hline
         &  DR3 & DR5 & GBTDS & DR3 & DR5 & GBTDS\\
         \hline
        Pre-interaction BH &  $323\pm61$ & $623\pm64$ & $0$ & $94\pm29$ & $295\pm51$ & $0$\\
        Pre-interaction NS &  $3070\pm405$ & $8237\pm806$ & $587\pm50$ & $579\pm90$ & $2077\pm254$ & $185\pm21$\\
        Post-interaction BH &  $248\pm52$ & $422\pm79$ & $32\pm6$ & $99\pm27$ & $182\pm42$ & $19\pm4$\\
        Post-interaction NS & $50\pm10$ & $91\pm22$ & $10\pm3$ & $13\pm4$ & $35\pm8$ & $5\pm2$\\
        X-ray faint BH &  $76\pm4$ & $217\pm13$ & $4\pm2$ & $10\pm2$ & $59\pm3$ & $2\pm1$\\
        X-ray faint NS &  $2\pm1$ & $10\pm3$ & $6\pm3$ & $0$ & $0$ & $2\pm1$\\
        \hline
        
    \end{tabular}
    \vspace{0.15cm} 
    \caption{Results for the number of detectable quiescent binary systems from each survey. The population is the total number of pre-interaction, post-interaction, and X-ray faint RLOF systems that have a period less than $10.5$ years. For the surveys themselves, the results are the mean of across the 16 positions we surveyed in our simulation. The optimistic population where the wobble is greater than $\sigma_{ast}$ and the pessimistic population is where the wobble is greater than $\rm{3\times\sigma_{ast}}$.}
    \label{tab:appendixgaiaromantable} 
\end{table}
% \clearpage
\begin{figure}[H]
\includegraphics[width=1\columnwidth]{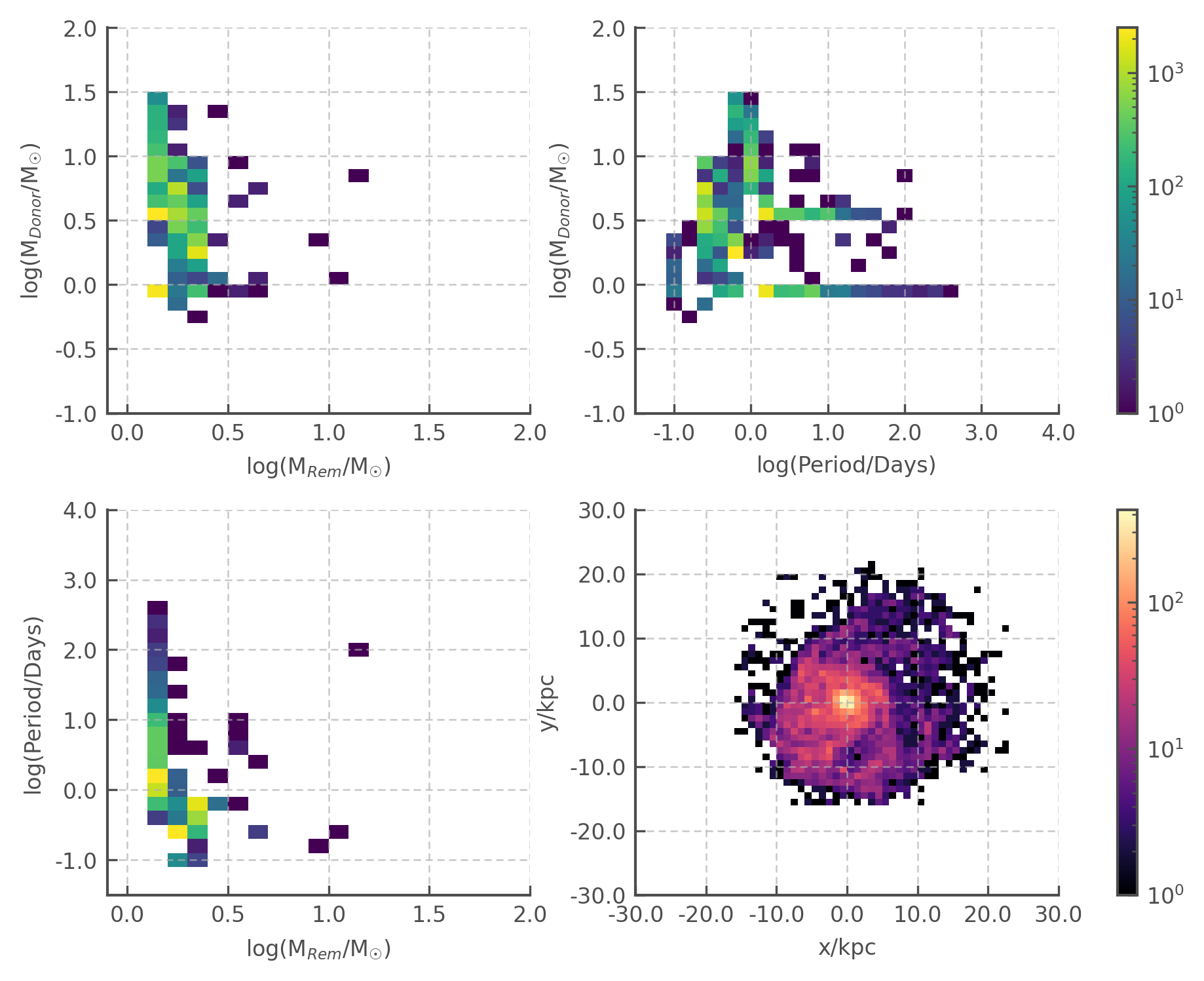}
\caption{CEE: These are systems where the donor radius is greater than or equal to the separation of the two objects masses. With a lack of Galactic observational data, we present the \textsc{BPASS} predictions alone.}
\label{fig:CEE}
\end{figure}

\begin{figure}
\includegraphics[width=1\columnwidth]{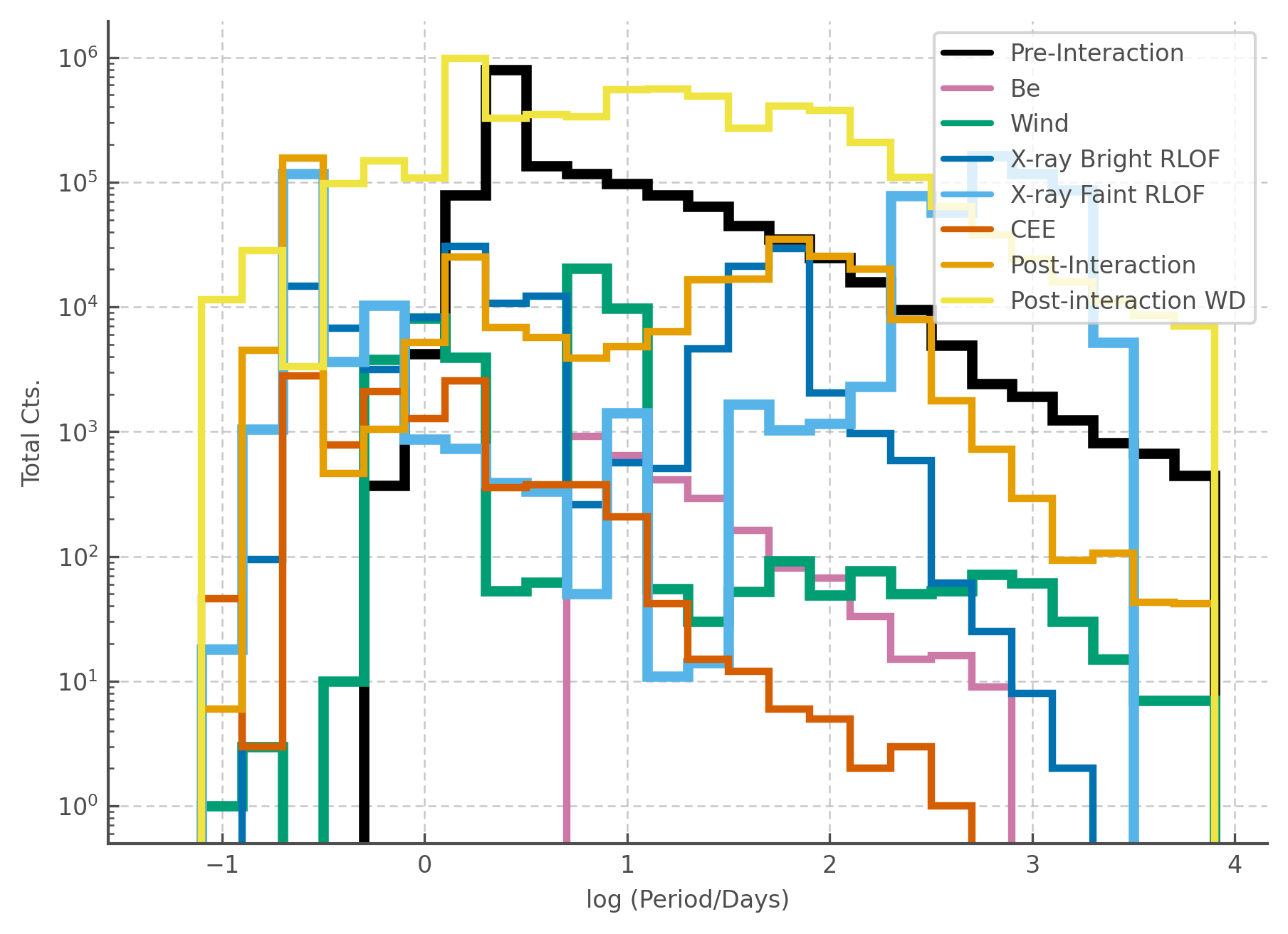}
\caption{Period distribution of the \textsc{BPASS} synthetic populations.}
\label{fig:BPASSperiod}
\end{figure}

\begin{figure}
\includegraphics[width=1\columnwidth]{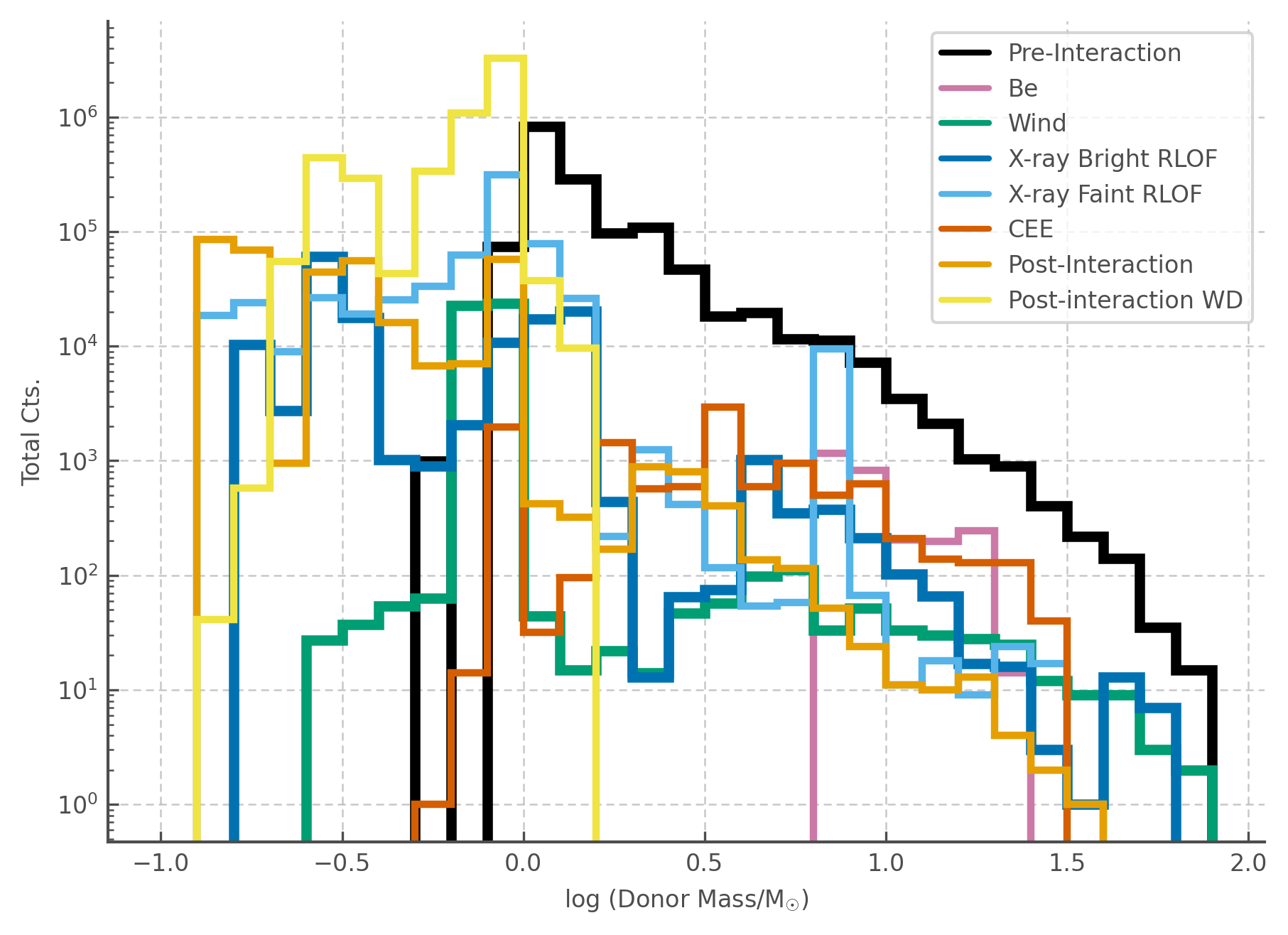}
\caption{Donor mass distribution of the \textsc{BPASS} synthetic populations.}
\label{fig:BPASSdonor}
\end{figure}

\begin{figure}
\includegraphics[width=1\columnwidth]{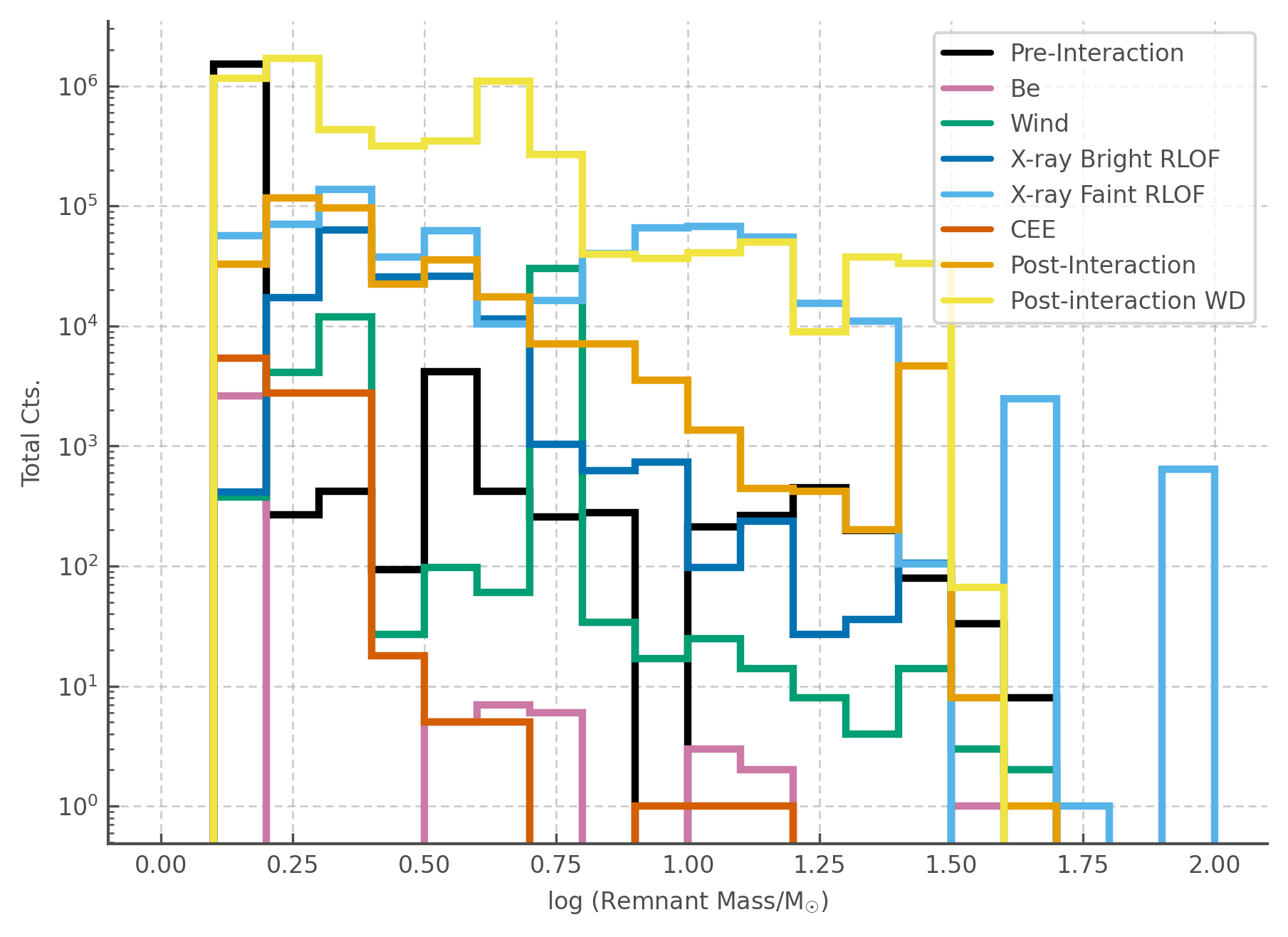}
\caption{Remnant mass distribution of the \textsc{BPASS} synthetic populations.}
\label{fig:BPASSdem}
\end{figure}

\begin{figure}
\includegraphics[width=1\columnwidth]{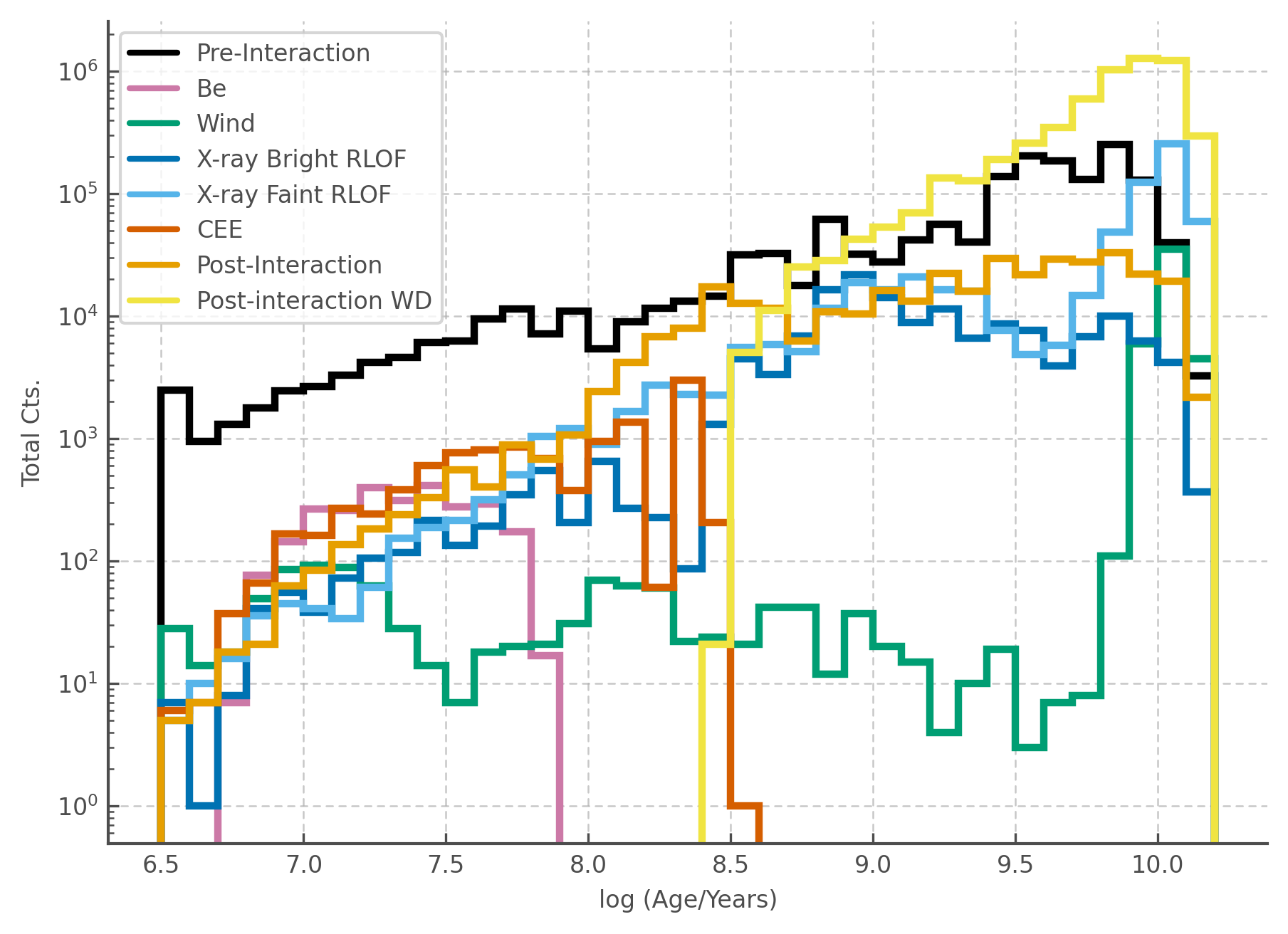}
\caption{Age distribution of the \textsc{BPASS} synthetic populations.}
\label{fig:bpassage}
\end{figure}

\begin{figure}
\includegraphics[width=1\columnwidth]{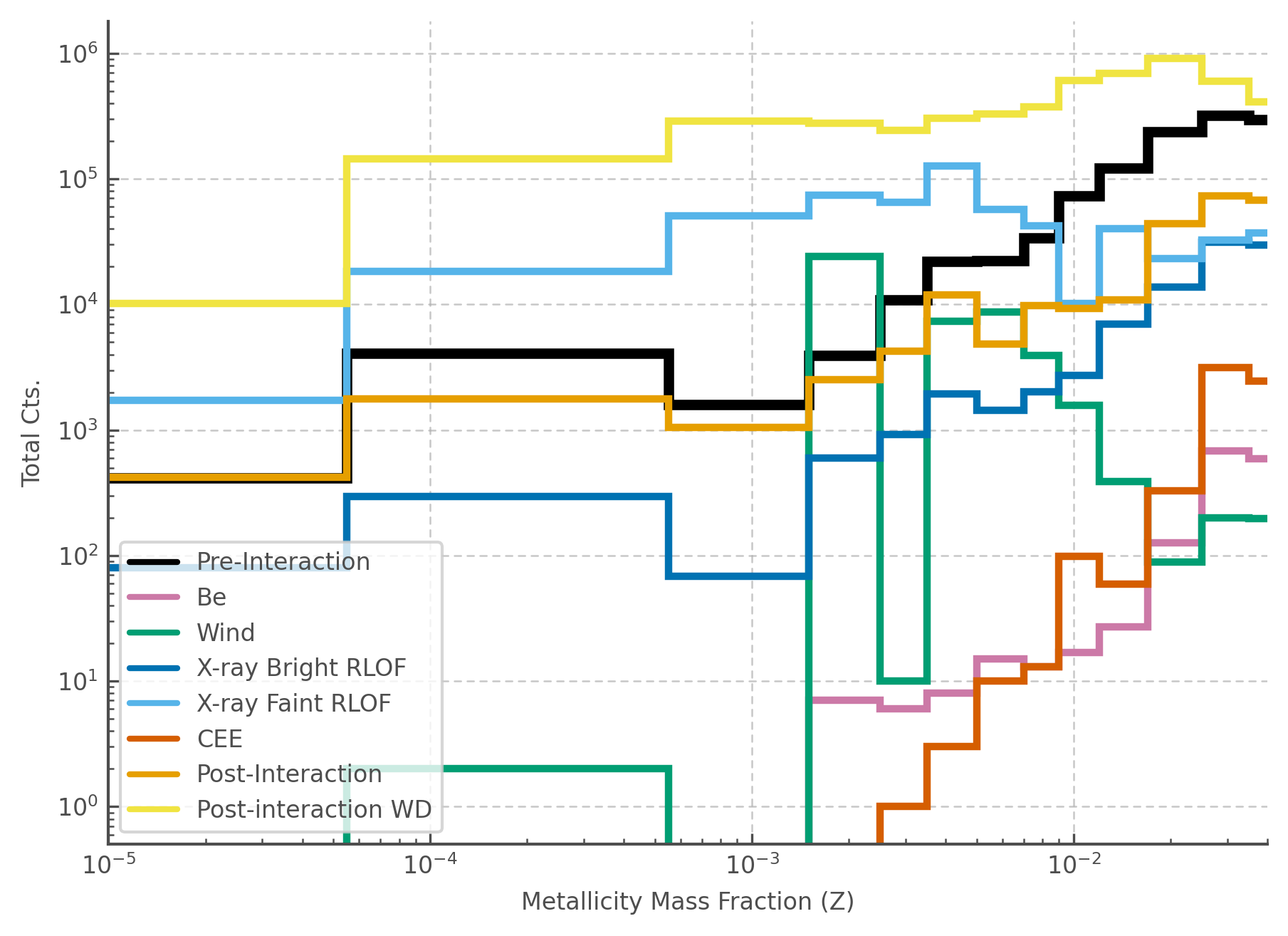}
\caption{Metallicity distribution of the \textsc{BPASS} synthetic populations.}
\label{fig:bpassz}
\end{figure}

\begin{figure}
\includegraphics[width=1\columnwidth]{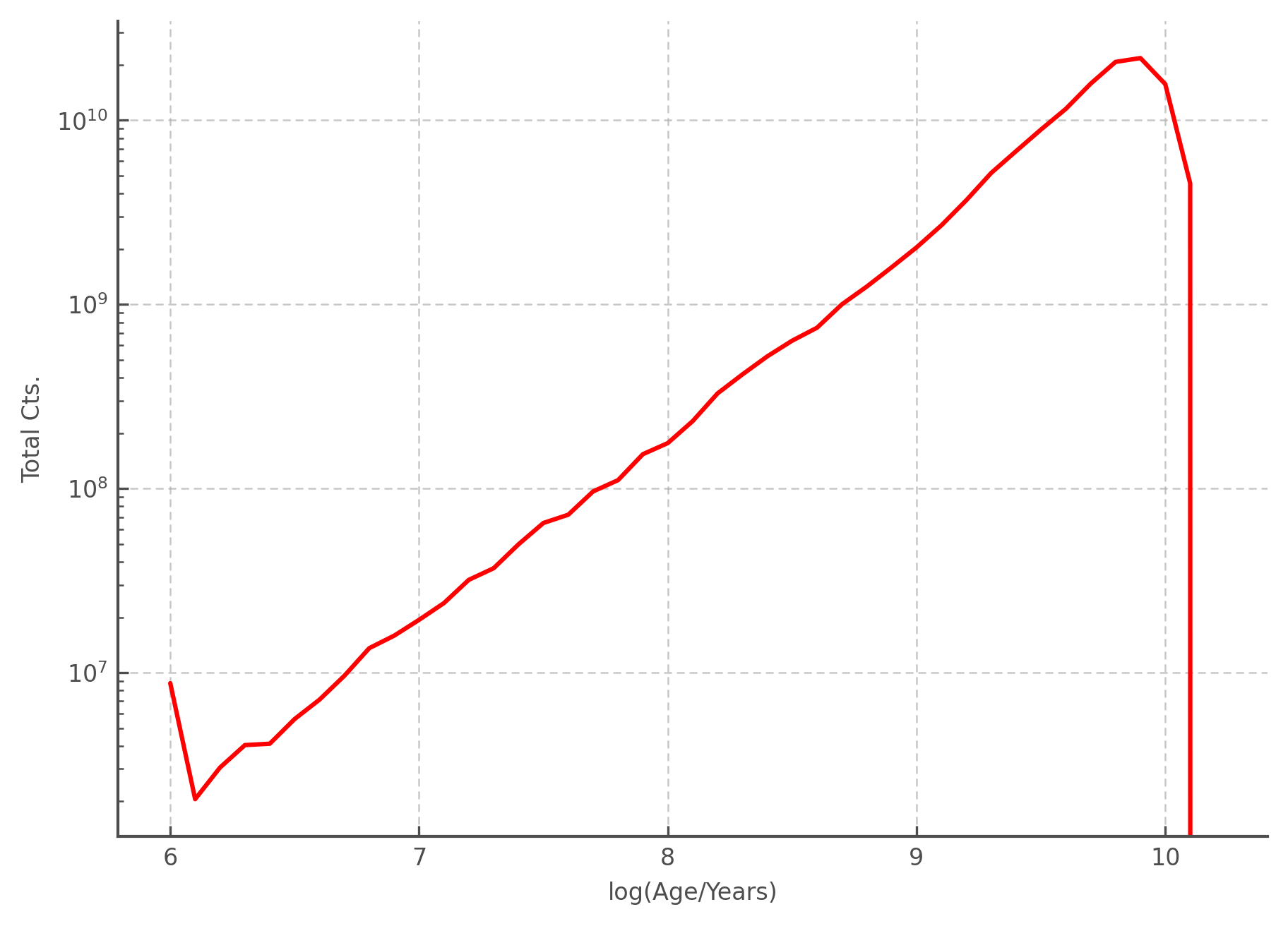}
\caption{Age distribution within the \textsc{FIRE} \textsc{m12i} simulation alone.}
\label{fig:fireage}
\end{figure}

\begin{figure}
\includegraphics[width=1\columnwidth]{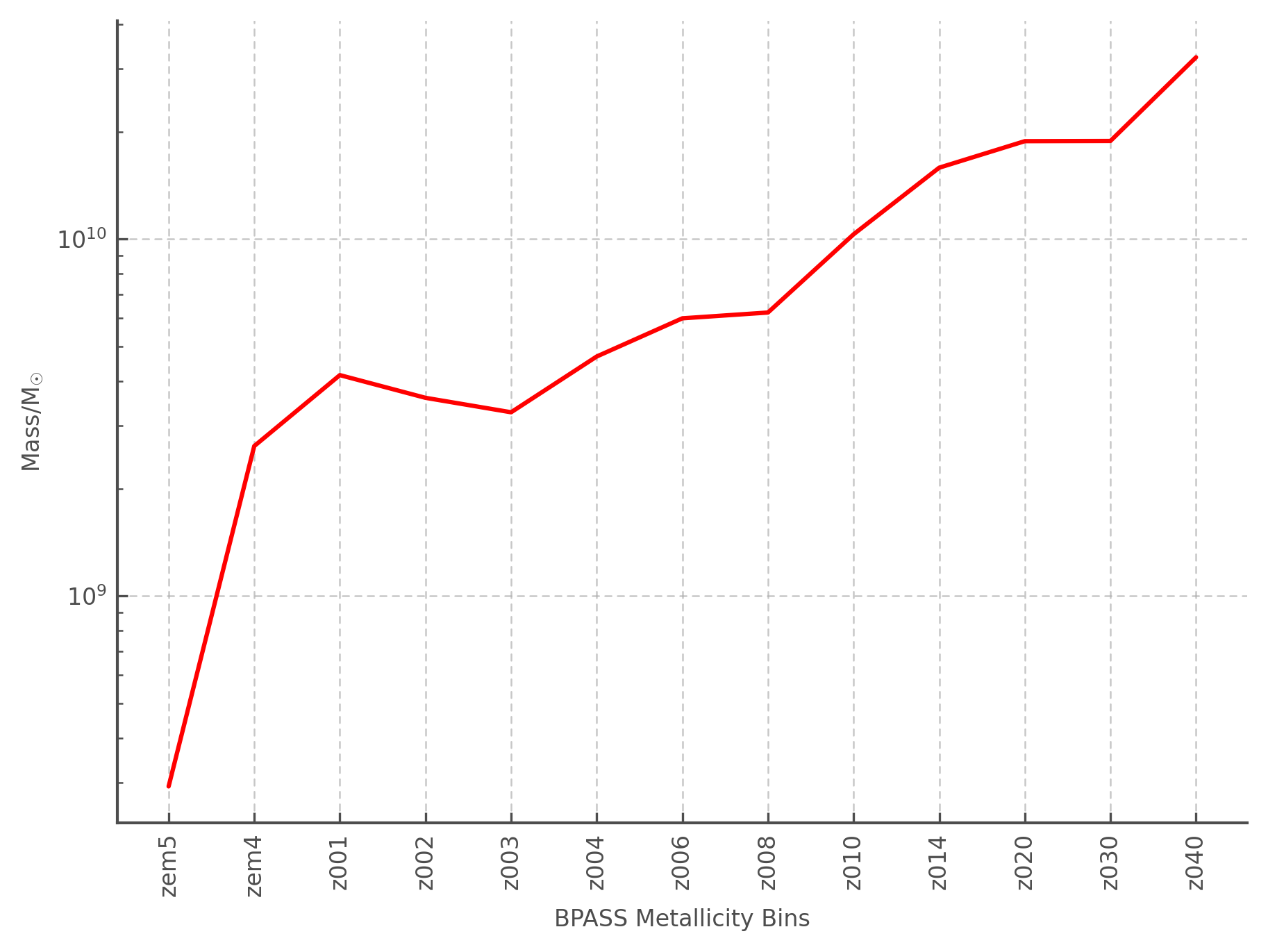}
\caption{Metallicity distribution within the \textsc{FIRE} \textsc{m12i} simulation alone.}
\label{fig:firez}
\end{figure}
\end{appendix}
\end{document}